\documentclass{aa}

\usepackage{natbib}
\bibpunct{(}{)}{;}{a}{}{,} 

\usepackage{latexsym,exscale,amssymb,amsmath}
\usepackage{eurosym}
\usepackage{color}
\usepackage{graphicx}
\usepackage[percent]{overpic}
\usepackage{hyperref}

\title{\bf [C\,{\sc ii}] $158\,\mu\mathrm{m}$ line emission from Orion A.\\ II. Photodissociation region physics}
\titlerunning{[C\,{\sc ii}] $158\,\mu\mathrm{m}$ line emission from Orion A II}
\author{C. H. M. Pabst\inst{\ref{inst1}} \and J. R. Goicoechea\inst{\ref{inst2}} \and A. Hacar\inst{\ref{inst1},\ref{inst11}} \and D. Teyssier\inst{\ref{inst3}} \and O. Bern\'{e}\inst{\ref{inst4}} \and M. G. Wolfire\inst{\ref{inst5}} \and R. D. Higgins\inst{\ref{inst6}} \and E. T. Chambers\inst{\ref{inst7}} \and S. Kabanovic\inst{\ref{inst6}} \and R. G\"{u}sten\inst{\ref{inst8}} \and J. Stutzki\inst{\ref{inst6}} \and C. Kramer\inst{\ref{inst9}} \and A. G. G. M. Tielens\inst{\ref{inst1},\ref{inst5}} }

\institute{Leiden Observatory, Leiden University, Niels Bohrweg 2, 2333 CA Leiden, Netherlands\label{inst1}; \href{mailto:pabst@strw.leidenuniv.nl}{\texttt{pabst@strw.leidenuniv.nl}}
\and Instituto de F\'{\i}sica Fundamental, CSIC, Calle Serrano 121-123, 28006 Madrid, Spain \label{inst2} 
\and University of Vienna, Department of Astrophysics, T\"{u}rkenschanzstrasse 17, 1180 Vienna, Austria\label{inst11}
\and Telespazio Vega UK Ltd. for ESA/ESAC, Urbanizacion Villafranca del Castillo, 28691 Madrid, Spain \label{inst3}
\and IRAP, Universit\'{e} de Toulouse, CNRS, CNES, UPS, 9 Av. colonel Roche, 31028 Toulouse Cedex 4, France \label{inst4}
\and Department of Astronomy, University of Maryland, College Park, MD 20742, USA\label{inst5}
\and I. Physikalisches Institut der Universit\"{a}t zu K\"{o}ln, Z\"{u}lpicher Strasse 77, 50937 K\"{o}ln, Germany\label{inst6}
\and USRA/SOFIA, NASA Ames Research Center, Mail Stop 232-12, Building N232, P.O. Box 1, Moffett Field, CA 94035-0001, USA\label{inst7}
\and Max-Planck-Institut f\"{u}r Radioastronomie, Auf dem H\"{u}gel 69, 53121 Bonn, Germany\label{inst8}
\and Institut de Radioastronomie Millim\'etrique, 300 rue de la Piscine, 38406 Saint Martin d'H\`{e}res, France\label{inst9}
}
\date{Received 13 March 2021, Accepted 22 November 2021}

\abstract{The [C\,{\sc ii}] $158\,\mu\mathrm{m}$ fine-structure line is the dominant cooling line of moderate-density photodissociation regions (PDRs) illuminated by moderately bright far-ultraviolet (FUV) radiation fields. This makes this line a prime diagnostic for extended regions illuminated by massive stars.}{We aim to understand the origin of [C\,{\sc ii}] emission and its relation to other tracers of gas and dust in PDRs. One focus is a study of the heating efficiency of interstellar gas as traced by the [C\,{\sc ii}] line to test models of the photoelectric heating of neutral gas by polycyclic aromatic hydrocarbon (PAH) molecules and very small grains.}{We make use of a one-square-degree map of velocity-resolved [C\,{\sc ii}] line emission toward the Orion Nebula complex, and split this out into the individual spatial components, the expanding Veil Shell, the surface of OMC4, and the PDRs associated with the compact H\,{\sc ii} region of M43 and the reflection nebula NGC 1977. We employed {\it Herschel} far-infrared photometric images to determine dust properties. Moreover, we compared with {\it Spitzer} mid-infrared photometry to trace hot dust and large molecules, and velocity-resolved IRAM 30m CO(2-1) observations of the molecular gas.}{The [C\,{\sc ii}] intensity is tightly correlated with PAH emission in the IRAC $8\,\mu\mathrm{m}$ band and far-infrared emission from warm dust, with small variations between the four studied subregions (Veil Shell, OMC4, M43, and NGC 1977). The correlation between [C\,{\sc ii}] and CO(2-1) is very different in the four subregions and is very sensitive to the detailed geometry of the respective regions. Constant-density PDR models are able to reproduce the observed [C\,{\sc ii}], CO(2-1), and integrated far-infrared (FIR) intensities. The physical conditions in the Veil Shell of the Orion Nebula, M43, and NGC 1977 reveal a constant ratio of thermal pressure $p_{\mathrm{th}}$ over incident FUV radiation field measured by $G_0$. We observe strong variations in the photoelectric heating efficiency in the Veil Shell behind the Orion Bar and these variations are seemingly not related to the spectral properties of the PAHs.}{The [C\,{\sc ii}] emission from the Orion Nebula complex stems mainly from moderately illuminated PDR surfaces. The correlations of the different tracers ([C\,{\sc ii}], FIR, CO, $70\,\mu\mathrm{m}$, and $8\,\mu\mathrm{m}$ emission) show small variations that are not yet understood. Future observations with the James Webb Space Telescope can shine light on the PAH properties that may be linked to these variations.}

\vspace{0.5cm}

\begin{document}

\maketitle

\section{Introduction}

The [C\,{\sc ii}] $158\,\mu\mathrm{m}$ fine-structure line is the dominant cooling line of neutral atomic gas at moderate densities ($n_{\rm H}\lesssim 10^4\,\mathrm{cm^{-3}}$) and temperatures \citep[$T_{\rm k} \lesssim 300\,\mathrm{K}$;][]{DalgarnoMcCray1972}. This includes diffuse clouds in the general interstellar medium (ISM) of galaxies as well as regions illuminated by nearby, newly formed, massive stars \citep{HollenbachTielens1999}. In these regions carbon atoms are rapidly ionized by penetrating far-ultraviolet (FUV) photons \mbox{$E<13.6$\,eV)} and molecules such as CO are quickly photodissociated. Hence, most of the gaseous carbon is in the form of C$^+$ ions. The ground state of C$^+$  has two fine structure levels separated by $\Delta E/k_{\rm B}=91.2\,\mathrm{K}$. This transition has a critical density of $(2\text{-}6)\times10^3\,\mathrm{cm^{-3}}$ for collisions with atomic or molecular hydrogen. Since the upper fine-structure level \mbox{$^{2}P_{3/2}$} is easy to excite collisionally at typical ISM densities and as the interstellar carbon abundance is high, the \mbox{[C\,{\sc ii}] $158\,\mu\mathrm{m}$} line dominates the cooling of the neutral ISM.
 
Extreme ultraviolet (EUV) radiation \mbox{$E>13.6$\,eV)} from massive stars ionize hydrogen atoms and create an H\,{\sc ii}~region in their environment. This ionized gas is separated from the surrounding nascent molecular cloud by a photodissociation region (PDR) where penetrating FUV photons dissociate molecules and ionize low ionization potential atoms. These FUV photons ultimately heat the PDR gas to temperatures of 200-1000\,K. The neutral gas (i.e., hydrogen in neutral form) couples to the FUV photon field through the photoelectric effect on large polycyclic aromatic hydrocarbon molecules (PAHs) and very small grains \citep[VSGs;][]{BakesTielens1994}. This same process heats the gas in diffuse interstellar clouds \citep{Wolfire1995}. Dense PDRs in star-forming regions are bright in the far-infrared (FIR) atomic cooling lines, e.g. the [C\,{\sc ii}] $158\,\mu\mathrm{m}$ line at moderate densities and warmish temperatures \citep{TielensHollenbach1985, HollenbachTT1991}, as well as the PAH emission features \citep[][and references therein]{Tielens2008}, fluorescent ro-vibrational lines \citep[e.g.,][]{Sellgren1986, Field1998, Kaplan2017}, pure rotational H$_2$ lines \citep[e.g.,][]{Sheffer2013, Habart2011, Allers2005}, warm dust \citep[e.g.,][]{Berne2007,Arab2012}, and a plethora of molecular radicals and reactive ions \citep[e.g.,][]{Pety2005, Fuente2003, Nagy2013, Goicoechea2017}. Their high surface brightness and compact size make dense PDRs very appropriate targets for observational studies on the interaction of massive stars with their interstellar environment. This allows detailed studies of the physics and chemistry relevant to radiative feedback by massive stars. In particular, observations of the dominant atomic fine structure lines, through the energy balance, provide a direct probe of the heating processes of interstellar neutral atomic gas \citep[e.g.,][]{Tielens2008, Okada2013, Pabst2017, Salas2019}.

The [C\,{\sc ii}] $158\,\mu\mathrm{m}$ line is the brightest FIR line in the spectrum of the Milky Way and, in general, of star-forming galaxies, typically radiating about 0.3\% of the IR dust continuum emission \citep[e.g.,][]{Crawford1985, Bennett1994, Malhotra2001, Luhman2003, Stacey2010, Diaz-Santos2013}. As ionization of carbon requires FUV photons with energies in excess of 11.2 eV, and such photons are only emitted by short-lived massive stars, the [C\,{\sc ii}] emission line is also considered a tracer of the star formation rate \citep[SFR; e.g.,][]{Pineda2014, DeLooze2011, Herrera-Camus2015, Herrera-Camus2018b}. With ALMA and NOEMA, the use of the [C\,{\sc ii}] line as a SFR tracer has now been extended to the high redshift universe \citep[e.g.,][]{Walter2012, Venemans2012, Knudsen2016, Bischetti2018, Khusanova2020}. The use of the [C\,{\sc ii}] $158\,\mu\mathrm{m}$ line as a SFR indicator requires a reliable conversion factor and extensive observational studies have focused on determining this factor through detailed studies of the [C\,{\sc ii}] line in samples of nearby star forming galaxies \citep[e.g.,][]{Malhotra2001, Herrera-Camus2015, Herrera-Camus2018b, Chevance2016, Pineda2018}. These observational studies have revealed a so-called C$^+$-deficit issue: the [C\,{\sc ii}]/FIR luminosity ratio is systematically lower in regions characterized by warm dust, possibly due to the onset of other cooling processes of neutral dense gas (e.g., the [O\,{\sc i}] $63\,\mu\mathrm{m}$ line), changes in the coupling between the neutral atomic gas and the FUV photons, reduced gas-heating efficiency when the strong FIR field leads to positively charged PAHs and small grains making it more difficult for photoelectrons to escape and heat the gas, or the presence of energy sources other than FUV photons \citep[e.g., AGNs or deeply embedded star formation;][]{Luhman2003, Abel2009, Gracia2011}. Aside from the photoelectric effect heating the gas H$_2$ formation heating and collisional de-excitation of vibrationally excited H$_2$ may play a role \citep[e.g.,][]{Roellig2006}. Besides this observational validation of the use of the [C\,{\sc ii}] line as a SFR indicator, we also need to develop a deep understanding of the underlying physics involved in the heating and cooling of neutral atomic gas if we want to confidently extrapolate the local results over cosmological relevant timescales.

The recent rapid development of receiver technology has allowed the construction of multi-element heterodyne arrays at FIR frequencies. In particular, the upGREAT instrument has been specifically designed to map the [C\,{\sc ii}] $158\,\mu\mathrm{m}$ line \citep{Risacher2016}. This sensitive instrument coupled with the nimble telescope of the Stratospheric Observatory For Infrared Astronomy (SOFIA) allows for the first time wide-field  [C\,{\sc ii}] observations of regions of massive star formation at sub-km\,s$^{-1}$ spectral resolution \citep{Pabst2017, Bally2018}. The C+SQUAD SOFIA Large Program has surveyed one square degree of the molecular cloud cores in Orion  in the [C\,{\sc ii}] $158\,\mu\mathrm{m}$ line to probe the radiative and mechanical energy feedback of  massive stars with their environment. The results revealed the presence of a $\sim 2$\,pc radius massive ($\sim 1500\,M_{\odot}$) shell of neutral atomic gas expanding at 13\,km\,s$^{-1}$ \citep{Pabst2019}. The rapid expansion of this bubble is driven by the hot plasma generated by the stellar wind from the O7V star $\theta^1$ Ori C \citep{Guedel2008} as envisioned by \citet{Weaver1977}. The data also revealed bubbles of neutral gas expanding at a slower pace around the B1 stars powering the H\,{\sc ii} regions M43 and NGC 1977 \citep{Pabst2020}. In this case, expansion is driven by the overpressure of the ionized gas, a so-called Spitzer expansion. While these studies focused on the kinematics of the region and the mechanical feedback by massive stars, this data also allows an in depth study of the radiative interaction of massive stars with neutral atomic gas on unprecedented spatial scales.

In Paper I \citep{Pabst2021}, we used the C+SQUAD results to study the global characteristics of the [C\,{\sc ii}] emission in Orion as a template to understand the use of the [C\,{\sc ii}] emission as a SFR indicator and to investigate the [C\,{\sc ii}] deficit. In this paper, we separate the surveyed region in its spatial components, compare the observations with specific PDR models, and examine the radiative interaction of the powering stars with their environment in detail. This paper is organized as follows: In Section 2, we summarize the observations used in this study. Section 3 discusses the global morphology of the [C\,{\sc ii}] line emission and presents the correlations we find of the [C\,{\sc ii}] emission with other tracers of gas and dust in the Orion Nebula (M42), M43, and NGC 1977, separately. In Section 4, we compare the observations to PDR models, discuss the implications thereof, and examine the heating efficiency of the [C\,{\sc ii}]-emitting gas. We summarize our results in Section 5.

\begin{figure}[tb]
\includegraphics[width=0.5\textwidth, height=0.5\textwidth]{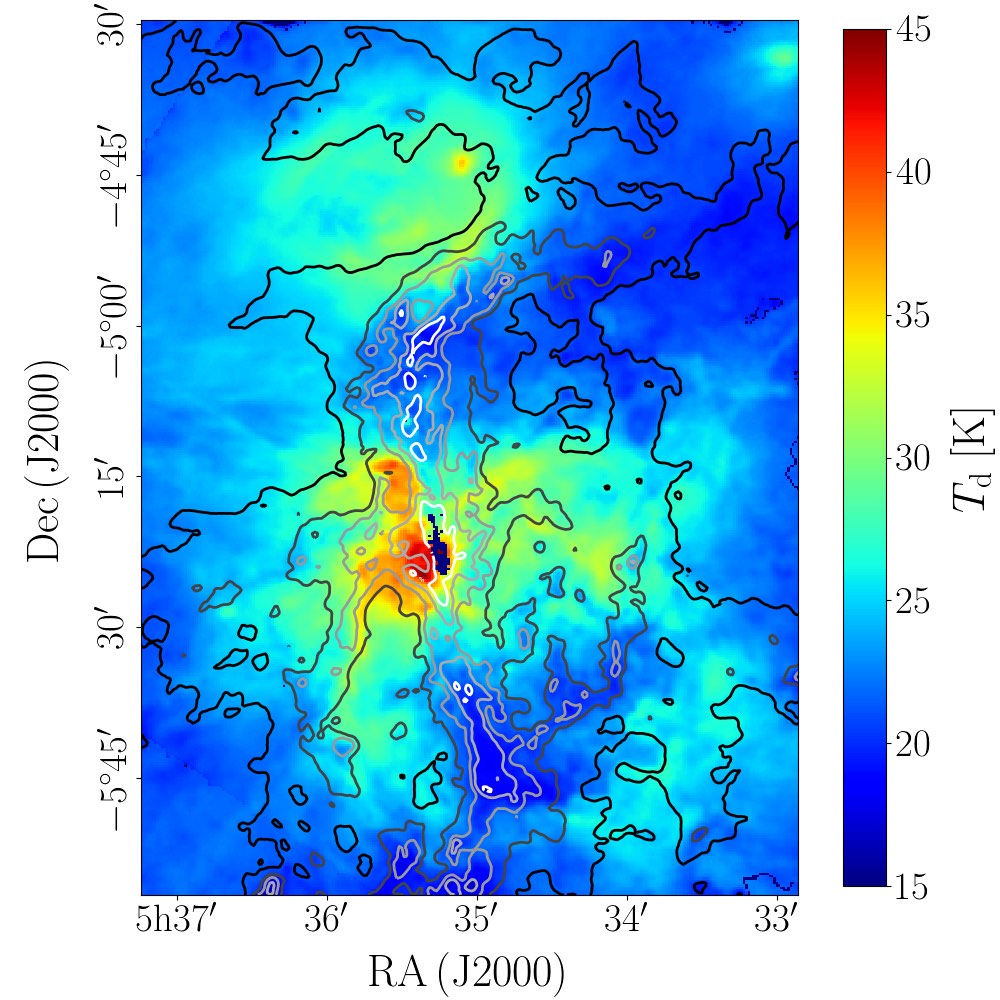}
\caption{Dust temperature (color scale) and optical depth at 160\,$\mu$m (contours, from black to white: $3\times 10^{-3}$, $5\times 10^{-3}$, $1\times 10^{-2}$, $3\times 10^{-2}$, $5\times 10^{-2}$) from SED fit ($\beta=2$).}
\label{Fig.T_tau}
\end{figure}


\begin{figure}[ht]
\includegraphics[width=0.5\textwidth, height=0.5\textwidth]{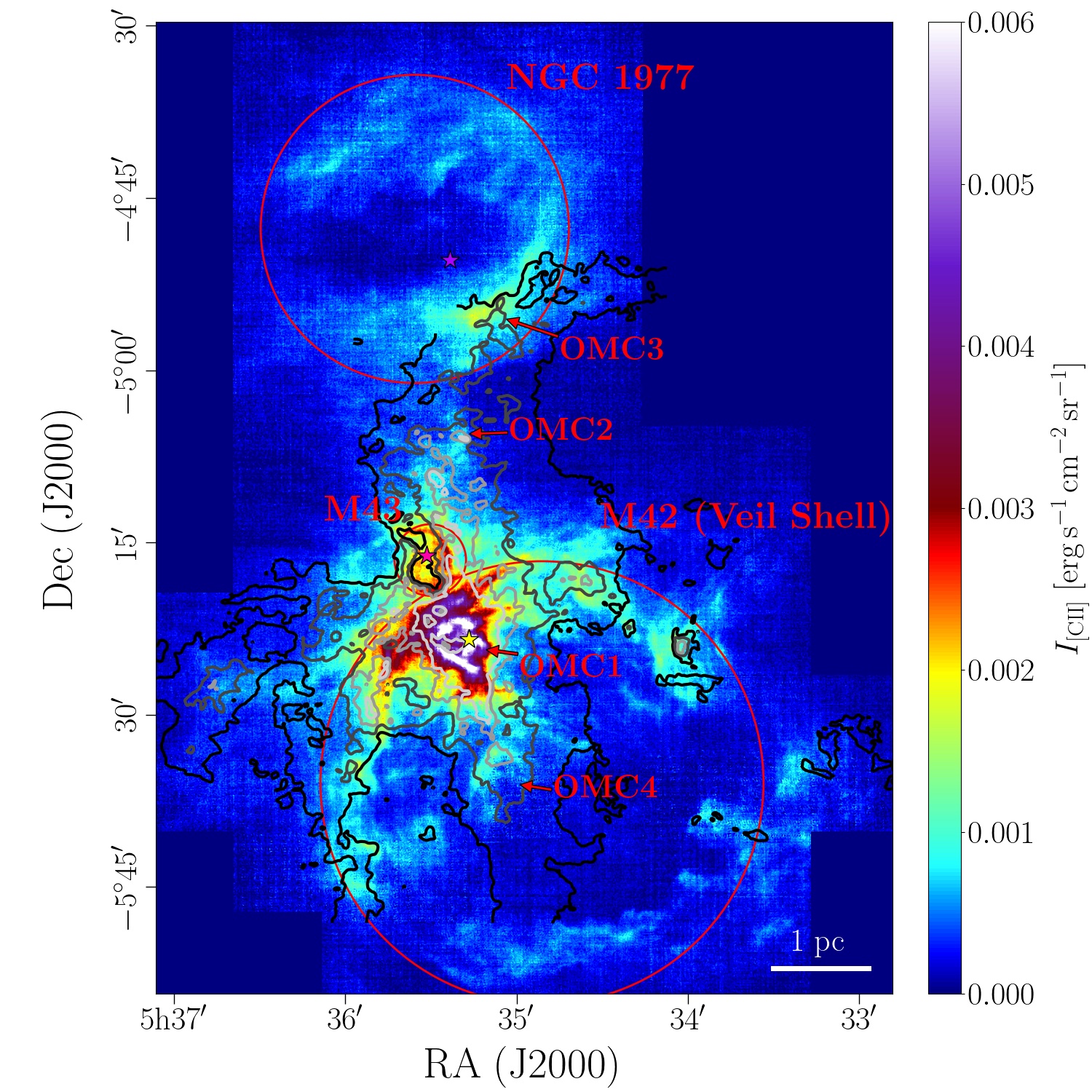}
\caption{[C\,{\sc ii}] line-integrated intensity ($v_{\mathrm{LSR}}=-10\,\text{-}\,+20\,\mathrm{km\,s^{-1}}$) from Orion A (at a resolution of $16\arcsec$) with CO(2-1) line-integrated intensity ($v_{\mathrm{LSR}}=-10\,\text{-}\,+20\,\mathrm{km\,s^{-1}}$, coverage not complete, cf. Paper I) in contours (from black to light gray: $1,2,3,4\times 10^{-6}\,\mathrm{erg\,s^{-1}\,cm^{-2}\,sr^{-1}}$). The red circles delineate the three distinct shells of M42, M43, and NGC 1977. The stars indicate the most massive stars within each region: $\theta^1$ Ori C (yellow) in M42, NU Ori (pink) in M43, and 42 Orionis (purple) in NGC 1977. Arrows indicate the positions of the four molecular cores OMC1-4 along the Integral-Shaped Filament.}
\label{Fig.map}
\end{figure}

\begin{figure*}[tb]
\includegraphics[width=\textwidth, height=0.75\textwidth]{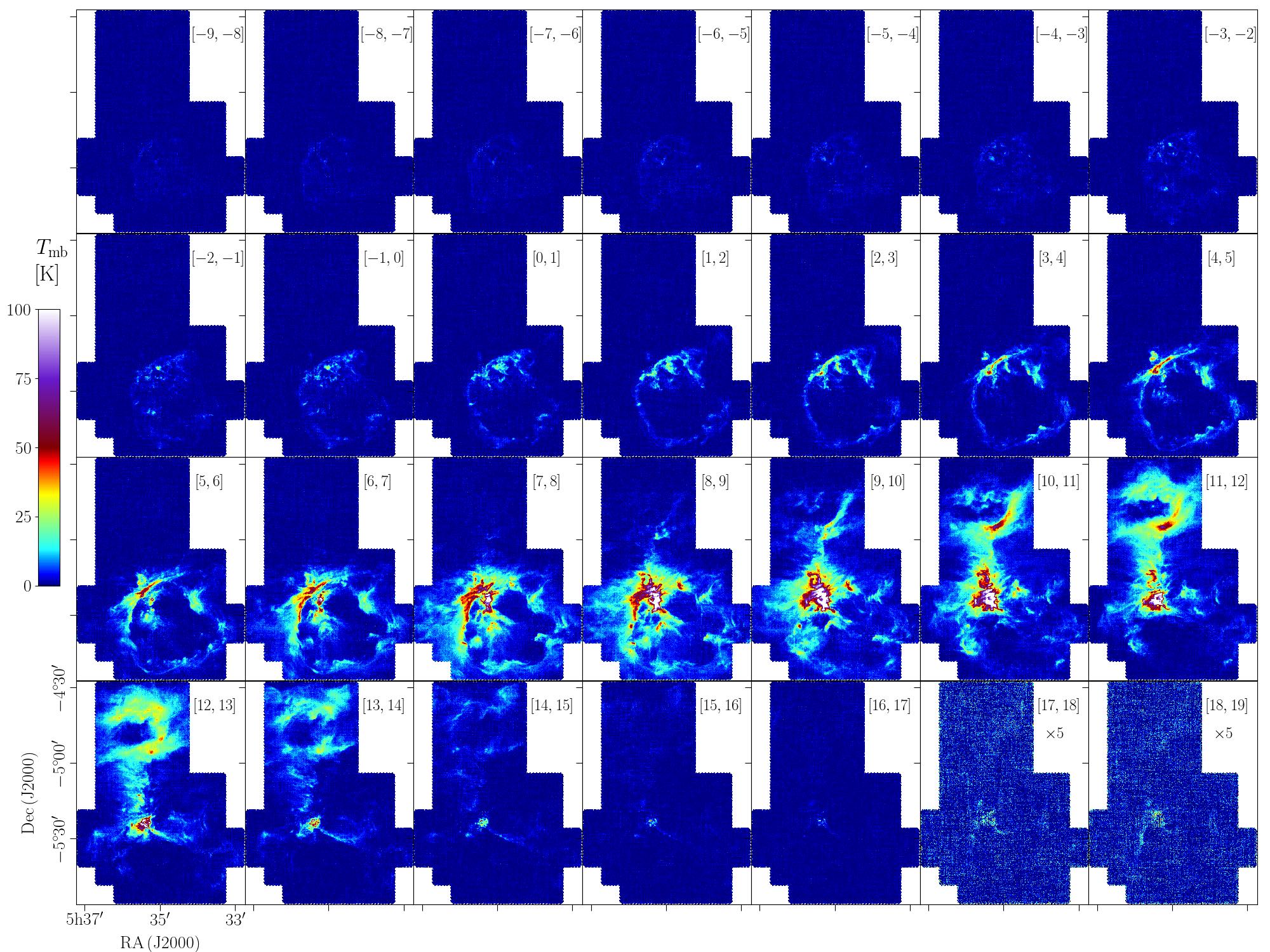}
\caption{[C\,{\sc ii}] intensity per $1\,\mathrm{km\,s^{-1}}$ channel in the range $v_{\mathrm{LSR}}=-9\,\text{-}\, 19\,\mathrm{km\,s^{-1}}$. The intensities in the last two channels ($v_{\mathrm{LSR}}=17\text{-}19\,\mathrm{km\,s^{-1}}$) contain [$^{13}$C\,{\sc ii}] $F=2\text{-}1$ emission and are multiplied by 5.}
\label{Fig.channel_maps}
\end{figure*}

\section{Observations}

We make use of velocity-resolved [C\,{\sc ii}] line observations toward Orion A, obtained by upGREAT onboard SOFIA, IRAM 30m velocity-resolved $^{12}$CO \mbox{$J$\,=\,2-1} (230.5\,GHz) and $^{13}$CO \mbox{$J$\,=\,2-1} (220.4\,GHz) line observations, {\it Herschel} dust continuum photometry, {\it Spitzer} observations in the Multi-band Imaging Photometer for {\it Spitzer} (MIPS) $24\,\mu\mathrm{m}$ band and the Infrared Array Camera (IRAC) 8\,$\mu$m band, ESO/Wide Field Imager (WFI) H$\alpha$ observations, and Digitized Sky Survey 2 (DSS-2) red-band observations converted to H$\alpha$ intensity (by dividing by 81.3). The observations are described in Paper I. We convolve all images to the same angular resolution of $36\arcsec$ with a pixel size of $14\arcsec$. In the analysis of the [C\,{\sc ii}] data we include only points above the $3\sigma$ threshold, that is $I_{\mathrm{[C\,\textsc{ii}]}}>5\times 10^{-5}\,\mathrm{erg\,s^{-1}\,cm^{-2}\,sr^{-1}}$ at $36\arcsec$ angular resolution\footnote{The conversion factor from line-integrated brightness temperature ($W\;\mathrm{[K\,km\,s^{-1}]}$) to surface brightness ($I\;\mathrm{[erg\,s^{-1}\,cm^{-2}\,sr^{-1}]}$) is $2k_{\rm B} \nu^3/c^3$. This is $I\mathrm{(erg\,s^{-1}\,cm^{-2}\,sr^{-1})} \simeq 7.0\times 10^{-6}\,W\mathrm{(K\,km\,s^{-1})}$ for the [C\,{\sc ii}] line, and $I\mathrm{(erg\,s^{-1}\,cm^{-2}\,sr^{-1})} \simeq 1.3\times 10^{-8}\,W\mathrm{(K\,km\,s^{-1})}$ for the CO(2-1) lines.}.

From the dust continuum photometry, we determine the dust effective temperature and the dust optical depth by fitting a modified blackbody with dust emissivity index $\beta=2$,
\begin{align}
I_{\lambda} = B(\lambda,T_{\rm d})\, \left[1-\exp\left(-\tau_{160}\left(\frac{160\,\mu\mathrm{m}}{\lambda}\right)^{\beta}\right)\right], \label{eq.I}
\end{align}
to the PACS $70\,\mu\mathrm{m}$, $100\,\mu\mathrm{m}$, and $160\,\mu\mathrm{m}$ bands, and the SPIRE $250\,\mu\mathrm{m}$, $350\,\mu\mathrm{m}$, and $500\,\mu\mathrm{m}$ bands (cf. Paper I and \citet{Pabst2019} for details). Where the coverage of the PACS $100\,\mu\mathrm{m}$ band is not sufficient, we complete the SED output maps with an SED fit of the remaining five PACS and SPIRE bands, which we use for illustrative purposes only. The dust temperature in the five-band fit is slightly higher than in the six-band fit (by about $0.3\text{-}0.5\,\mathrm{K}$), and the dust optical depth is somewhat lower, but negligibly so. The differences are largest in the low-intensity regions. The brightest regions in the BN/KL region and Orion S, the star-forming cores close to the massive Trapezium stars, are saturated in the {\it Herschel} images and hence excluded from our analysis. By integrating Eq. \ref{eq.I} from $\lambda=40\,\mu\mathrm{m}$ to 500\,$\mu$m, we obtain the integrated far-infrared (FIR) intensity.

Figure \ref{Fig.T_tau} shows the dust temperature and dust optical depth resulting from the SED fits ($\beta=2$). The dust optical depth and dust temperature depend sensitively on the exact value of $\beta$. In Appendix \ref{Sec.SED_Lombardi} and \ref{Sec.SED_beta_free} we compare the SED fits used here with the SED fits of \cite{Lombardi2014}, who employ $\beta\simeq 1.6$, and to SED fits, where we let $\beta$ be a free parameter. The general behavior of the correlations identified in this study is not affected.

\section{Analysis}

\subsection{Global morphology and channel maps}
\label{Sec.global-morpholgy}

Figure \ref{Fig.map} shows the line-integrated [C\,{\sc ii}] intensity from the Orion Nebula complex in the Orion A molecular cloud. The mapped area comprises three distinct regions: M42, M43, and NGC 1977. The most massive stars in the Orion Nebula complex, the Trapezium stars, are found close to the surface of OMC1. The heavily irradiated PDR at the surface of the molecular cloud radiates bright [C\,{\sc ii}] emission. Also the Veil Shell, the expanding shell that is created by the stellar wind of the O7V star $\theta^1$ Ori C (the most massive of the Trapezium stars), is readily observed in [C\,{\sc ii}] emission. The H\,{\sc ii} region of M43 is encircled by a [C\,{\sc ii}]-emitting shell, as well. In the background of the B0.5V star NU Ori, M43 is bounded by the molecular cloud with a [C\,{\sc ii}]-emitting PDR at its surface. The shell surrounding the H\,{\sc ii} regions NGC 1973, 1975, and 1977 also emits substantially in the [C\,{\sc ii}] line\footnote{We refer to the region encompassing the three reflection nebulae NGC 1973, 1975, and 1977 as NGC 1977 for short. These regions are also collectively known as Sharpless 279 (Sh2-279).}. The brightest part associated with NGC 1977 and irradiated by the B1V star 42 Orionis, however, is the PDR at the surface of the molecular core OMC3. The three shells associated with M42, M43, and NGC 1977 are subject to ongoing dynamical evolution due to over-pressurized hot plasma (M42) and ionized gas (M43 and NGC 1977) created by the central stars \citep{Pabst2019, Pabst2020}.

Figure \ref{Fig.channel_maps} shows the [C\,{\sc ii}] intensity of the Orion Nebula complex per $1\,\mathrm{km\,s^{-1}}$ channel in the range $v_{\mathrm{LSR}}=-9\text{-}19\,\mathrm{km\,s^{-1}}$. Channel maps of velocity-resolved observations allow to disentangle spatially distinct structures within the line of sight and thus to obtain insight into the global morphology of a region. We can distinguish several components: For $v_{\mathrm{LSR}}< -2\,\mathrm{km\,s^{-1}}$, we observe [C\,{\sc ii}] emission that fills the space within the shell filaments of the Orion Nebula. This emission is associated with the expanding Veil Shell, that is moving toward us \citep{Pabst2019}. The edge-on shell filaments form a coherent structure and are most distinctly visible in the velocity range $v_{\mathrm{LSR}}=0\text{-}9\,\mathrm{km\,s^{-1}}$. In the $v_{\mathrm{LSR}}=8\text{-}12\,\mathrm{km\,s^{-1}}$ range, the bright OMC1/Huygens Region reveals its structure, studied in detail by \cite{Goicoechea2015}. M43 also becomes apparent in this velocity range. In $v_{\mathrm{LSR}}=12\text{-}15\,\mathrm{km\,s^{-1}}$, a structure that seems to be an extension of the Orion Bar becomes visible, that was also noted in optical observations \citep{Henney2007}. Emission contributed by the [$^{13}$C\,{\sc ii}] $F=2\text{-}1$ hyperfine component\footnote{The [$^{13}$C\,{\sc ii}] line splits into three hyperfine components (the frequency of the [$^{12}$C\,{\sc ii}] line is 1900.537\,GHz and $\Delta v$ is given with respect to this frequency, \citet{Ossenkopf2013}):
\def\arraystretch{1.2}
\addtolength{\tabcolsep}{2pt}
\begin{tabular}{l|cccc}
\hline\hline
component & $\nu$ [GHz] & $\Delta v$ [km\,s$^{-1}$] & rel. intensity \\ \hline
$F=1\text{-}0$ & 1900.950 & -65.2 & 0.250 \\
$F=2\text{-}1$ & 1900.466 & +11.2 & 0.625 \\
$F=1\text{-}1$ & 1900.136 & +63.3 & 0.125\\ \hline
\end{tabular}
} can be discerned in the last two channels, $v_{\mathrm{LSR}}=17\text{-}19\,\mathrm{km\,s^{-1}}$, toward the brightest part of the map, that is OMC1 and the bright eastern arm of the Veil Shell, the Eastern Rim. The northern part of the map with NGC 1973, 1975, and 1977, is visible in the range $v_{\mathrm{LSR}}=10\text{-}14\,\mathrm{km\,s^{-1}}$, offset by $\sim 2 \,\mathrm{km\,s^{-1}}$ from the emission peak of the OMC1 region. Many small-scale structures also possess distinct dynamic morphology, that can be observed in the channel maps, but this is outside the scope of this study.

\begin{figure*}[tb]
\includegraphics[width=\textwidth, height=0.48\textwidth]{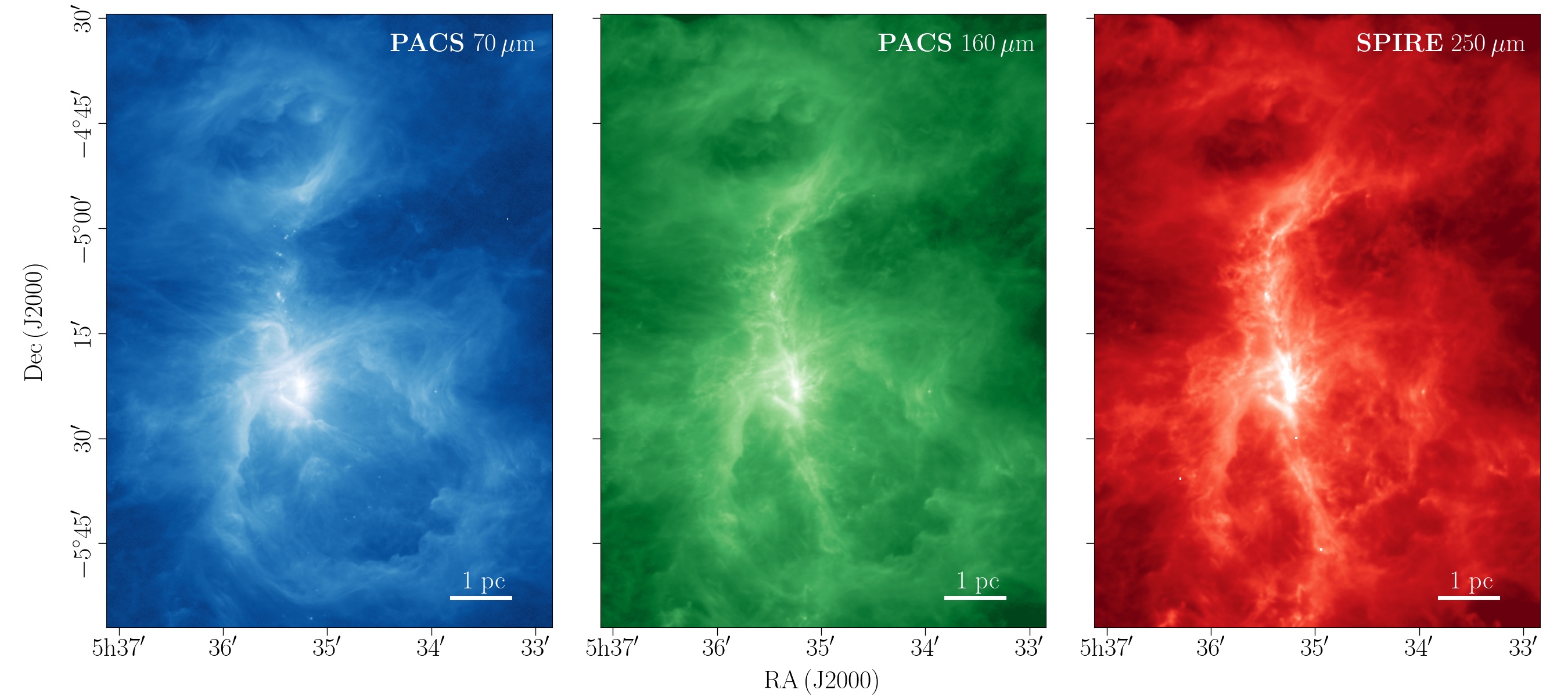}
\caption{PACS $70\,\mu\mathrm{m}$, PACS $160\,\mu\mathrm{m}$ and SPIRE $250\,\mu\mathrm{m}$ bands in their respective native resolutions (see PACS Handbook) and on logarithmic scale (units $\log_{10}\mathrm{MJy\,sr^{-1}}$).}
\label{Fig.pacs_spire_panels}
\end{figure*}

Figure \ref{Fig.pacs_spire_panels} shows the PACS $70\,\mu\mathrm{m}$, PACS $160\,\mu\mathrm{m}$ and SPIRE $250\,\mu\mathrm{m}$ photometric images. We observe that colder structures become more prominent in wavelengths longwards of $160\,\mu\mathrm{m}$. At submillimeter wavelengths, the dust emission is dominated by the so-called integral-shaped filament \citep[ISF,][]{Johnstone1999}. The ISF is a cold and dense structure that runs through the entire Orion Nebula complex and has fragmented into molecular cores, OMC1, 2, 3, and 4, that are sites of active star formation.

On large scales ($\sim 25\,\mathrm{pc}$) along the Orion molecular cloud, a global north-south velocity gradient was determined from observations of the molecular gas \citep{Bally1987}. In the region observed in [C\,{\sc ii}], we cannot distinguish a significant global velocity gradient. Rather, the velocity gradient is dominated by the separation of structures that move at distinct velocities, the M42 Veil Shell in the south and the northern bubble of NGC 1977. The local velocity gradient between OMC1 and OMC3 is also observed in molecular observations of \cite{Hacar2017}. This abrupt velocity change, as noted by \cite{Bally1987}, is larger than the global velocity gradient ($\sim 0.5\,\mathrm{km\,s^{-1}\,pc^{-1}}$) from molecular observations would suggest.

\begin{figure}[!tb]
\includegraphics[width=0.5\textwidth, height=0.625\textwidth]{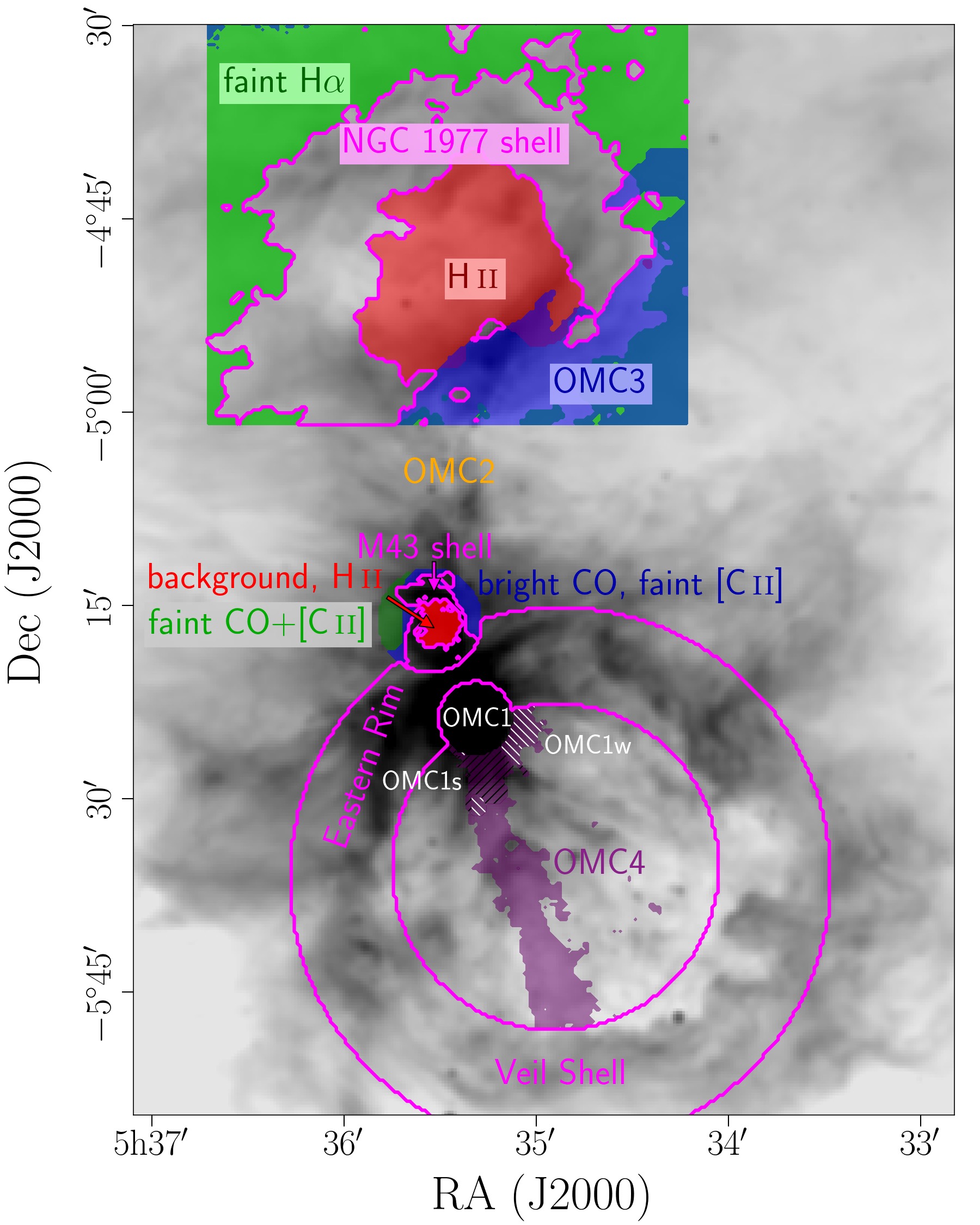}
\caption{Regions in correlations plots discussed in Sections \ref{FIR-vs-d} to \ref{CII-vs-CO} on IRAC $8\,\mu\mathrm{m}$ emission convolved to an angular resolution of $36\arcsec$.}
\label{Fig.map_regions}
\end{figure}

\subsection{Definition of regions}
\label{Sec.subregions}

In Paper I, we have discussed the global correlations. In this paper, we study the correlations in more detail by dividing them into subregions. Figure 3 of Paper I shows the correlation plots disregarding individual (sub)regions, including all areas that fall outside of the later-defined regions, as point-density plots. Table 1 of Paper I summarizes the power-law fits we find for each correlation, using the ordinary least-squares (OLS) bisector method \citep{Isobe1990}.

Because some of the spread in the correlation plots may stem from differences in the characteristics of the sources and their individual components, we have subdivided the survey area into four regions, that are further subdivided by their emission components. Figure \ref{Fig.map_regions} gives an overview of the regions we define in order to study the correlations in the following plot in greater detail. We have divided the Orion Nebula into two regions, the region that is viewed edge on, which comprises the Eastern Rim, the edge-on Veil Shell and the rim separating M42 from M43, and the region that presumably has face-on geometry, that is the surface of the ISF in the region of OMC4. We have defined the latter as regions situated inside the Veil Shell with $\tau_{160}>8\times 10^{-3}$, which includes areas close to OMC1 (purple hatched areas in Fig. \ref{Fig.map_regions}). Those latter areas are defined by $I_{\mathrm{FIR}}>2\times 10^{-1}\,\mathrm{erg\,s^{-1}\,cm^{-2}\,sr^{-1}}$. Anticipating the discussion on the [C\,{\sc ii}]-FIR correlation in section \ref{CII-vs-FIR}, we have subdivided them into two subregions, OMC1s and OMC1w. Points in OMC1s lie above the regression curve computed from all OMC4 points, points in OMC1w lie below this average regression curve. We encircle the points in OMC1w with black circles in the correlation plots. In the panels showing the OMC4 correlation, we include the points lying inside the OMC1 region in gray scale.

We have divided M43 into four subregions: The interior H\,{\sc ii} region with the molecular background and foreground expanding shell ($I_{\mathrm{H}\alpha}>4\times 10^3\,\mathrm{MJy\,sr^{-1}}$), the bright rim around it, that is the shell, a region where [C\,{\sc ii}] emission is faint ($I_{\mathrm{[C\,\textsc{ii}]}}<1.4\times 10^{-3}\,\mathrm{erg\,s^{-1}\,cm^{-2}\,sr^{-1}}$) but CO(2-1) emission is still bright ($I_{\mathrm{CO(2\text{-}1)}}>1.5\times 10^{-6}\,\mathrm{erg\,s^{-1}\,cm^{-2}\,sr^{-1}}$), and a region where both [C\,{\sc ii}] and CO(2-1) emission are faint. NGC 1977 is divided into four subregions, as well: The H\,{\sc ii} region with the expanding shell ($I_{\mathrm{H}\alpha}>150\,\mathrm{MJy\,sr^{-1}}$), the shell surrounding it, the region of OMC3, defined by $\tau_{160}>6\times 10^{-3}$ and $I_{\mathrm{CO(2\text{-}1)}}/I_{\mathrm{FIR}}>3\times 10^{-5}$, and the outward areas with faint H$\alpha$ emission ($I_{\mathrm{H}\alpha}<90\,\mathrm{MJy\,sr^{-1}}$).

\subsection{FIR versus distance}
\label{FIR-vs-d}

\begin{figure*}[tb]
\begin{minipage}{0.49\textwidth}
\includegraphics[width=\textwidth, height=0.67\textwidth]{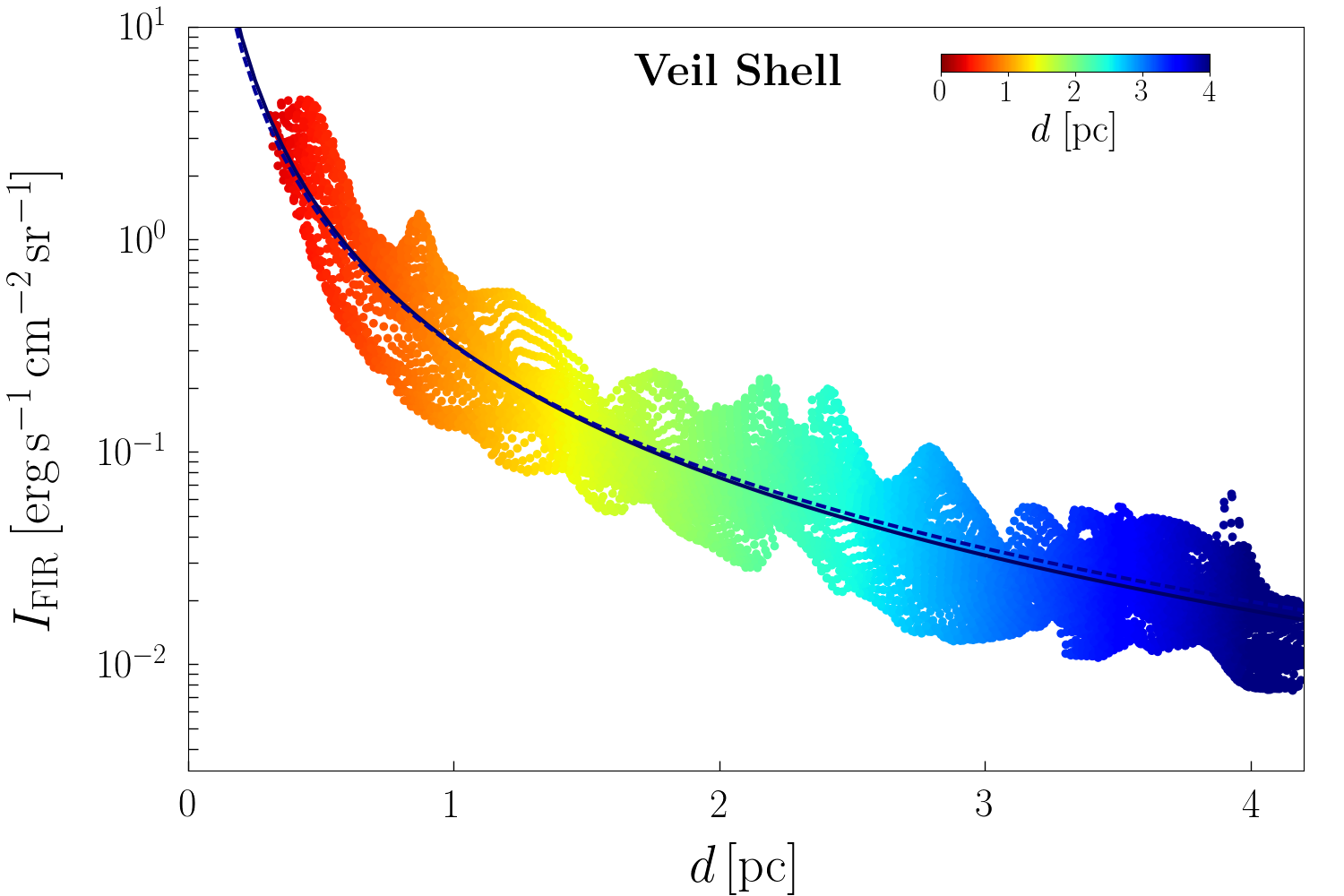}
\end{minipage}
\begin{minipage}{0.49\textwidth}
\includegraphics[width=\textwidth, height=0.67\textwidth]{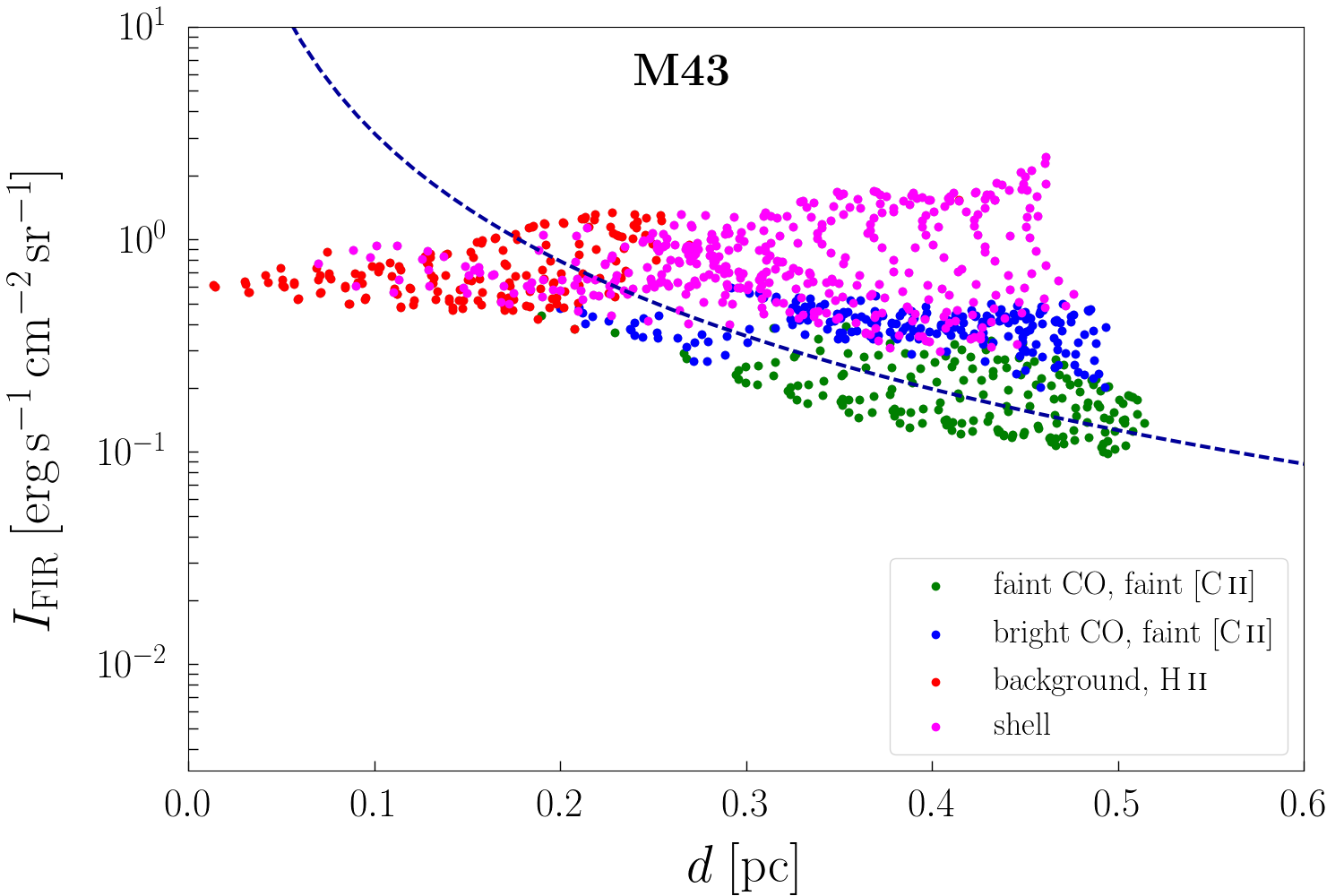}
\end{minipage}

\begin{minipage}{0.49\textwidth}
\includegraphics[width=\textwidth, height=0.67\textwidth]{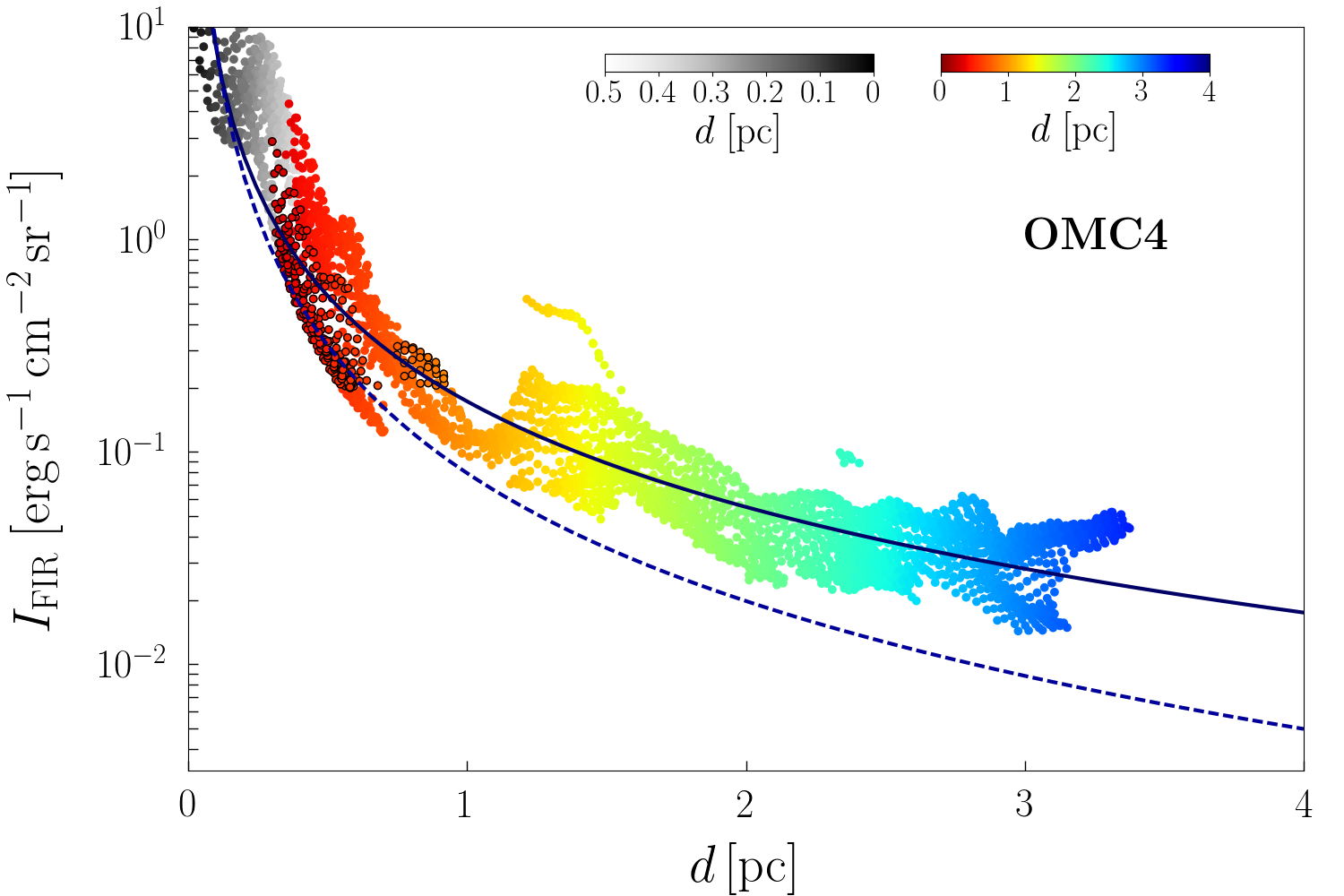}
\end{minipage}
\begin{minipage}{0.49\textwidth}
\includegraphics[width=\textwidth, height=0.67\textwidth]{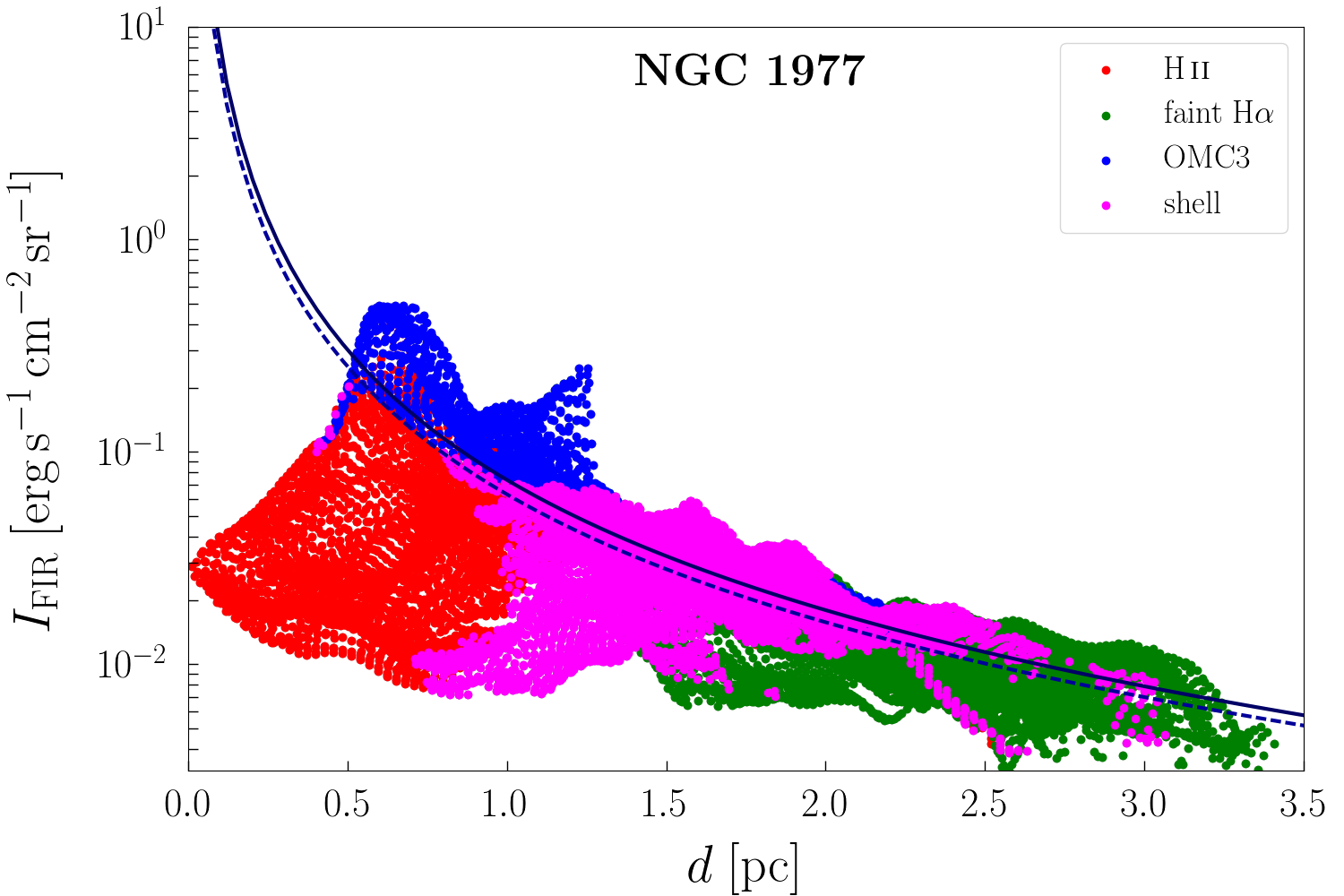}
\end{minipage}
\caption{FIR intensity versus (projected) distance from respective central stars in M42 (top left: Veil Shell, bottom left: OMC4), M43 (top right) and NGC 1977 (bottom right). The division of M43 and NGC 1977 into the regions given in the legend is described in Section \ref{Sec.subregions}. The blue solid lines are the fits summarized in Table \ref{Tab.FIR-d}. The blue dashed lines indicate a FIR intensity scaling with $d^{-2}$.}
\label{Fig.FIR-d}
\end{figure*}

Figure \ref{Fig.FIR-d} shows the dependence of the FIR intensity on (projected) distance from the respective central star in M42, M43 and NGC 1977. The FIR intensity scales with the incident FUV radiation field. However, geometry and line-of-sight effects are important. While, in principle, the incident radiation field can be estimated from the stellar luminosity and the (true) distance from the illuminating source, oftentimes only the projected distance is known.

We find that both in the limb-brightened edge of the Veil Shell and in OMC4 at $d<1\,\mathrm{pc}$ the FIR intensity scales with $d^{-2}$ approximately. The incident radiation field at a distance of $1\,\mathrm{pc}$ is $G_0\sim 500$, derived from the presumed face-on geometry of the ISF in the OMC4 region using $G_0\simeq I_{\mathrm{FIR}}/2/1.3\times 10^{-4}\,\mathrm{erg\,s^{-1}\,cm^{-2}\,sr^{-1}}$ \citep{HollenbachTielens1999}. This is a third of the value derived from the total FIR luminosity from the Orion Nebula (cf. Paper I) and reflects the geometry of the region. In fact, this discrepancy corresponds to an angle of $70^{\circ}$ between the incident radiation and the surface normal. At $d>1\,\mathrm{pc}$, the FIR intensity in OMC4 scales with an exponent somewhat smaller than 2. This region is characterized by large columns of cold dust that produce a FIR excess (cf. Paper I). 

In M43 and NGC 1977, we do not find a strong dependence of the FIR intensity on distance. We attribute this to the geometry of these regions. M43 is a half-spherical shell situated in front of a dense background cloud close to the ISF. NGC 1977 is an expanding H\,{\sc ii} region \citep{Pabst2020}, whose limb-brightened shell is readily observed in surface tracers. Considering only points in the NGC 1977 shell with $d>1.5\,\mathrm{pc}$, the FIR intensity drops with $d^{-2}$ approximately. However, this fit also describes points with $d<1.5\,\mathrm{pc}$ to a certain extent.

\begin{table}[ht]
\def\arraystretch{1.2}
\addtolength{\tabcolsep}{-2pt}
\caption{Summary of FIR-distance correlation in Fig. \ref{Fig.FIR-d}.}
\begin{tabular}{l|cccc}
\hline\hline
Region & $a$ & $b$ & $\rho$ & $rms\,\mathrm{[dex]}$ \\ \hline
Veil Shell & $0.32\pm 0.07$ & $-2.08\pm 0.31$ & $-0.94$ & 0.19 \\
OMC4 & $0.17\pm 0.04$ & $-1.65\pm 0.27$ & $-0.88$ & 0.19 \\
M43 & -- & -- & $-0.41$ & 0.27 \\
NGC 1977 & $0.07\pm 0.05$ & $-2.04\pm 1.3$ & $-0.68$ & 0.13 \\ \hline
\end{tabular}
\tablefoot{$I_{\mathrm{FIR}} = a (d/1\,\mathrm{pc})^b$. In NGC 1977 fit only for $d>1.5\,\mathrm{pc}$. $\rho$ is the Pearson correlation coefficient, $rms$ is the root-mean-square of the residual of the least-squares fit.}
\label{Tab.FIR-d}
\end{table}

In summary, we observe geometric dilution of the radiation field in M42, but not in M43 and in NGC 1977 only partly. Table \ref{Tab.FIR-d} summarizes the FIR-distance correlations discussed above. The Veil Shell and OMC4 exhibit a strong correlation, measured by the correlation coefficient $\rho$, while this is less pronounced in M43 and NGC 1977. The spread in the correlations is non-negligible, with $0.19\,\mathrm{dex}$ in the Orion Nebula being the smallest ($\rho=-0.69$, $rms=0.44\,\mathrm{dex}$ for the entire of NGC 1977). The scatter around the fits can be due to geometry effects and the deviation of the real distance from the projected distance.

The intensity of the incident radiation field, $G_0$ in Habing units, can be expressed as
\begin{align}
G_0\simeq 200\chi \frac{L_{\star}}{1\times 10^4\,L_{\sun}} \left(\frac{1\,\mathrm{pc}}{d}\right)^2,
\end{align}
where $\chi=L_{\mathrm{FUV}}/L_{\star}$ \citep[cf. eq. 5.43 in][]{Tielens}. For early B stars and late O stars ($T_{\mathrm{eff}}\simeq 2.5\text{-}4.0\times 10^4\,\mathrm{K}$), $\chi\simeq 0.5$ \citep{Spaans1994}. We estimate the central star's luminosity assuming it is re-radiated in dust continuum emission. In M42, we have $L_{\mathrm{FIR}}\simeq 1.5\times 10^5\,L_{\sun}$, hence theory predicts $G_0\simeq 1.5\times 10^3(1\,\mathrm{pc}/d)^2$. From the data associated with OMC4, we estimate $G_0\simeq 500(1\,\mathrm{pc}/d)^2$ (Fig. \ref{Fig.FIR-d} and Table \ref{Tab.FIR-d}). As we have shown in Paper I, from the data for the edge-on PDR in the Eastern Rim we also estimate $G_0\simeq 500(1\,\mathrm{pc}/d)^2$. This discrepancy between the theoretical and observed relations is likely caused by the geometry of the Orion Nebula. The close adherence to a $d^{-2}$ law illustrates that the continuum emission is due to dust predominately heated by the central massive stars.

In M43, we obtain $L_{\mathrm{FIR}}\simeq 1.4\times 10^4\,L_{\sun}$, hence we expect $G_0\simeq 1.0\text{-}1.5\times 10^3$ in the shell at $d\simeq 0.3\text{-}0.4\,\mathrm{pc}$ from NU Ori. In NCG 1977, $L_{\mathrm{FIR}}\simeq 1.8\times 10^4\,L_{\sun}$. For OMC3 at a distance of $d\simeq 0.4\,\mathrm{pc}$ from 42 Orionis, we compute $G_0\simeq 1.1\times 10^3$. In the shell at a distance of $d\simeq 1.0\text{-}1.5\,\mathrm{pc}$, we estimate $G_0\simeq 100\text{-}300$.

\subsection{[C\,{\sc ii}] versus $70\,\mu\mathrm{m}$}
\label{CII-vs-70}

\begin{figure*}[tb]
\begin{minipage}{0.49\textwidth}
\includegraphics[width=\textwidth, height=0.67\textwidth]{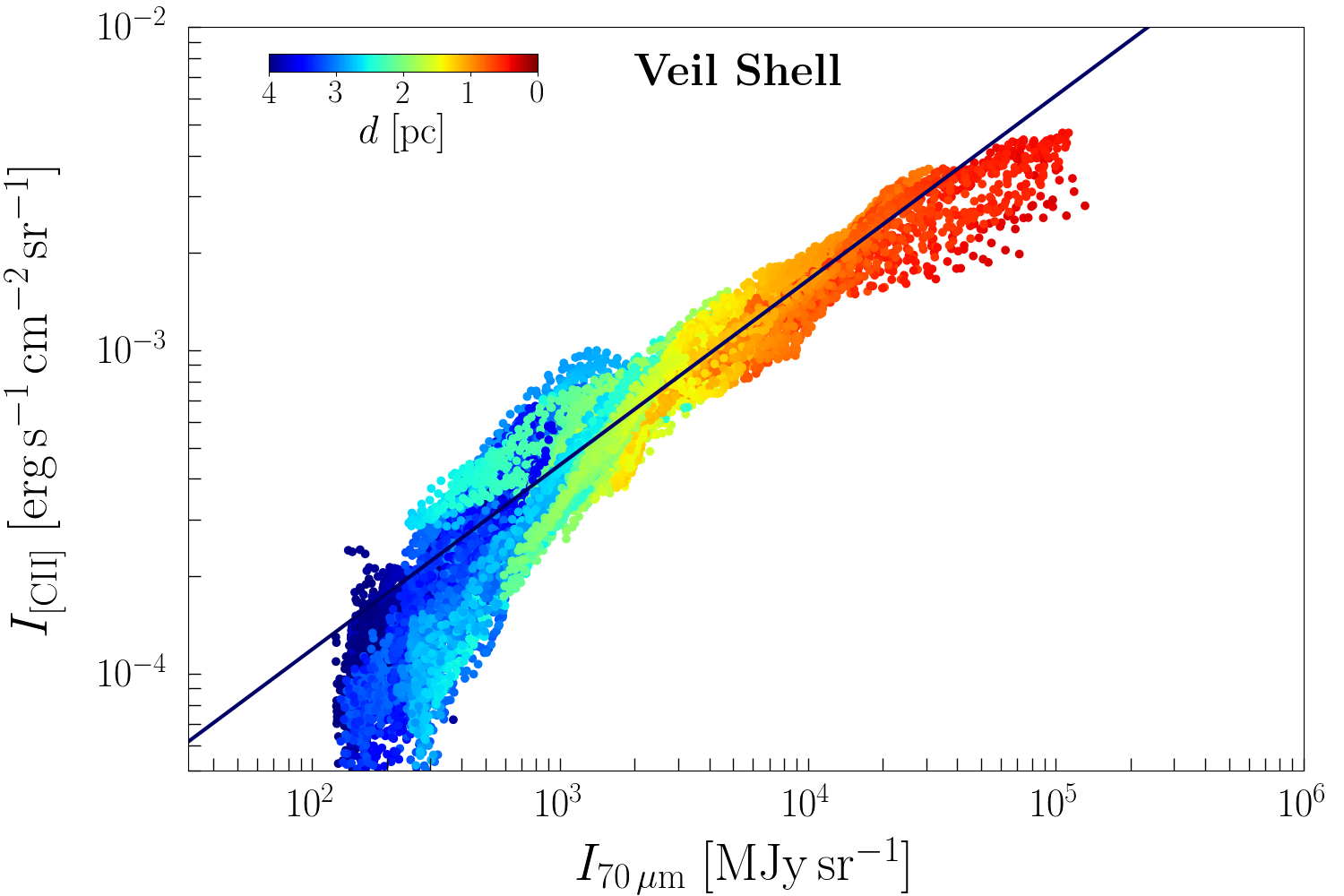}
\end{minipage}
\begin{minipage}{0.49\textwidth}
\includegraphics[width=\textwidth, height=0.67\textwidth]{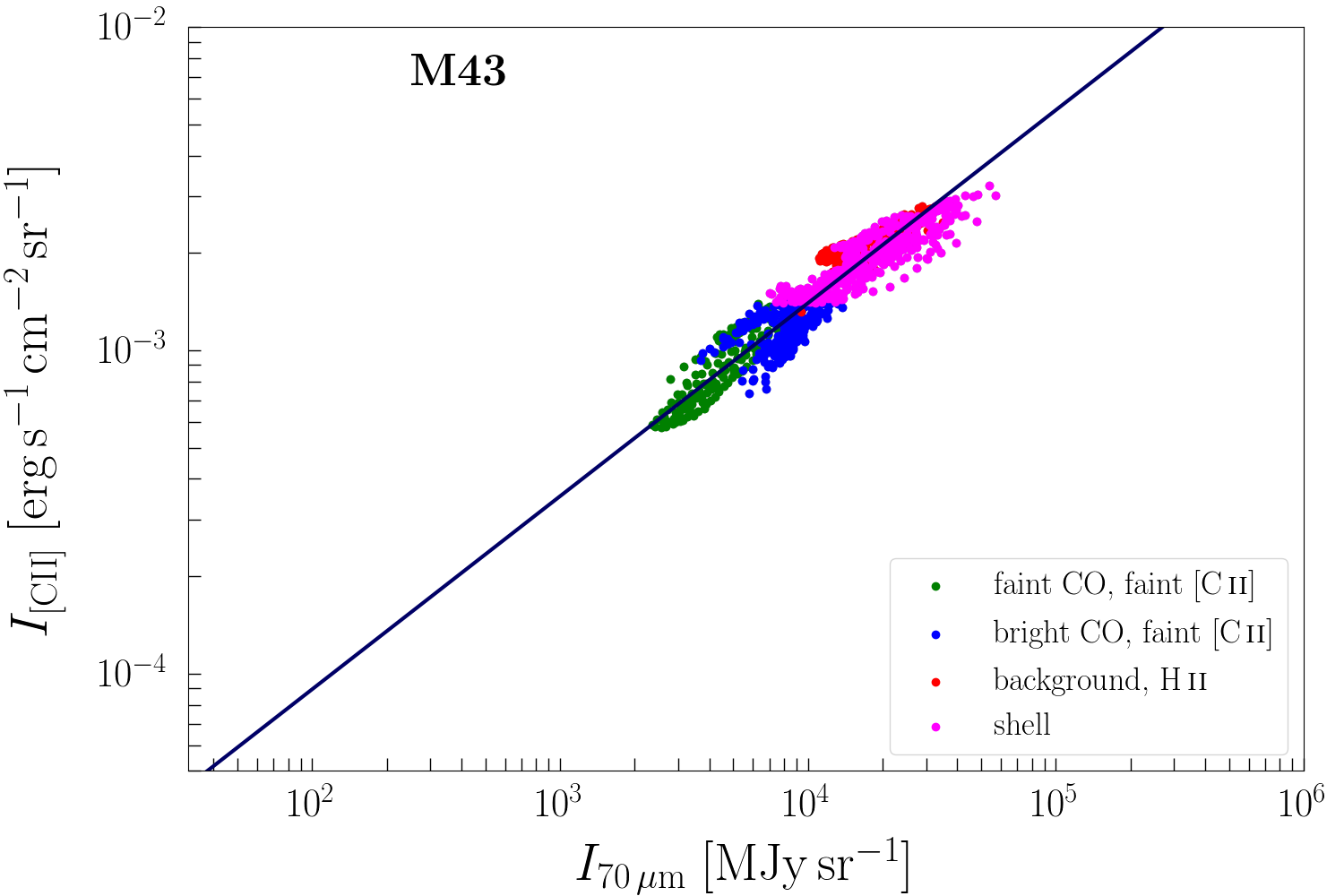}
\end{minipage}

\begin{minipage}{0.49\textwidth}
\includegraphics[width=\textwidth, height=0.67\textwidth]{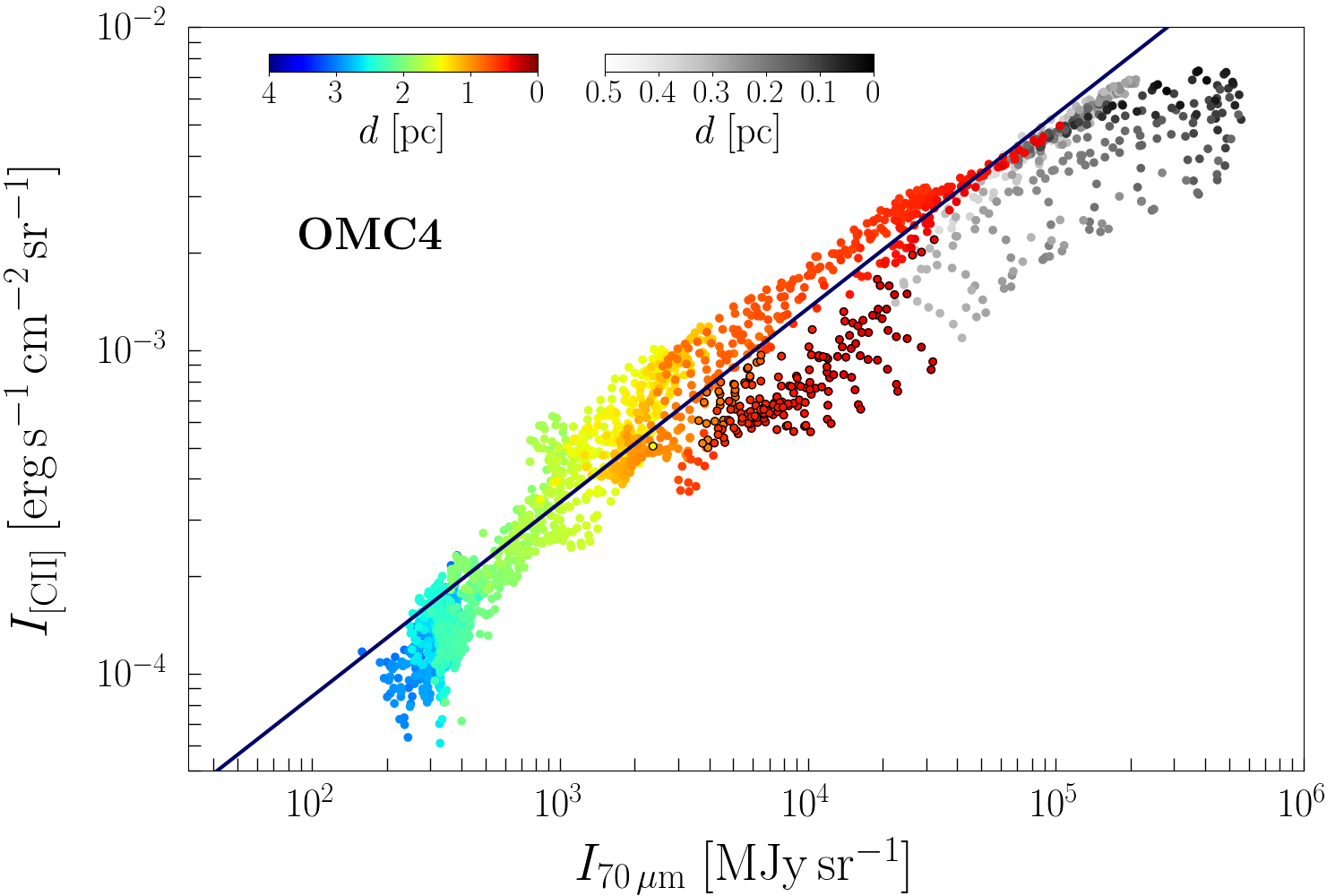}
\end{minipage}
\begin{minipage}{0.49\textwidth}
\includegraphics[width=\textwidth, height=0.67\textwidth]{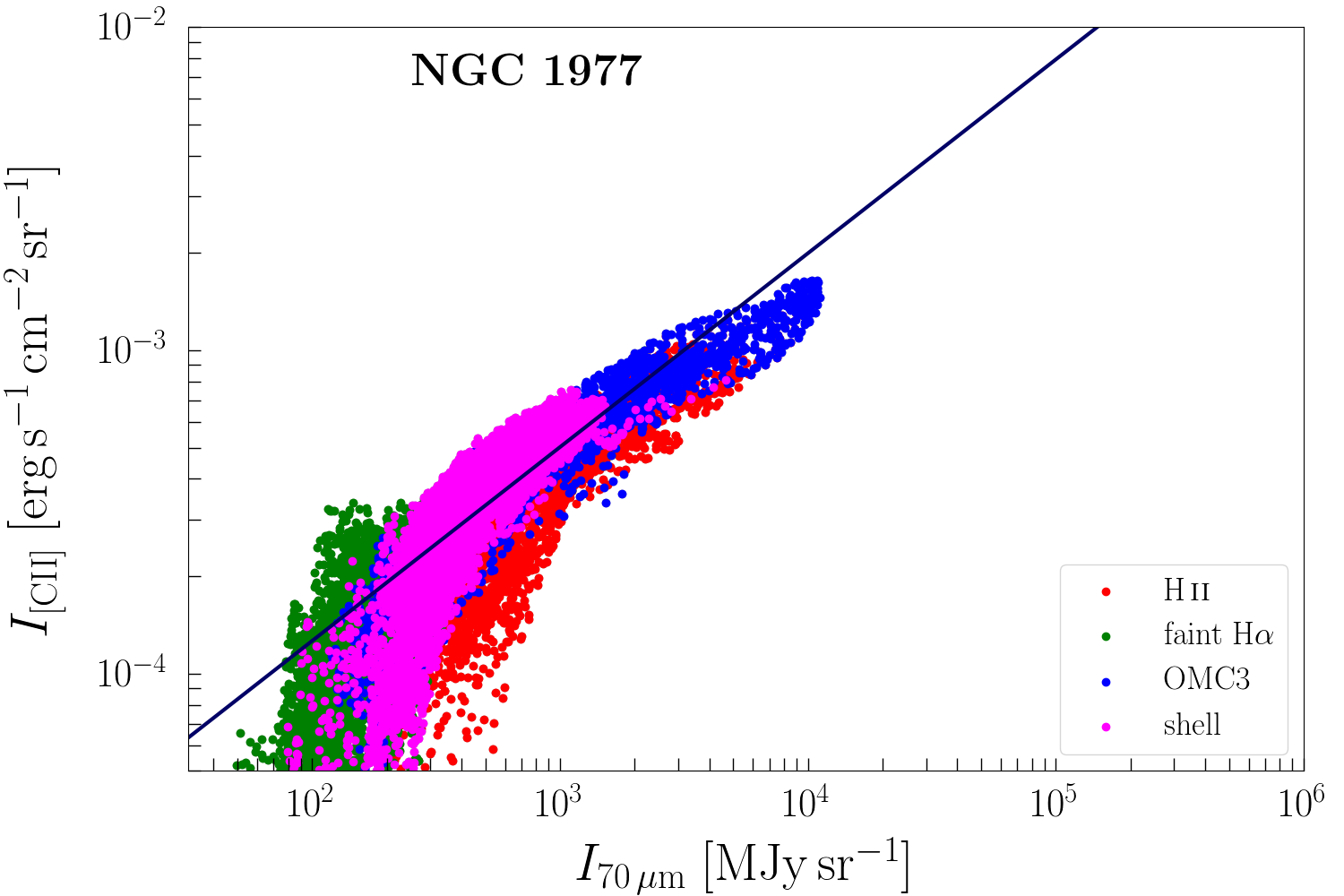}
\end{minipage}
\caption{[C\,{\sc ii}] intensity versus PACS $70\,\mu\mathrm{m}$ intensity in M42 (top left: Veil Shell, bottom left: OMC4 with OMC1 (in gray scale; black-edged red points lie to the west of Orion S)), M43 (top right) and NGC 1977 (bottom right). The color scale in M42 indicates the distance from $\theta^1$ Ori C. The division of M43 and NGC 1977 into the regions given in the legend is described in Section \ref{Sec.subregions}. The blue solid lines are the fits summarized in Table \ref{Tab.CII-70}.}
\label{Fig.CII-70}
\end{figure*}

The observed [C\,{\sc ii}] intensity is compared to the $70\,\mu\mathrm{m}$ intensity in Fig. \ref{Fig.CII-70}. All four regions show a tight correlation of these two quantities. These correlations are summarized in Table \ref{Tab.CII-70}.

\begin{table}[ht]
\def\arraystretch{1.2}
\addtolength{\tabcolsep}{-1.5pt}
\caption{Summary of [C\,{\sc ii}]-$70\,\mu\mathrm{m}$ correlation in Fig. \ref{Fig.CII-70}.}
\begin{tabular}{l|cccc}
\hline\hline
Region & $a$ & $b$ & $\rho$ & $rms\,\mathrm{[dex]}$ \\ \hline
Veil Shell & $0.57\pm 0.23$ & $-5.07\pm 0.76$ & 0.94 & 0.10 \\
OMC4 & $0.60\pm 0.17$ & $-5.27\pm 0.61$ & 0.91 & 0.13 \\
M43 & $0.60\pm 0.19$ & $-5.24\pm 0.77$ & 0.94 & 0.06 \\
NGC 1977 & $0.60\pm 0.62$ & $-5.10\pm 1.86$ & 0.77 & 0.10 \\ \hline
\end{tabular}
\tablefoot{$\log_{10} I_{\mathrm{[C\,\textsc{ii}]}} = a \log_{10} I_{\mathrm{70\,\mu\mathrm{m}}} + b$. Fit for $I_{70\,\mu\mathrm{m}}>5\times 10^{2}\,\mathrm{MJy\,sr^{-1}}$. $\rho$ is the Pearson correlation coefficient, $rms$ is the root-mean-square of the residual of the least-squares fit.}
\label{Tab.CII-70}
\end{table}

As observed in Paper I, the correlations exhibit two regimes, with decreasing slope from low to high $70\,\mu\mathrm{m}$ intensity. We fit and discuss only the regime with $I_{70\,\mu\mathrm{m}}>5\times 10^{2}\,\mathrm{MJy\,sr^{-1}}$, since the low-intensity regime is problematic because of calibration uncertainties in the PACS bands. The [C\,{\sc ii}] intensity scales with the $70\,\mu\mathrm{m}$ intensity with a power of $0.57\text{-}0.60$ (for $I_{70\,\mu\mathrm{m}}>5\times 10^{2}\,\mathrm{MJy\,sr^{-1}}$). The largest scatter about the regression curve (and the lowest correlation coefficient) is observed in NGC1977. Also, the spread is higher in OMC4 as compared to the Veil Shell. The [C\,{\sc ii}] intensity tends to turn off the regression curve toward lower values at the high-intensity end. As bright-emission regions tend to be optically thick in the [C\,{\sc ii}] line (while cooling mainly through the front side viewed edge-on), and [O\,{\sc i}] $63\,\mu\mathrm{m}$ and $145\,\mu\mathrm{m}$ line cooling becomes significant, we expect less [C\,{\sc ii}] emission as compared to the intensity in the $70\,\mu\mathrm{m}$ band. In M43, the high-intensity points (toward the shell and the H\,{\sc ii} region and background cloud) also tend to follow a slightly shallower slope, likely because of the same effect.

\subsection{[C\,{\sc ii}] versus FIR}
\label{CII-vs-FIR}

\begin{figure*}[tb]
\begin{minipage}{0.49\textwidth}
\includegraphics[width=\textwidth, height=0.67\textwidth]{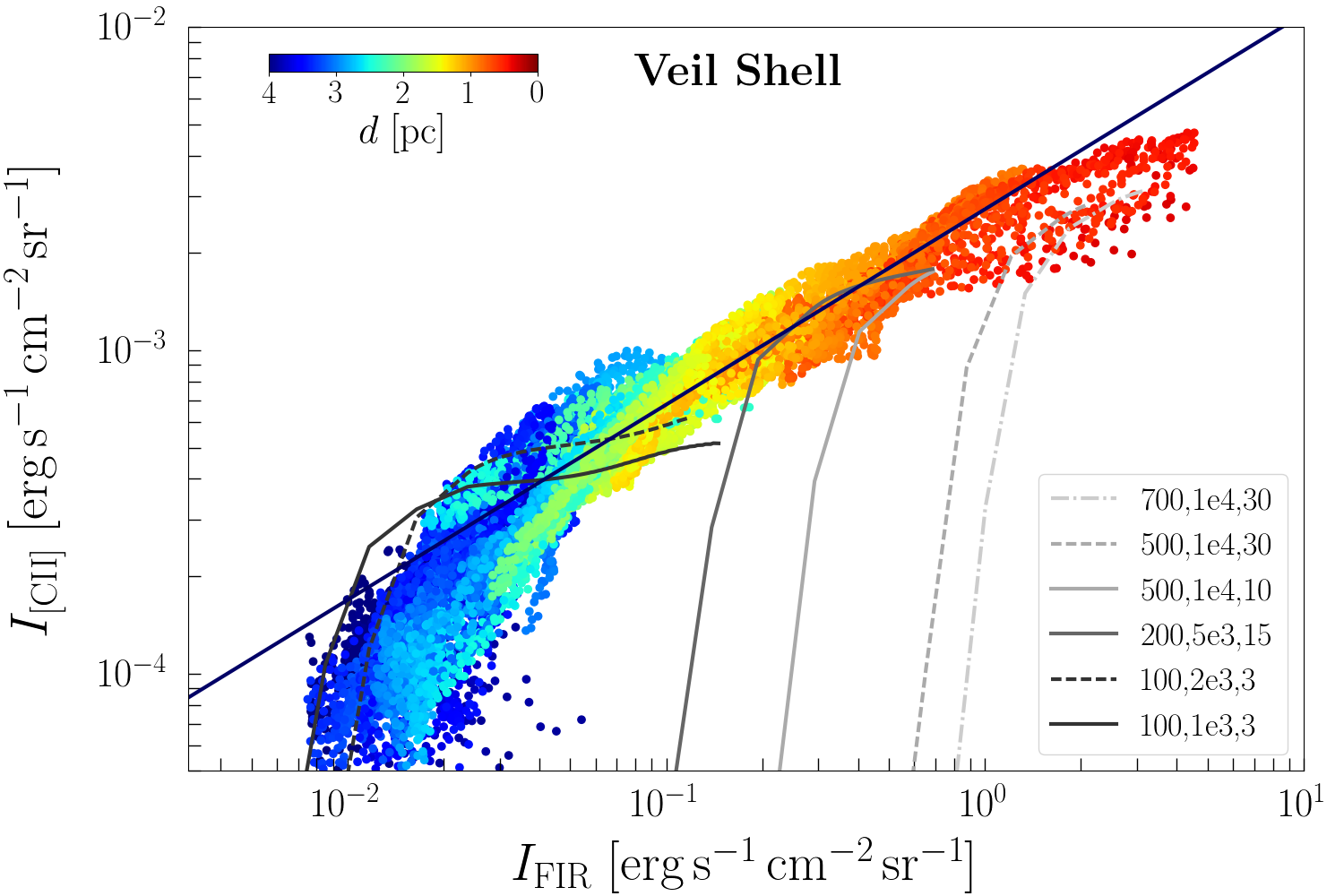}
\end{minipage}
\begin{minipage}{0.49\textwidth}
\includegraphics[width=\textwidth, height=0.67\textwidth]{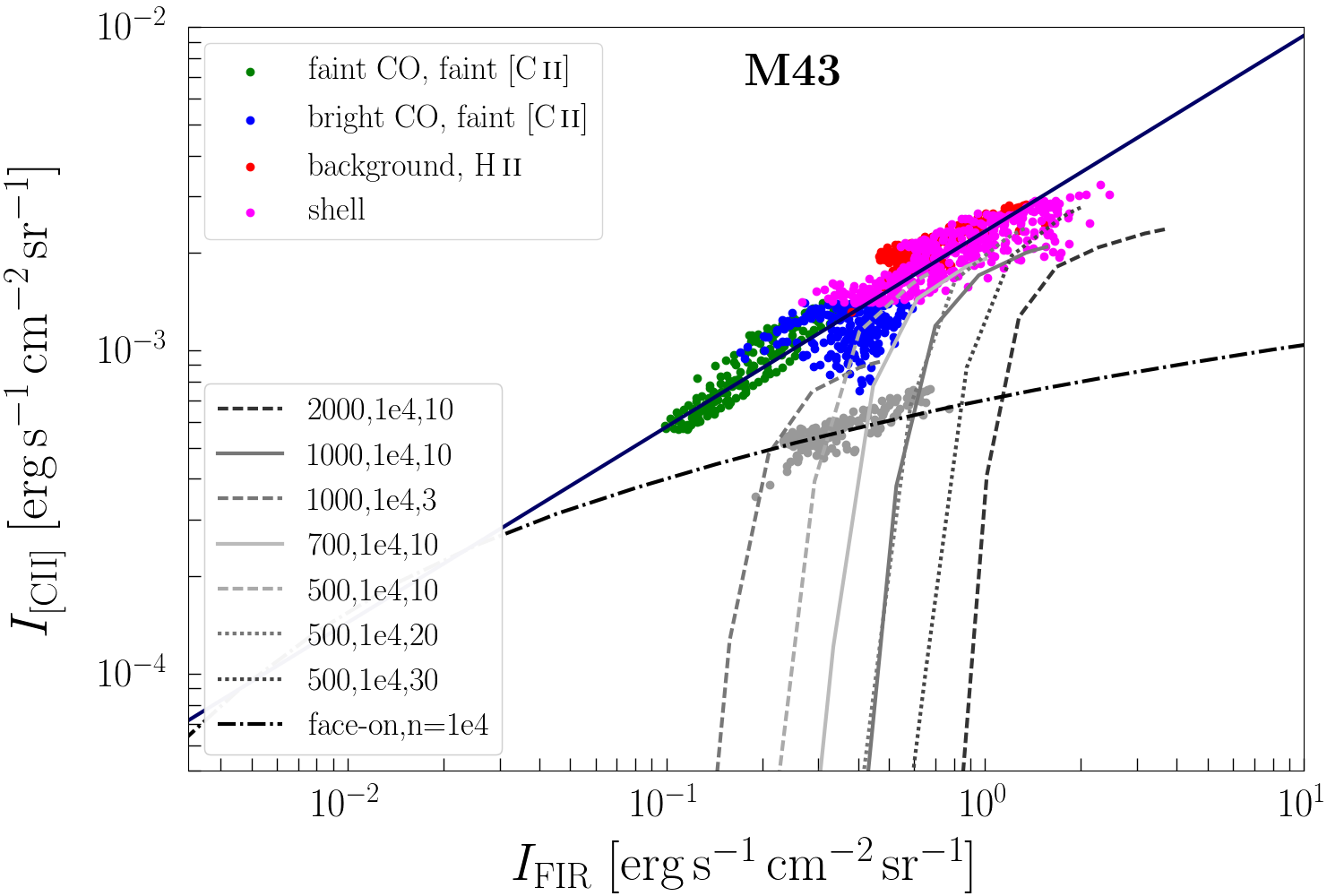}
\end{minipage}

\begin{minipage}{0.49\textwidth}
\includegraphics[width=\textwidth, height=0.67\textwidth]{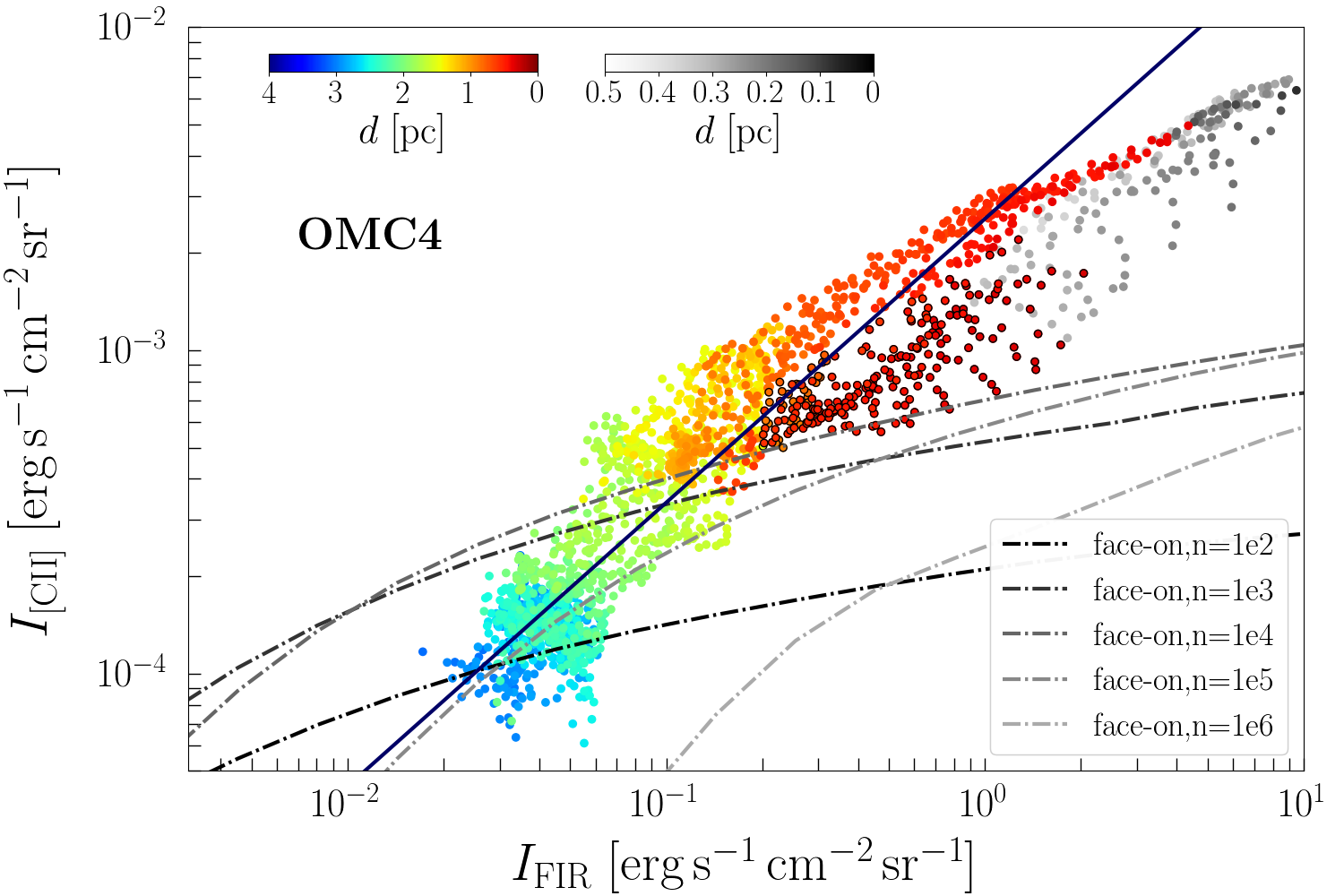}
\end{minipage}
\begin{minipage}{0.49\textwidth}
\includegraphics[width=\textwidth, height=0.67\textwidth]{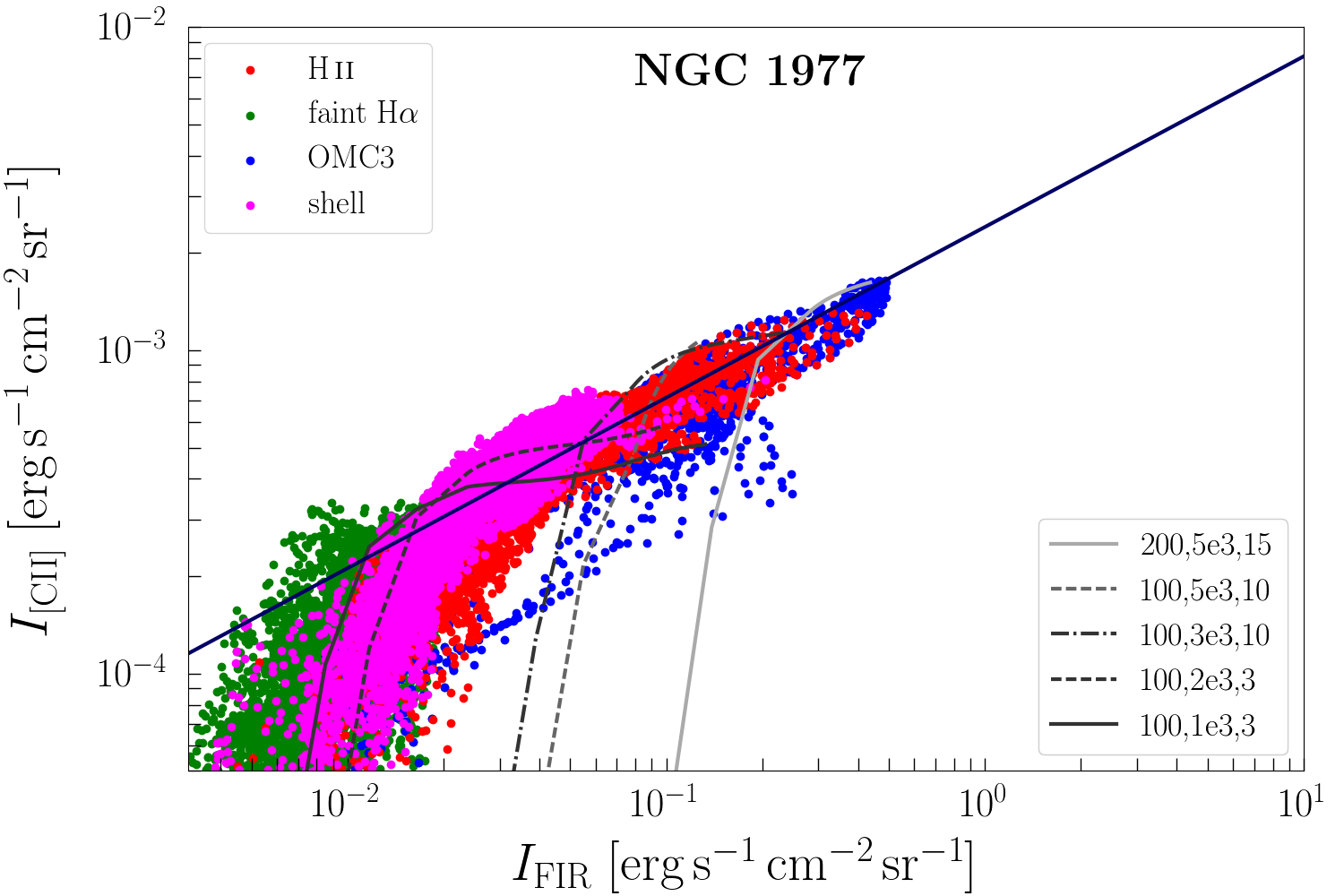}
\end{minipage}
\caption{[C\,{\sc ii}] intensity versus FIR intensity in M42 (top left: Veil Shell, bottom left: OMC4 with OMC1 (in gray scale; black-edged red points lie to the west of Orion S)), M43 (top right) and NGC 1977 (bottom right). The color scale in M42 indicates the distance from $\theta^1$ Ori C. The division of M43 and NGC 1977 into the regions given in the legend is described in Section \ref{Sec.subregions}. The blue solid lines are the fits summarized in Table \ref{Tab.CII-FIR}. Dashed lines are edge-on model outputs with the parameters ($G_0$, $n$, $A_{\mathrm{V,los}}$) given in the legend. Dashed dotted lines in the OMC4 and the M43 panels are face-on model outputs for the densities given in the legend. gray points in M43 correspond to the estimated intensities of the background PDR toward the center of M43 (cf. Sec. \ref{Sec.PDR-comparison}).}
\label{Fig.CII-FIR}
\end{figure*}

The observed [C\,{\sc ii}] intensity is compared to the FIR intensity\footnote{Again, we exclude the very bright (in FIR emission) regions of the BN/KL and Orion S cores, that are saturated in our SEDs.} in Fig. \ref{Fig.CII-FIR}. All four regions show a tight correlation of these two quantities. These correlations are summarized in Table \ref{Tab.CII-FIR}.

\begin{table}[ht]
\def\arraystretch{1.2}
\addtolength{\tabcolsep}{-1.5pt}
\caption{Summary of [C\,{\sc ii}]-FIR correlation in Fig. \ref{Fig.CII-FIR}.}
\begin{tabular}{l|cccc}
\hline\hline
Region & $a$ & $b$ & $\rho$ & $rms\,\mathrm{[dex]}$ \\ \hline
Veil Shell & $0.60\pm 0.27$ & $-2.56\pm 0.26$ & 0.94 & 0.09 \\
OMC4 & $0.88\pm 0.33$ & $-2.59\pm 0.29$& 0.92 & 0.17 \\
M43 & $0.61\pm 0.21$ & $-2.63\pm 0.06$ & 0.92 & 0.07 \\
NGC 1977 & $0.53\pm 0.39$ & $-2.62\pm 0.48$ & 0.79 & 0.08 \\ \hline
\end{tabular}
\tablefoot{$\log_{10} I_{\mathrm{[C\,\textsc{ii}]}} = a \log_{10} I_{\mathrm{FIR}} + b$. Fit for $I_{\mathrm{FIR}}>3\times 10^{-2}\,\mathrm{erg\,s^{-1}\,cm^{-2}\,sr^{-1}}$. $\rho$ is the Pearson correlation coefficient, $rms$ is the root-mean-square of the residual of the least-squares fit.}
\label{Tab.CII-FIR}
\end{table}

Similar to the [C\,{\sc ii}]-$70\,\mu\mathrm{m}$ correlation, the [C\,{\sc ii}]-FIR correlation exhibits two regimes, but we fit only the regime with $I_{\mathrm{FIR}}>3\times 10^{-2}\,\mathrm{erg\,s^{-1}\,cm^{-2}\,sr^{-1}}$. In this regime, the [C\,{\sc ii}] intensity depends less than linear on the FIR intensity, the exponent lying in the range $0.53\text{-}0.88$ (for $I_{\mathrm{FIR}}>3\times 10^{-2}\,\mathrm{erg\,s^{-1}\,cm^{-2}\,sr^{-1}}$). The correlation coefficient in M42 and M43 is larger than 0.9, in NGC 1977 it is significantly lower (0.79).

The dust temperature in OMC1w is somewhat lower than in the region to the southeast of the Orion Bar. The dust optical depth at 160\,$\mu$m lies between $8\times 10^{-3}$ and $1\times 10^{-2}$. Thus, although this region is associated with OMC1, it is included in our OMC4 correlation. Points in the region south of OMC1 and north of OMC4, OMC1s, are also included in the OMC4 correlation. Those points have higher [C\,{\sc ii}] intensity at the same FIR intensity as points in OMC1w.

Most scatter about the regression curve is observed in OMC4, which also possesses a larger average slope. Also the bright-CO and faint-[C\,{\sc ii}] regions in M43, lying close to the spine of the ISF, deviate from the regression curve, which might be due to the FIR emission stemming from the colder molecular background. The same is true for OMC3, as part of the ISF. As the FIR intensity is dominated by the PACS $70\,\mu\mathrm{m}$ band, the [C\,{\sc ii}]-FIR correlation is very similar to the [C\,{\sc ii}]-$70\,\mu\mathrm{m}$ correlation. The low-intensity regime ($I_{\mathrm{FIR}}<3\times 10^{-2}\,\mathrm{erg\,s^{-1}\,cm^{-2}\,sr^{-1}}$) is again fitted with exponents larger than unity, but with large standard errors of the fit parameters and lower correlation coefficients ($\rho\simeq 0.8$).

The points in the cavity to the west of the molecular ridge of Orion S/BN-KL, that is in OMC1w, lie below the average regression of the OMC4 correlation. This region is just to the north of the extension of the Orion Bar and is characterized by a higher [C\,{\sc ii}] centroid velocity as compared to the average molecular background velocity. This region is subject to large-scale shocks identified by \cite{Henney2007}.

\subsection{[C\,{\sc ii}] versus $8\,\mu\mathrm{m}$}
\label{CII-vs-8}

\begin{figure*}[tb]
\begin{minipage}{0.49\textwidth}
\includegraphics[width=\textwidth, height=0.67\textwidth]{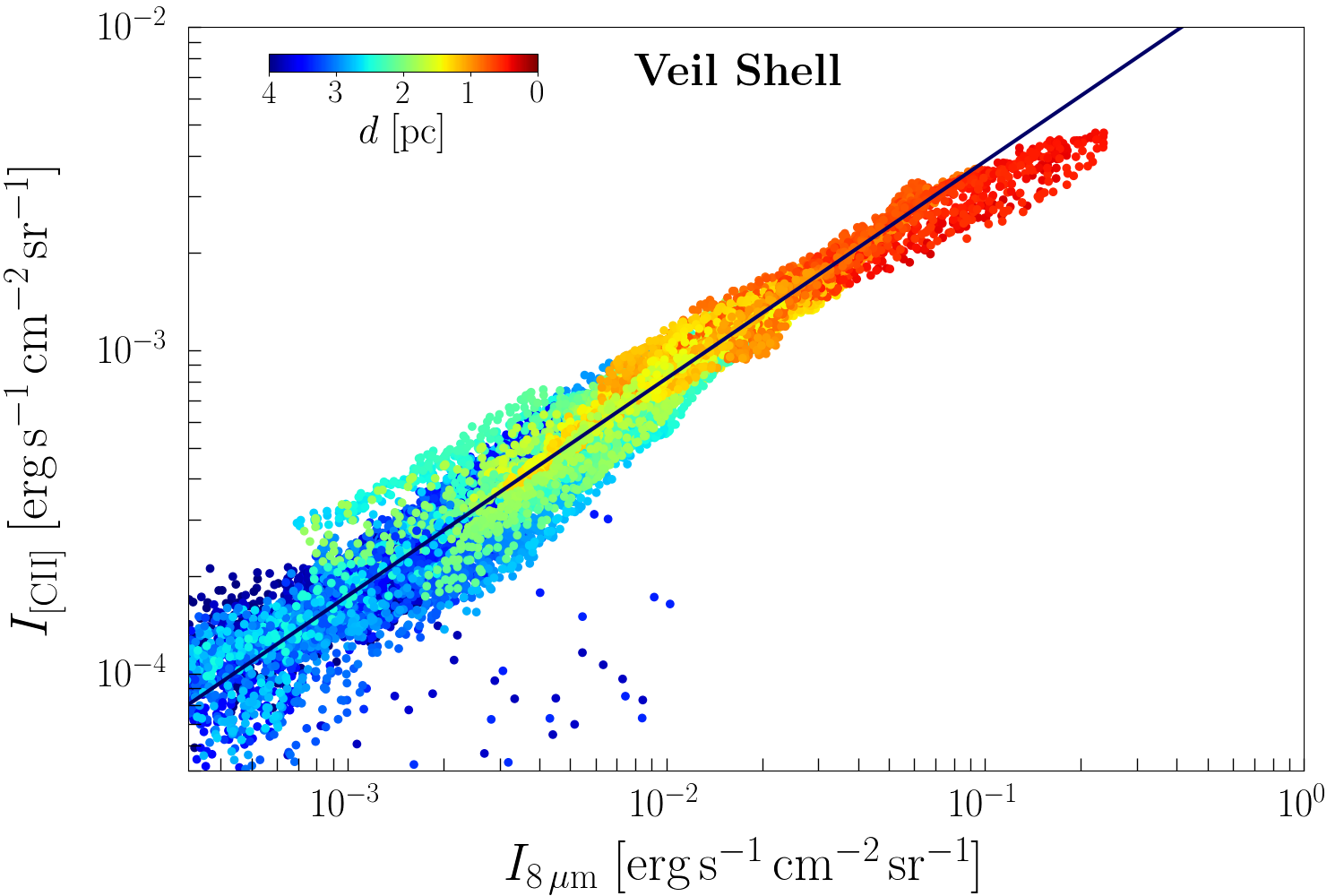}
\end{minipage}
\begin{minipage}{0.49\textwidth}
\includegraphics[width=\textwidth, height=0.67\textwidth]{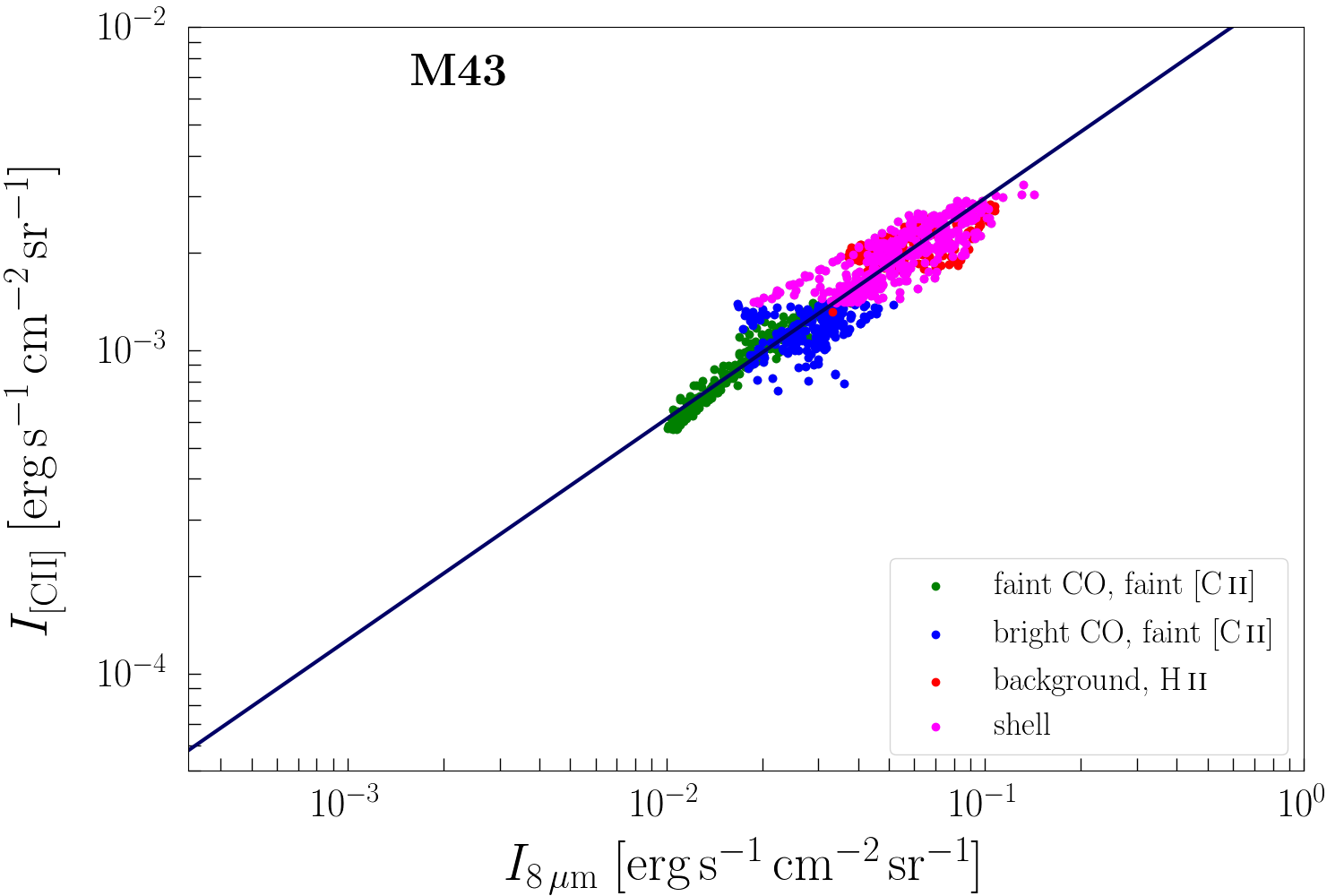}
\end{minipage}

\begin{minipage}{0.49\textwidth}
\includegraphics[width=\textwidth, height=0.67\textwidth]{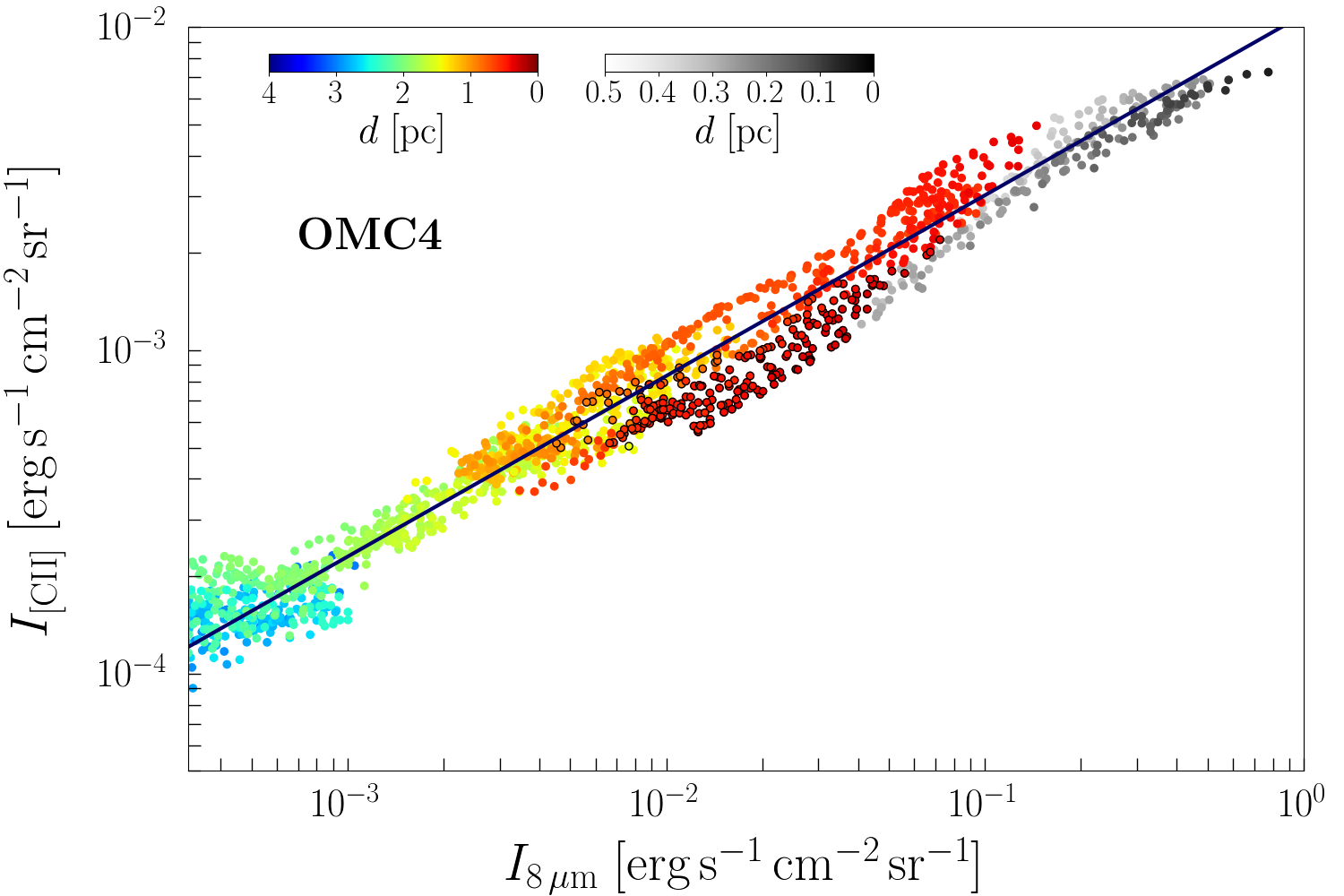}
\end{minipage}
\begin{minipage}{0.49\textwidth}
\includegraphics[width=\textwidth, height=0.67\textwidth]{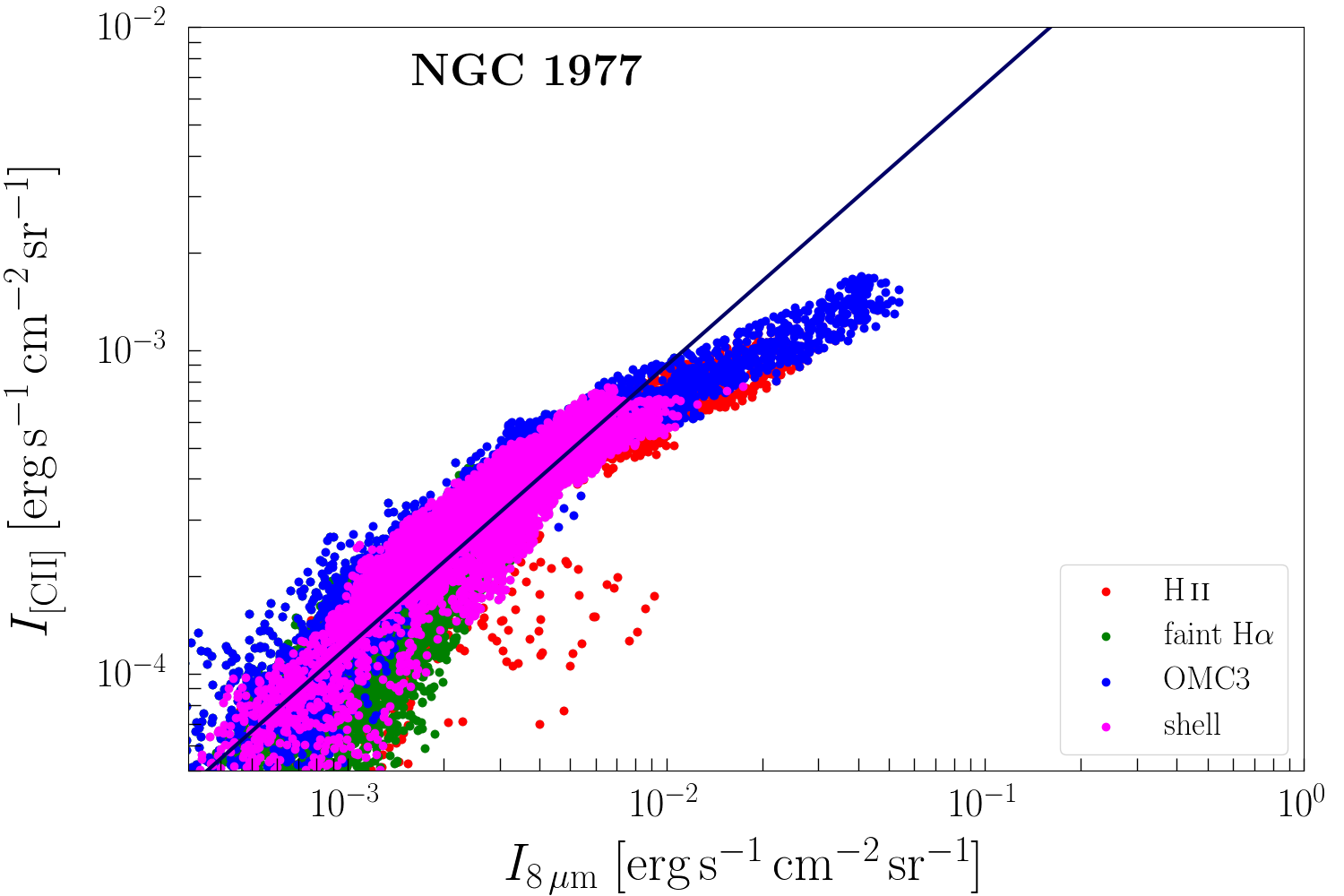}
\end{minipage}
\caption{[C\,{\sc ii}] intensity versus IRAC $8\,\mu\mathrm{m}$ intensity in M42 (top left: Veil Shell, bottom left: OMC4 with OMC1 (in gray scale; black-edged red points lie to the west of Orion S)), M43 (top right) and NGC 1977 (bottom right). The color scale in M42 indicates the distance from $\theta^1$ Ori C. The division of M43 and NGC 1977 into the regions given in the legend is described in Section \ref{Sec.subregions}. The blue solid lines are the fits summarized in Table \ref{Tab.CII-8}.}
\label{Fig.CII-8}
\end{figure*}

The [C\,{\sc ii}] intensity is compared to the $8\,\mu\mathrm{m}$ intensity in Fig. \ref{Fig.CII-8}. As this figure demonstrates, these intensities are tightly correlated. Table \ref{Tab.CII-8} summarizes the [C\,{\sc ii}]-$8\,\mu\mathrm{m}$ correlation in the four regions.

\begin{table}[ht]
\def\arraystretch{1.2}
\addtolength{\tabcolsep}{-1.5pt}
\caption{Summary of [C\,{\sc ii}]-$8\,\mu\mathrm{m}$ correlation in Fig. \ref{Fig.CII-8}.}
\begin{tabular}{l|cccc}
\hline\hline
Region & $a$ & $b$ & $\rho$ & $rms\,\mathrm{[dex]}$ \\ \hline
Veil Shell & $0.67\pm 0.27$ & $-1.74\pm 0.42$ & 0.97 & 0.09\\
OMC4 & $0.56\pm 0.10$ & $-1.96\pm 0.29$ & 0.98 & 0.09\\
M43 & $0.68\pm 0.20$ & $-1.85\pm 0.28$ & 0.93 & 0.06\\
NGC 1977 & $0.87\pm 0.45$ & $-1.31\pm 1.18$ & 0.92 & 0.11 \\ \hline
\end{tabular}
\tablefoot{$\log_{10} I_{\mathrm{[C\,\textsc{ii}]}} = a \log_{10} I_{\mathrm{8\,\mu m}} + b$. Fit for $I_{8\,\mu\mathrm{m}}>3\times 10^{-4}\,\mathrm{erg\,s^{-1}\,cm^{-2}\,sr^{-1}}$. $\rho$ is the Pearson correlation coefficient, $rms$ is the root-mean-square of the residual of the least-squares fit.}
\label{Tab.CII-8}
\end{table}

The overall [C\,{\sc ii}]-$8\,\mu\mathrm{m}$ relation is similar in all four regions. The [C\,{\sc ii}] intensity depends less than linear on the $8\,\mu\mathrm{m}$ intensity, with an exponent of $0.56\text{-}0.87$. The correlation coefficient for each region is larger than 0.9, and the scatter around the regression is usually small ($rms<0.11\,\mathrm{dex}$). The [C\,{\sc ii}] intensity tends to turn off the regression curve toward lower values at the high-intensity end. In the correlation plot of OMC4, points in OMC1w have lower [C\,{\sc ii}] intensity at the same $8\,\mu\mathrm{m}$ intensity compared to points in OMC1s. Points within OMC1 continue the general [C\,{\sc ii}]-$8\,\mu\mathrm{m}$ trend.

One difficulty in the interpretation of the correlation is the unknown additive intensity offset in the IRAC $8\,\mu\mathrm{m}$ band, a known IRAC calibration issue for extended emission (IRAC Instrument Handbood Sec. 8.2). We estimated the offset at $2\times 10^{-3}\,\mathrm{erg\,s^{-1}\,cm^{-2}\,sr^{-1}}$, but it might also be half as large. This does not affect the regression greatly if we restrict ourselves to the high-intensity end. It will affect, however, the estimate of the [C\,{\sc ii}] intensity from the $8\,\mu\mathrm{m}$ intensity at low intensities.

\subsection{FIR versus $8\,\mu\mathrm{m}$}
\label{FIR-vs-8}

\begin{figure*}[tb]
\begin{minipage}{0.49\textwidth}
\includegraphics[width=\textwidth, height=0.67\textwidth]{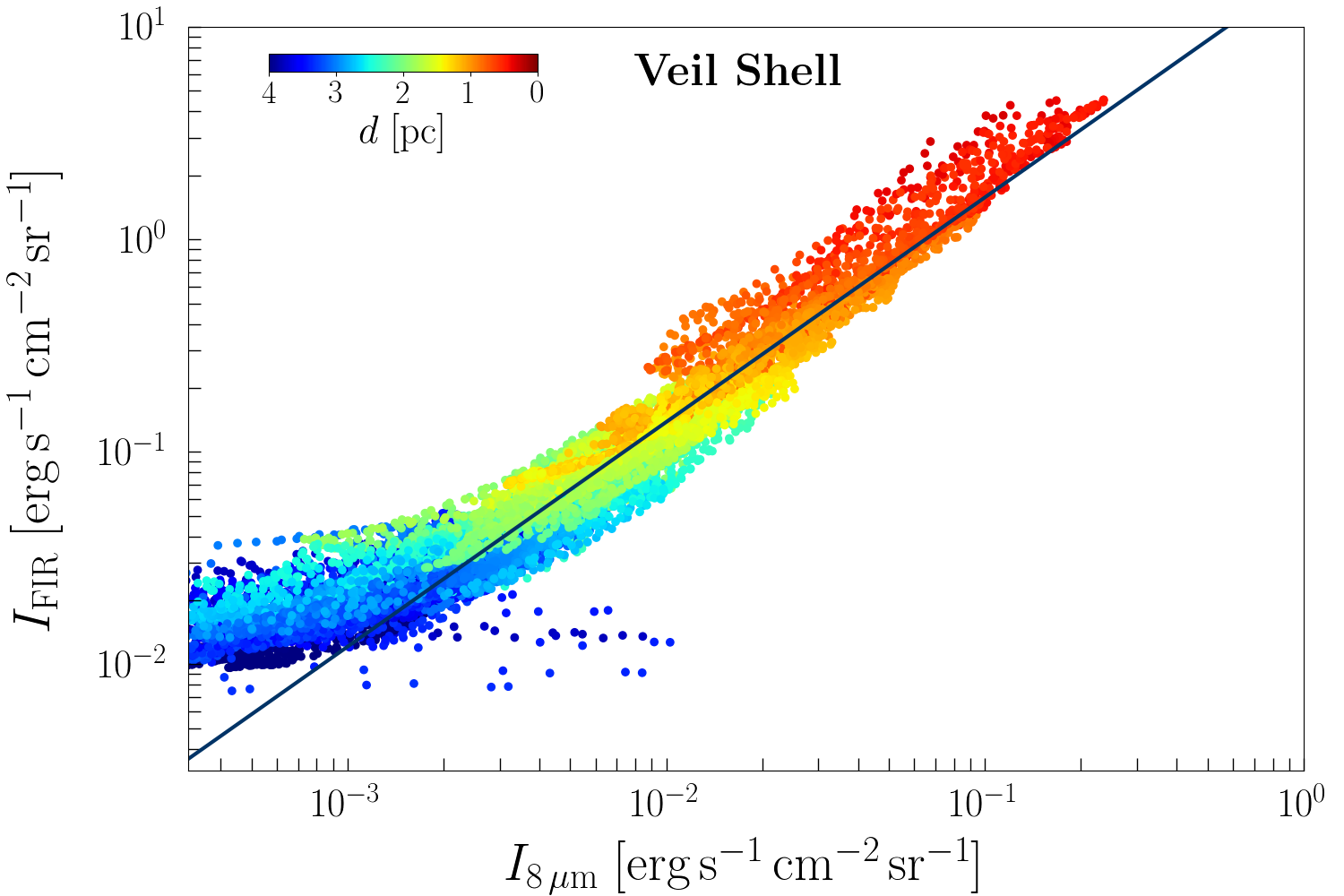}
\end{minipage}
\begin{minipage}{0.49\textwidth}
\includegraphics[width=\textwidth, height=0.67\textwidth]{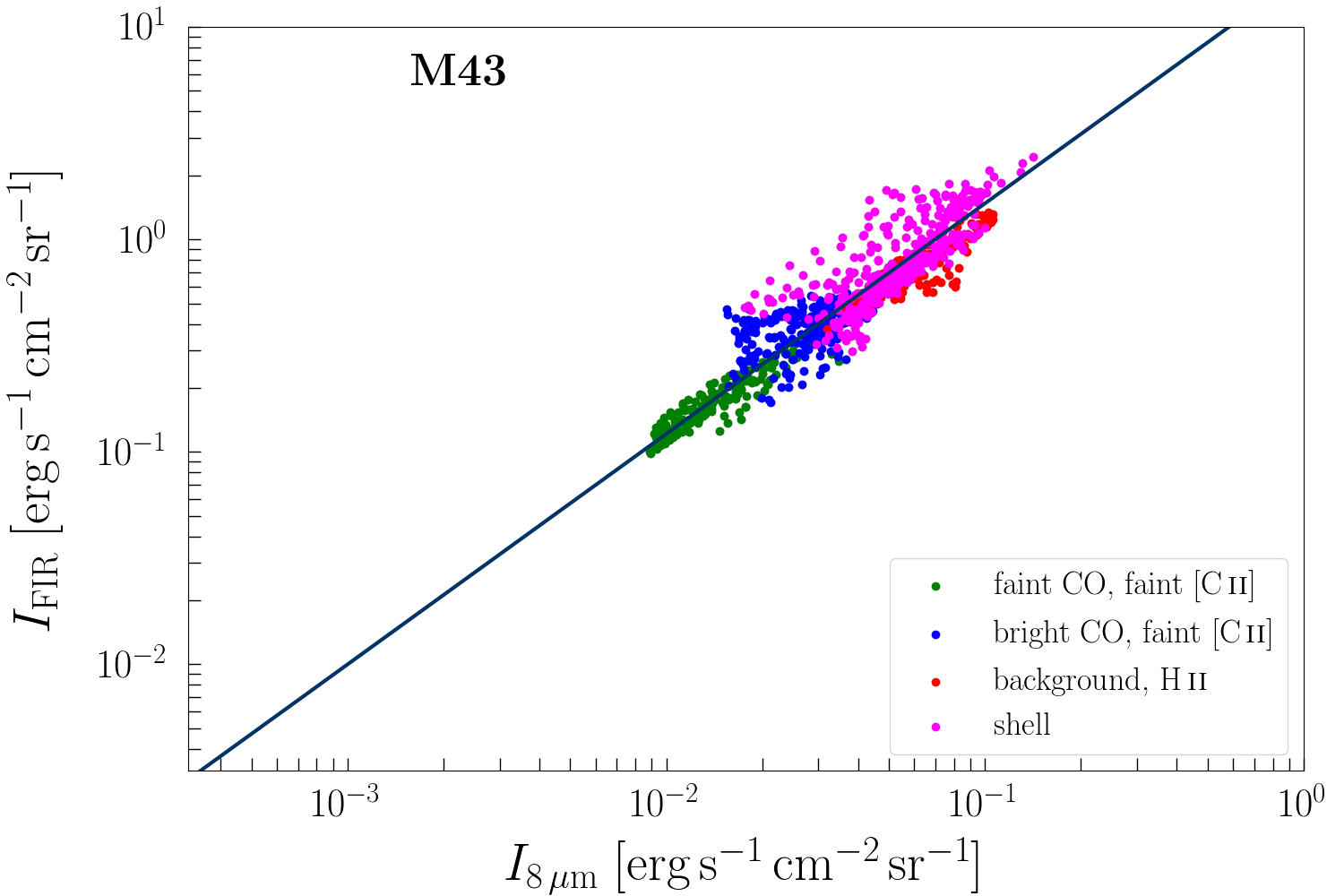}
\end{minipage}

\begin{minipage}{0.49\textwidth}
\includegraphics[width=\textwidth, height=0.67\textwidth]{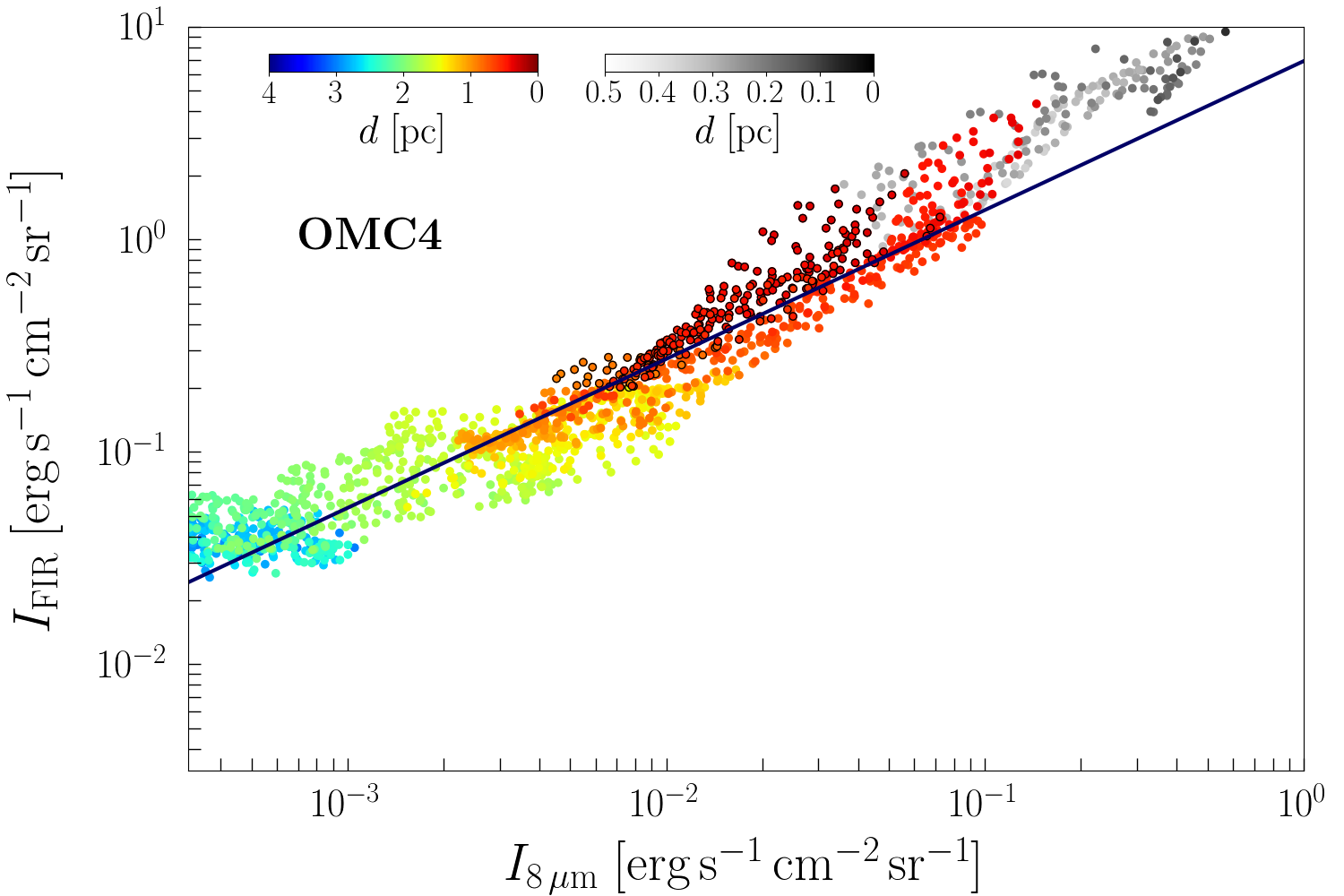}
\end{minipage}
\begin{minipage}{0.49\textwidth}
\includegraphics[width=\textwidth, height=0.67\textwidth]{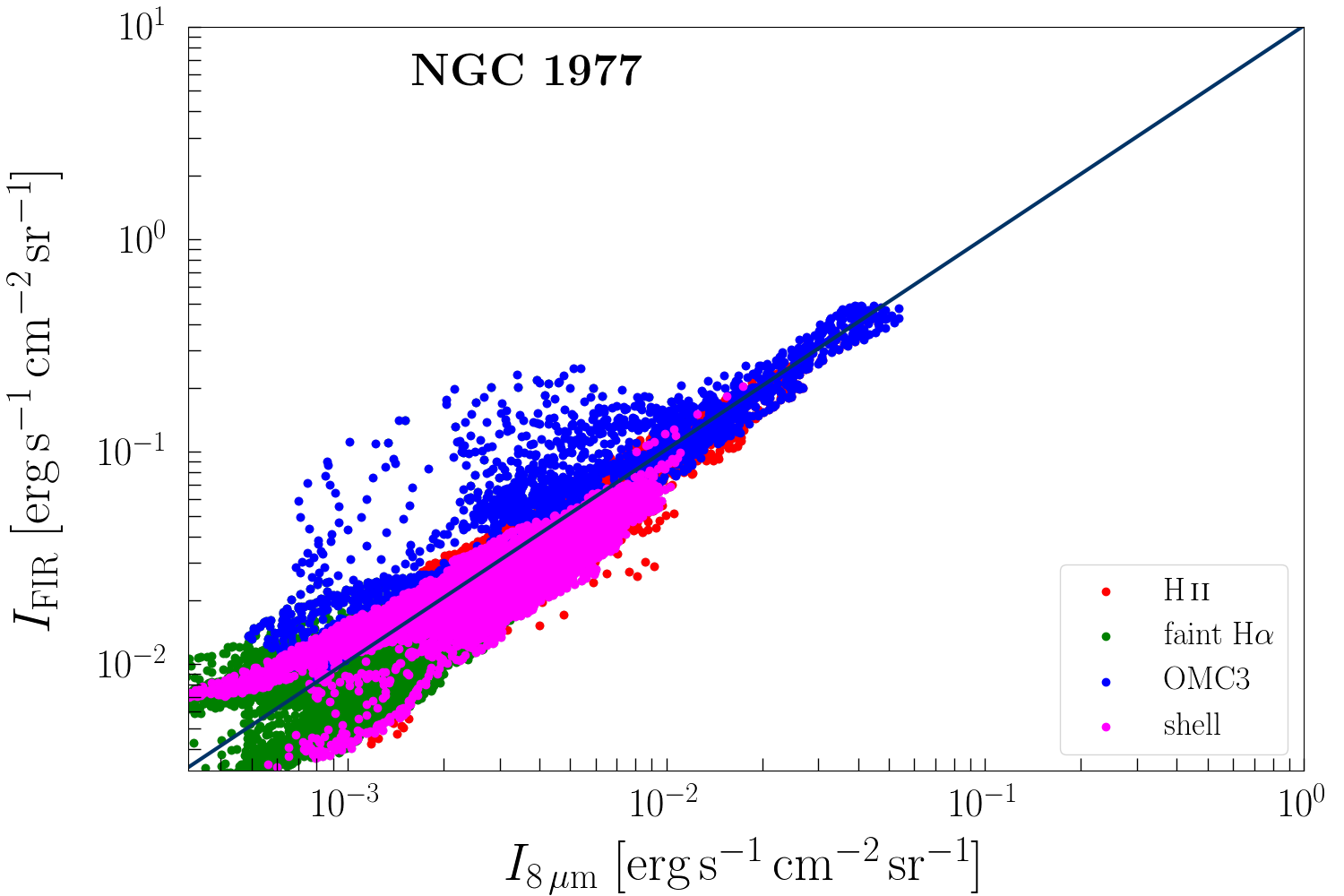}
\end{minipage}
\caption{FIR intensity versus IRAC $8\,\mu\mathrm{m}$ in M42 (top left: Veil Shell, bottom left: OMC4 with OMC1 (in gray scale; black-edged red points lie to the west of Orion S)), M43 (top right) and NGC 1977 (bottom right). The color scale in M42 indicates the distance from $\theta^1$ Ori C. The division of M43 and NGC 1977 into the regions given in the legend is described in Section \ref{Sec.subregions}. The blue solid lines are the fits summarized in Table \ref{Tab.FIR-8}.}
\label{Fig.FIR-8}
\end{figure*}

The FIR intensity is compared to the $8\,\mu\mathrm{m}$ intensity in Fig. \ref{Fig.FIR-8}. As this figure shows, these intensities are tightly correlated. Table \ref{Tab.FIR-8} summarizes the FIR-$8\,\mu\mathrm{m}$ correlations in the four regions. We consider only the high-intensity regime ($I_{\mathrm{FIR}}>3\times 10^{-2}\,\mathrm{erg\,s^{-1}\,cm^{-2}\,sr^{-1}}$).

\begin{table}[ht]
\def\arraystretch{1.2}
\addtolength{\tabcolsep}{-0.5pt}
\caption{Summary of FIR-$8\,\mu\mathrm{m}$ correlation in Fig. \ref{Fig.FIR-8}.}
\begin{tabular}{l|cccc}
\hline\hline
Region & $a$ & $b$ & $\rho$ & $rms\,\mathrm{[dex]}$\\ \hline
Veil Shell & $1.06\pm 0.44$ & $1.26\pm 0.92$ & 0.95 & 0.14 \\
OMC4 & $0.70\pm 0.24$ & $0.84\pm 0.56$ & 0.95 & 0.15 \\
M43 & $1.08\pm 0.36$ & $1.25\pm 0.51$ & 0.92 & 0.11 \\ 
NGC 1977 & $1.00\pm 0.50$ & $1.00\pm 1.12$ & 0.83 & 0.15 \\ \hline
\end{tabular}
\tablefoot{$\log_{10} I_{\mathrm{FIR}} = a \log_{10} I_{\mathrm{8\,\mu m}} + b$. Fit for $I_{\mathrm{FIR}}>3\times 10^{-2}\,\mathrm{erg\,s^{-1}\,cm^{-2}\,sr^{-1}}$. $\rho$ is the Pearson correlation coefficient, $rms$ is the root-mean-square of the residual of the least-squares fit.}
\label{Tab.FIR-8}
\end{table}

While both the [C\,{\sc ii}]-FIR and the [C\,{\sc ii}]-$8\,\mu\mathrm{m}$ relations are far from linear, the $8\,\mu\mathrm{m}$-FIR relation is approximately linear. The correlation coefficient is larger than 0.9 in M42 and M43, and 0.83 in NGC 1977. Scatter about the regression curve is significant, though, with $0.11\text{-}0.15\,\mathrm{dex}$. In OMC4, the FIR-$8\,\mu\mathrm{m}$ relation differs slightly from the relations found in the other regions, with a power-law exponent smaller than unity. Points in OMC1w have higher FIR intensity than others with the same $8\,\mu\mathrm{m}$ intensity. Points inside OMC1 continue the trend of those high-FIR points.

As the FIR measures the incident (FUV) radiation field, the good correlation between the $8\,\mu\mathrm{m}$ and FIR intensities implies that the PAH flux is proportional to the incident FUV field. Given that the IRAC $8\,\mu\mathrm{m}$ band measures the emission by PAH cations \citep{Peeters2002}, this implies that there is little variation in the PAH ionization balance within each of these regions and between the regions. Conversely, this implies that the observed $8\,\mu\mathrm{m}$ intensity is a good measure of the incident FUV field. Overall, we find that $I_{\mathrm{FIR}}/I_{\mathrm{8\,\mu m}} \simeq 10$.

\subsection{[C\,{\sc ii}] versus CO(2-1)}
\label{CII-vs-CO}

\begin{figure*}[tb]
\begin{minipage}{0.49\textwidth}
\includegraphics[width=\textwidth, height=0.67\textwidth]{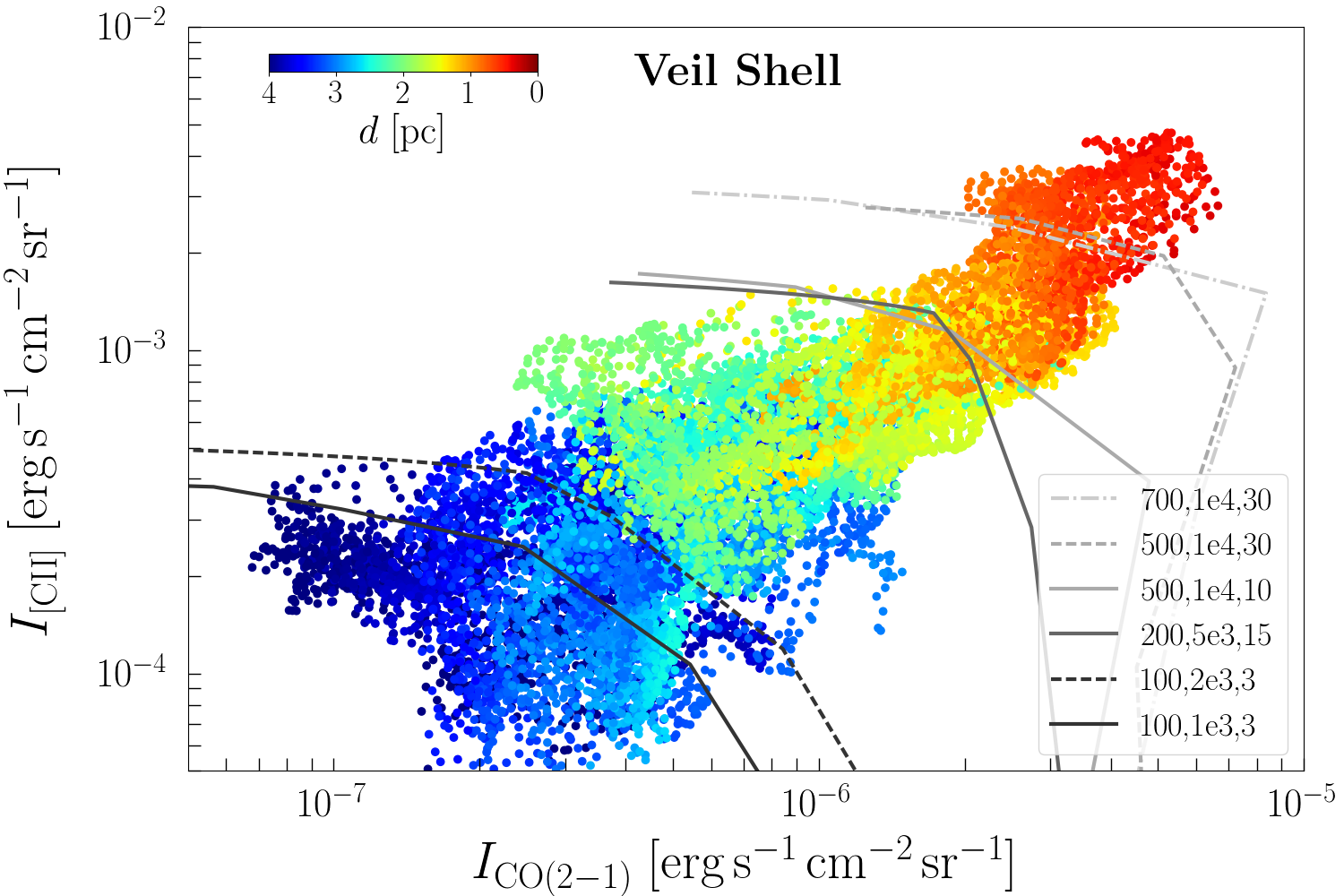}
\end{minipage}
\begin{minipage}{0.49\textwidth}
\includegraphics[width=\textwidth, height=0.67\textwidth]{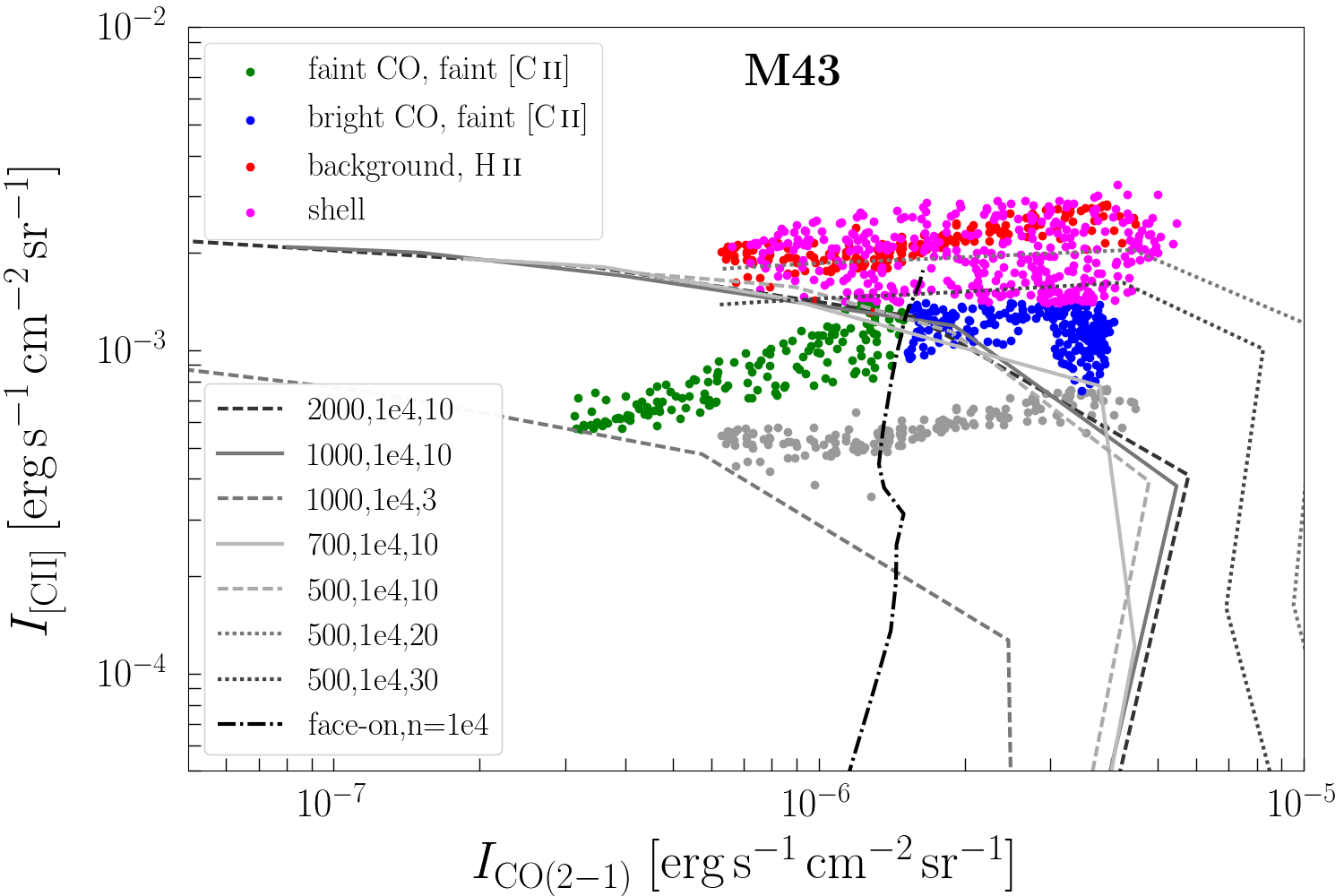}
\end{minipage}

\begin{minipage}{0.49\textwidth}
\includegraphics[width=\textwidth, height=0.67\textwidth]{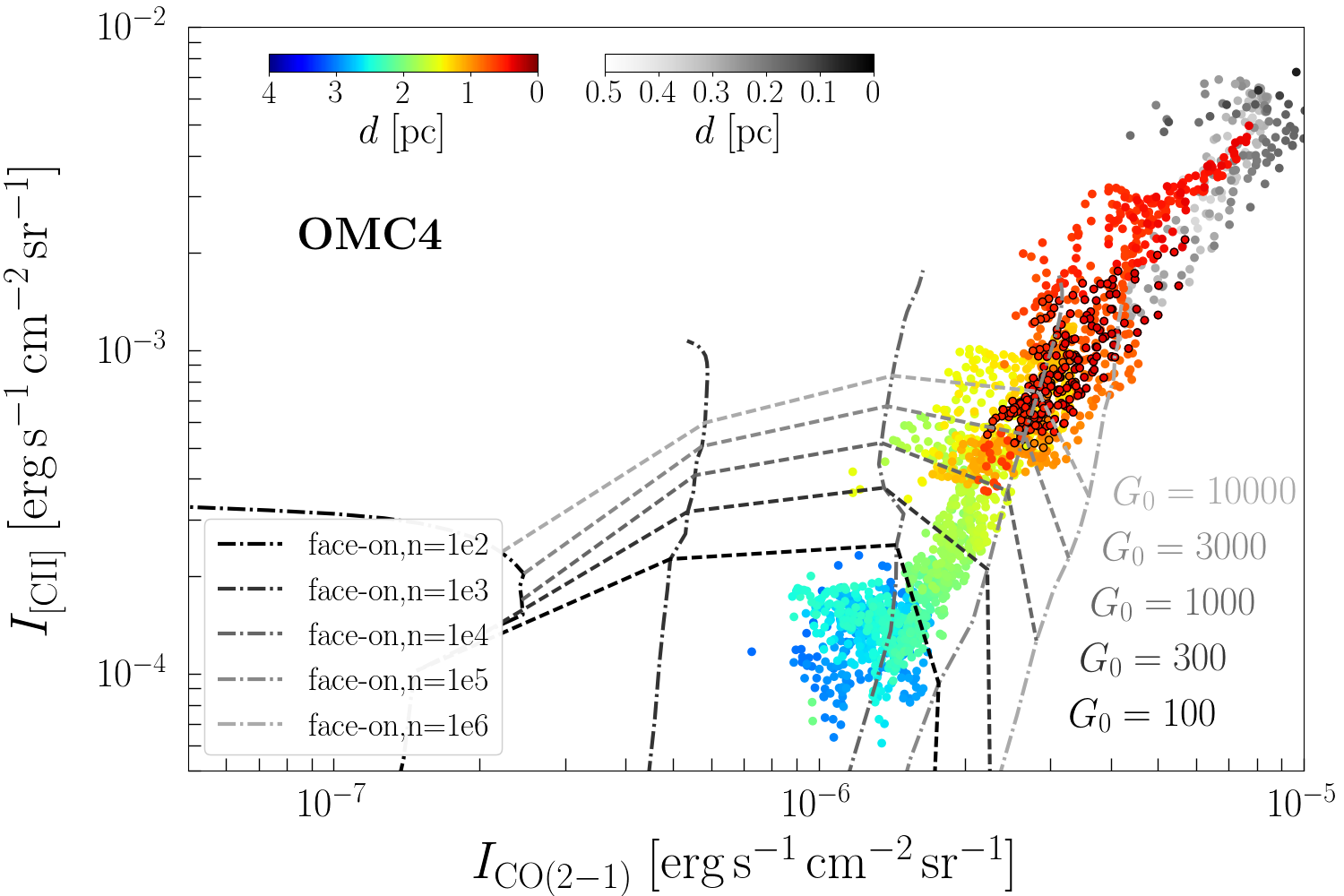}
\end{minipage}
\begin{minipage}{0.49\textwidth}
\includegraphics[width=\textwidth, height=0.67\textwidth]{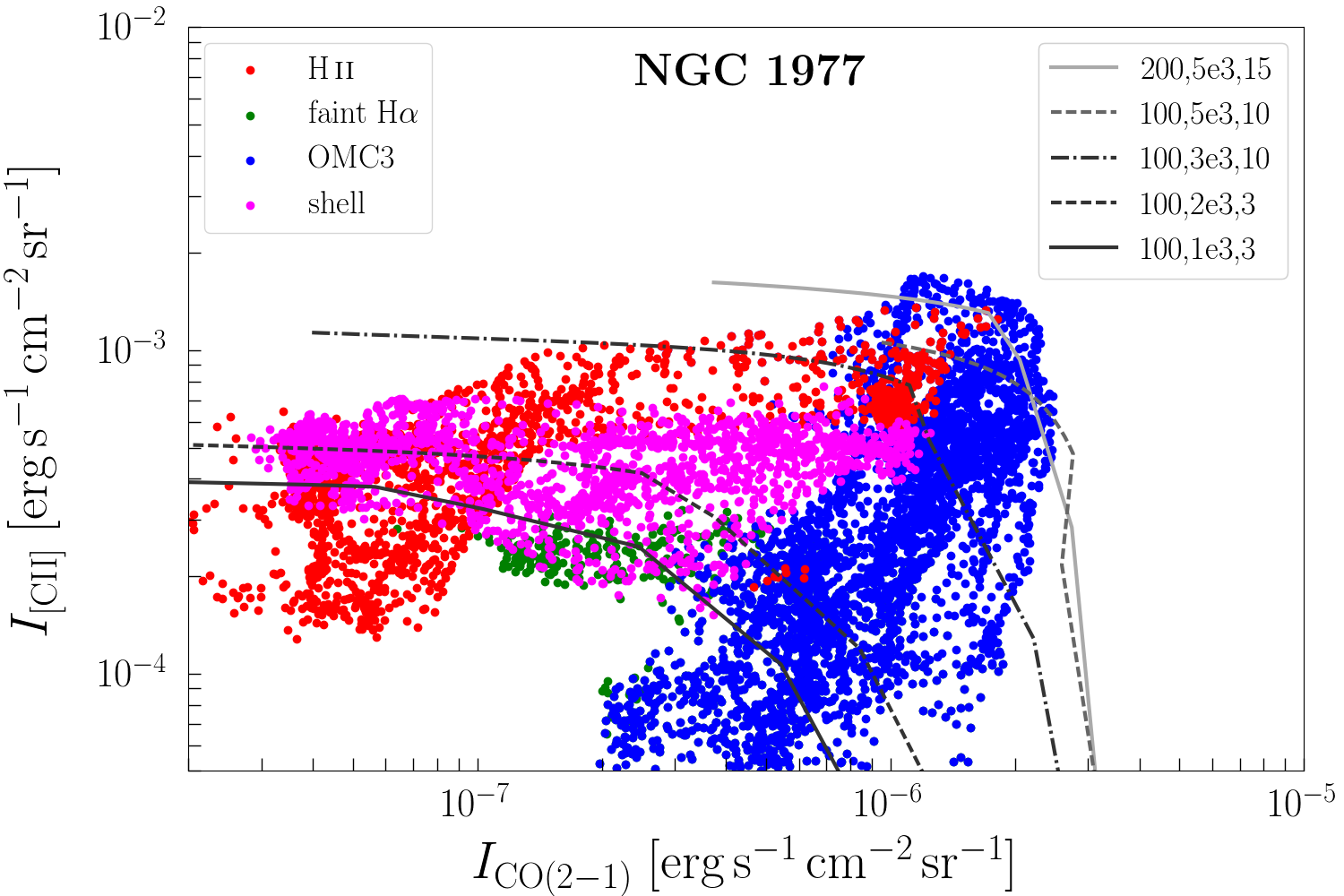}
\end{minipage}
\caption{[C\,{\sc ii}] versus CO(2-1) intensity in M42 (top left: Veil Shell, bottom left: OMC4 with OMC1 (in gray scale; black-edged red points lie to the west of Orion S)), M43 (top right) and NGC 1977 (bottom right). The color scale in M42 indicates the distance from $\theta^1$ Ori C. The division of M43 and NGC 1977 into the regions given in the legend is described in Section \ref{Sec.subregions}. Dashed lines are edge-on model outputs with the parameters ($G_0$, $n$, $A_{\mathrm{V,los}}$) given in the legend. Dashed dotted lines in the OMC4 and the M43 panels are face-on model outputs for the densities given in the legend. gray points in M43 correspond to the estimated intensities of the background PDR toward the center of M43.}
\label{Fig.CII-CO}
\end{figure*}

The [C\,{\sc ii}] intensity is compared to the CO (J=2-1) intensity in Fig. \ref{Fig.CII-CO}. While for the Veil Shell and OMC4, there are clear trends -- albeit with large scatter -- there is little correlation for M43 and NGC 1977. Specifically, for M43, the correlation coefficient is only 0.37. The global correlation coefficient for NGC 1977 is only 0.01. Limiting ourselves to the OMC 3 region in this source results in a much better correlation coefficient ($\rho=0.67$). toward OMC4, [C\,{\sc ii}] and CO(2-1) intensities also exhibit a correlation, with less scatter than in the Veil Shell ($\rho=0.93$). Despite the fact that the Veil Shell is a PDR viewed edge-on and we expect a spatial offset between the [C\,{\sc ii}]- and the CO-emitting gas layers, [C\,{\sc ii}] and CO(2-1) intensities show a correlation, albeit with large scatter ($\rho=0.79$). The Veil Shell itself is mostly CO-dark, however containing several CO-emitting globules \citep{Goicoechea2020}. CO emission is mostly stemming from the background molecular cloud and not the shell itself. Only in the Eastern Rim there is corresponding CO(2-1) emission at a small spatial offset.

In the center of M43, CO(2-1) emission stems mainly from the molecular cloud behind the H\,{\sc ii} region, but has an additional broad, weak component of yet undetermined origin \citep{Pabst2020}. The [C\,{\sc ii}]-bright edge is also bright in CO(2-1), albeit with a slight spatial separation. The shell of NGC 1977 has nearly constant [C\,{\sc ii}] intensity, but the CO(2-1) varies by a factor of about 30. In contrast, the CO(2-1) intensity is toward the OMC3 region varies little, while [C\,{\sc ii}] varies by two orders of magnitude.

\begin{figure*}[tb]
\begin{minipage}{0.49\textwidth}
\includegraphics[width=\textwidth, height=0.67\textwidth]{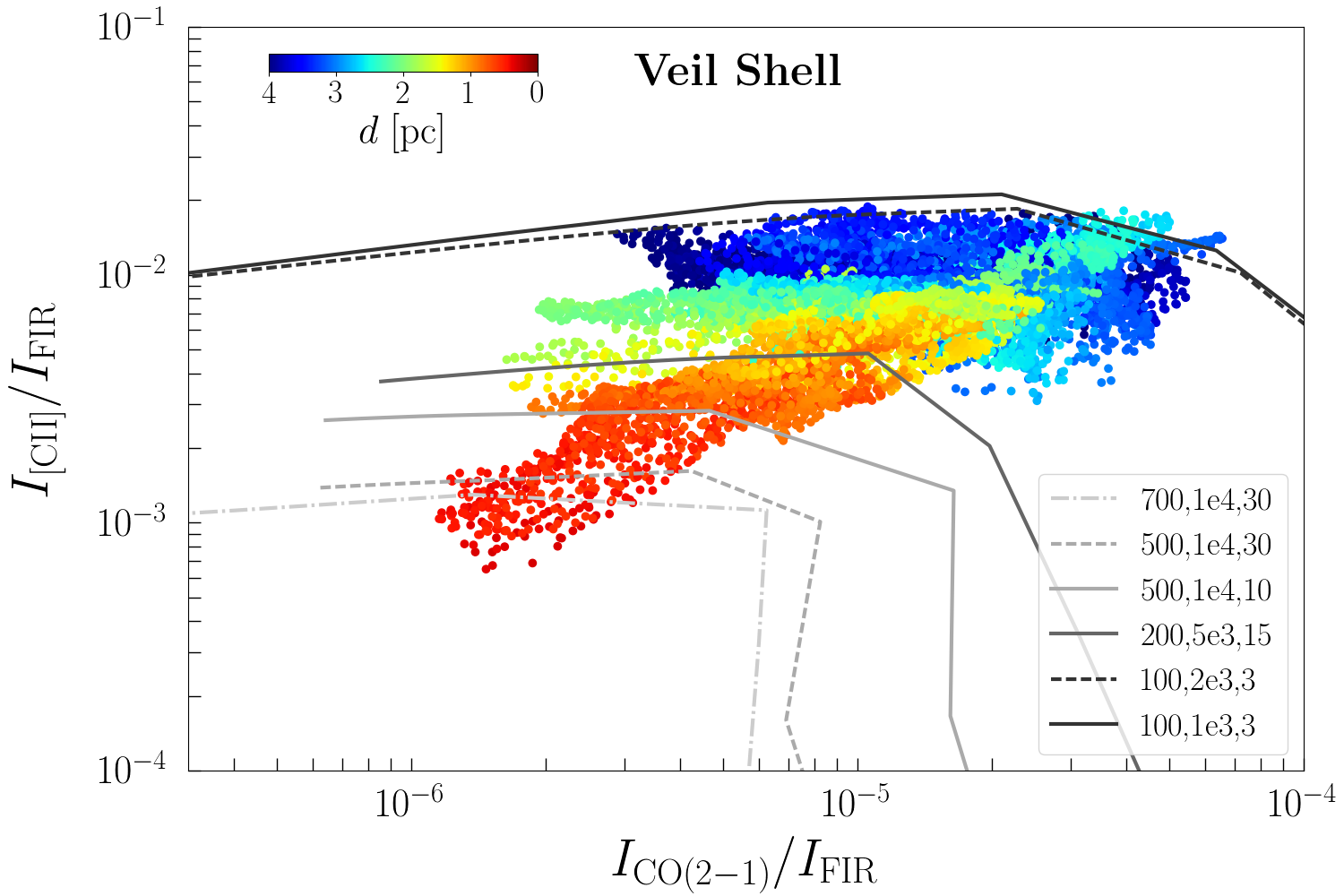}
\end{minipage}
\begin{minipage}{0.49\textwidth}
\includegraphics[width=\textwidth, height=0.67\textwidth]{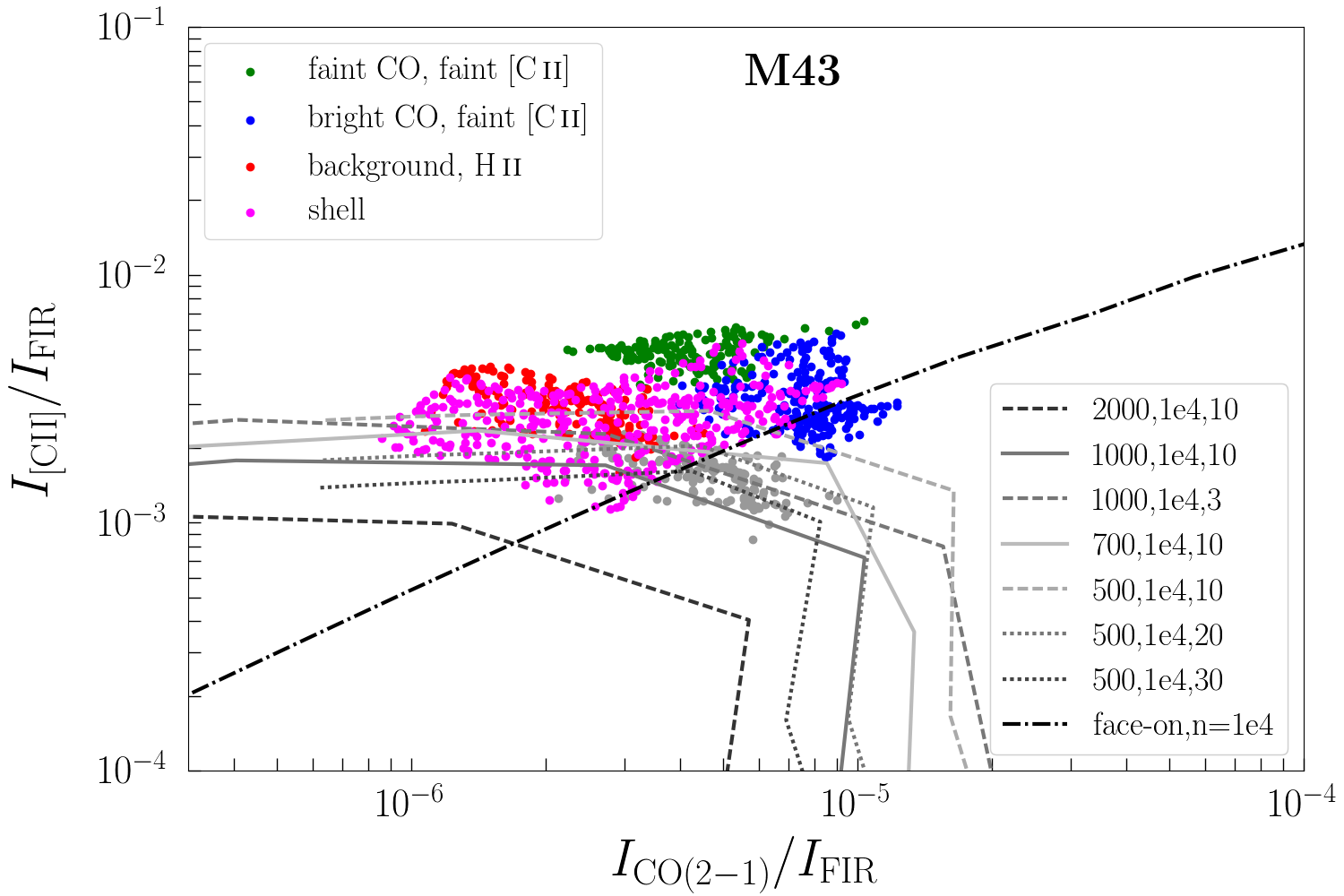}
\end{minipage}

\begin{minipage}{0.49\textwidth}
\includegraphics[width=\textwidth, height=0.67\textwidth]{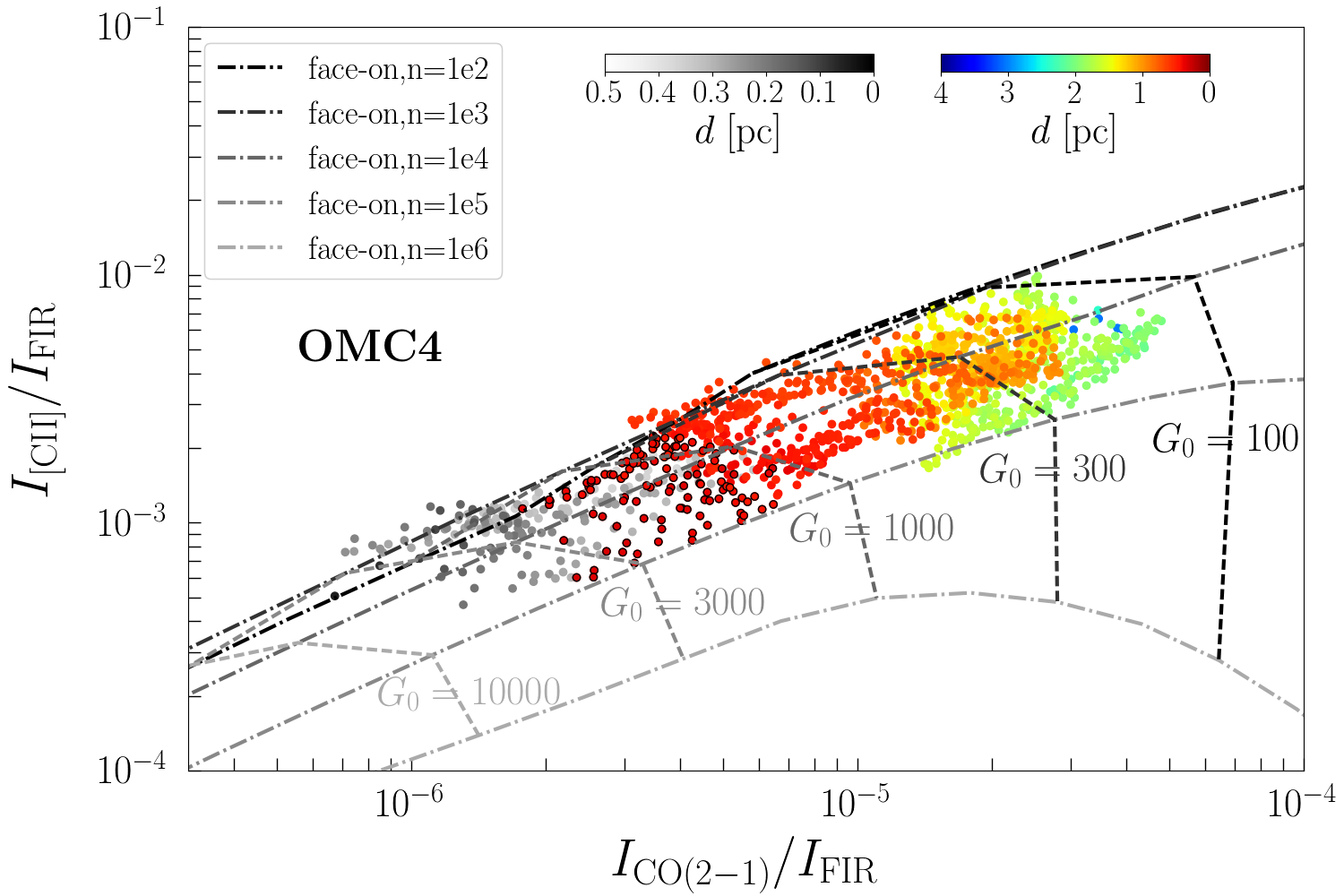}
\end{minipage}
\begin{minipage}{0.49\textwidth}
\includegraphics[width=\textwidth, height=0.67\textwidth]{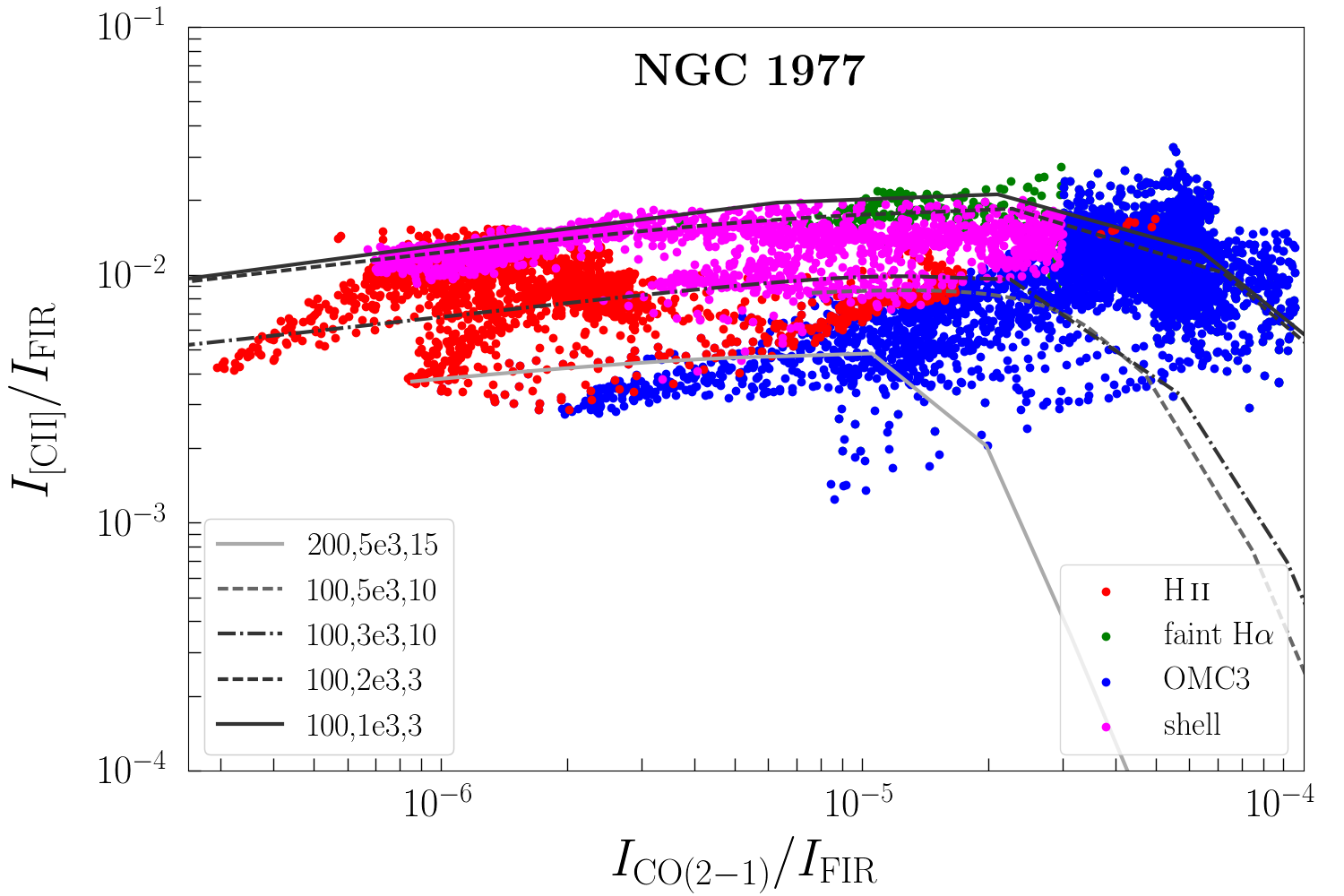}
\end{minipage}
\caption{[C\,{\sc ii}]/FIR versus CO(2-1)/FIR in M42 (top left: Veil Shell, bottom left: OMC4 with OMC1 (in gray scale; black-edged red points lie to the west of Orion S)), M43 (top right) and NGC 1977 (bottom right). The color scale in M42 indicates the distance from $\theta^1$ Ori C. The division of M43 and NGC 1977 into the regions given in the legend is described in Section \ref{Sec.subregions}. Dashed lines are edge-on model outputs with the parameters ($G_0$, $n$, $A_{\mathrm{V,los}}$) given in the legend. Dashed dotted lines in the OMC4 and the M43 panels are face-on model outputs for the densities given in the legend. gray points in M43 correspond to the estimated intensities of the background PDR toward the center of M43.}
\label{Fig.CII-CO-2}
\end{figure*}

The [C\,{\sc ii}]/FIR-CO(2-1)/FIR correlations, shown in Fig. \ref{Fig.CII-CO-2}, exhibit somewhat different behavior than the [C\,{\sc ii}]-CO(2-1) correlations. The normalized intensities are generally less correlated than the absolute intensities ($\rho=0.53$ in the Veil Shell, $\rho=0.77$ in OMC4, $\rho=0.22$ in M43, but $\rho=0.17$ in NGC 1977 and $\rho=0.67$ in OMC3 only). If one were to fit regression curves to the data (both absolute and normalized intensities), no consistent picture would emerge, the exponents of the power laws differing widely. This is reflected in the global correlations discussed in Paper I.

\subsection{Summary of the correlation studies}

In summary, the correlations of the [C\,{\sc ii}] intensity with the $70\,\mu\mathrm{m}$, FIR, and $8\,\mu\mathrm{m}$ intensity, respectively, are tight. The four regions, the M42 Veil Shell, OMC4, M43, and NGC 1977, behave in a similar manner, and the spread we observe is minor. The spread may be due to the dependency of the gas heating efficiency on the physical conditions and possibly subtle variations of the abundance of PAHs and VSGs through the regions. Most important to note is that the correlation of the [C\,{\sc ii}] intensity with each of the $70\,\mu\mathrm{m}$, FIR, and $8\,\mu\mathrm{m}$ intensity is not linear, but scales with those intensities to a power of less than unity. The FIR-$8\,\mu\mathrm{m}$ correlation deviates only slightly from linearity. In contrast, the [C\,{\sc ii}]-CO(2-1) correlations reveal a more complex behavior. This requires a more in-depth study which we pursue in section 4.

\section{Discussion}

\subsection{Edge-on PDR models}
\label{sec.model_description}

We complement the correlations of the [C\,{\sc ii}] emission with gas and dust tracers in the following with new PDR model runs that are based on the models of \cite{TielensHollenbach1985} with updates of \cite{Wolfire2010} and \cite{Hollenbach2012}. \cite{Pabst2017} describe the procedure by which the emission from an edge-on PDR of a given line-of-sight depth ($A_\mathrm{V,los}$) is computed from the output of a 1D face-on PDR model. Compared to those earlier models, we have updated the collisional excitation rates and chemical rates in the models. In particular, we adopt the [O\,{\sc i}] collisional rates with H and H$_2$ from \citet{Lique2018} (and F. Lique priv. comm.). Updates to chemical rates are noted in \cite{NeufeldWolfire2016}. More recently, we updated the photorates to those from \cite{Heays2017} and adopted the exponential integral formalism  for the depth dependence due to dust. Additional changes to the chemical rates include oxygen chemistry rates from \cite{Kovalenko2018} and \cite{Tran2018}, and carbon chemistry rates from \cite{Dagdigian2019}. In order to compute synthetic line emission intensities we assume a Doppler line width of $\Delta v = 1.5\,\mathrm{km\,s}^{-1}$.

\begin{figure*}[!tb]
\includegraphics[width=\textwidth, height=0.6\textwidth]{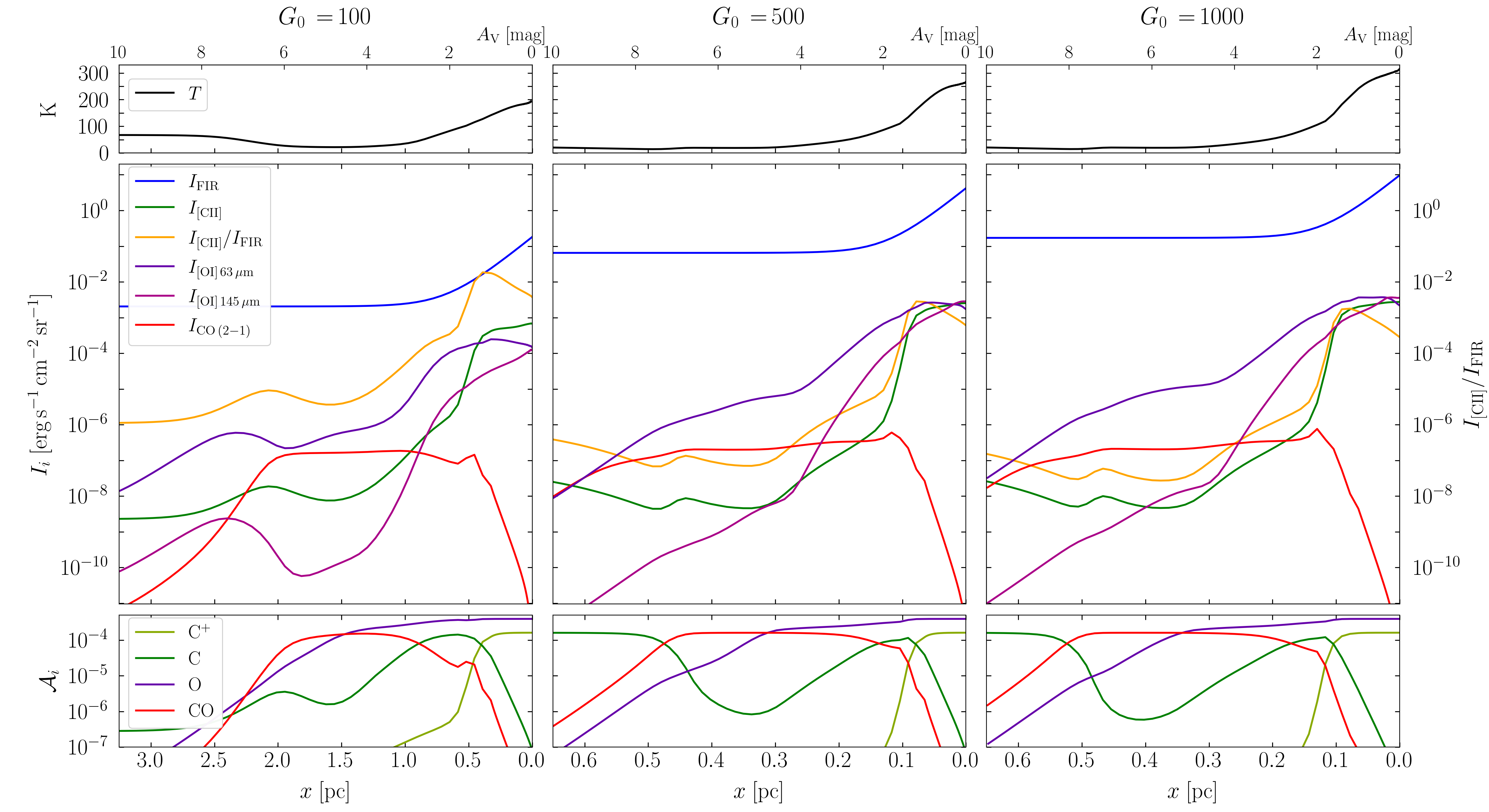}
\caption{Results of our edge-on models described in Section \ref{sec.model_description} for incident FUV intensities of $G_0=100,\;500,\,\mbox{ and }1000$ on physical scale, illuminated from the right. The panels show the gas temperature $T$ (upper panels), $I_{\mathrm{FIR}}$, $I_{\mathrm{[C\,\textsc{ii}]}}$, $I_{\mathrm{[C\,\textsc{ii}]}}/I_{\mathrm{FIR,}}$, $I_{\mathrm{[O\,\textsc{i}]\,63\,\mu m}}$, $I_{\mathrm{[O\,\textsc{i}]\,145\,\mu m}}$, and $I_{\mathrm{CO\,(2\text{-}1)}}$ (middle panels) and C$^+$, C, O, and CO fractional abundances (lower panels) versus physical scale. The gas densities are $n_{\mathrm{H}}=2.0\times 10^3\,\mathrm{cm^{-3}},\;1.0\times10^4\,\mathrm{cm^{-3}},\;1.0\times 10^4\,\mathrm{cm^{-3}}$ (left to right panels), and the respective line-of-sight depths are $A_\mathrm{V,los}=3,10,10$. We note that $I_{\mathrm{[CII]}}/I_{\mathrm{FIR}}$ is a unitless ratio, whose value is given by the y-axis.}
\label{Fig.model}
\end{figure*}

We compute models for several FUV intensities, gas densities, and line-of-sight visual extinctions $A_\mathrm{V,los}$ appropriate for different regions in the Orion Nebula, M43 and NGC 1977 (cf. Section \ref{Sec.PDR-comparison}). In Fig. \ref{Fig.model}, we present the results of the constant-density PDR models for incident FUV intensities of $G_0=100,500,\mbox{and }1000$, appropriate for selected regions in the Orion Nebula, versus physical scale. The $x$-axes share the same range of visual extinction, $A_\mathrm{V}=0\text{-}10\,\mathrm{mag}$ normal to the PDR surface.The three models in Fig. \ref{Fig.model} qualitatively produce similar results: A warm surface gas layer cools through the [C\,{\sc ii}] and [O\,{\sc i}] lines. The FIR dust emission also peaks at the surface. The colder gas deeper in the cloud emits mainly in low-J CO lines. Deep in the cloud, CO and water molecules freeze out, while cosmic ray ionization enhances the abundance of atomic carbon. The line-of-sight depth of the cloud slightly affects the ratios of FIR, [C\,{\sc ii}]-, and CO-line intensities. The [O\,{\sc i}] $145\,\mu\mathrm{m}$ intensity is significantly increased compared to face-on models, while [O\,{\sc i}] $63\,\mu\mathrm{m}$ is optically thick in both geometries. For higher radiation field and density, the contribution of the [O\,{\sc i}] lines to the cooling overtakes that of the [C\,{\sc ii}] line.

\subsection{Comparison of PDR models with observations}
\label{Sec.PDR-comparison}

For each region, we estimated the input parameters of our PDR models -- the gas density $n$, the incident radiation field $G_0$, and the line-of-sight depth $A_{\mathrm{V,los}}$ -- from the SED results and the [C\,{\sc ii}] and CO data. For the Eastern Rim of the Orion Nebula, \cite{Pabst2020} report a gas density, as estimated from the spatial separation of peak [C\,{\sc ii}] and CO(2-1) emission, of $n\simeq 1\times 10^4\,\mathrm{cm^{-3}}$, while the density drops to $n\simeq 2\times 10^3\,\mathrm{cm^{-3}}$ in the southern Veil Shell, estimated from the dust optical depth using the synthetic extinction curves of \citet{Weingartner2001} with $R_{\mathrm{V}}=5.5$. In the rim of M43, \cite{Pabst2020} find $n\simeq 1\times 10^4\,\mathrm{cm^{-3}}$ from the [C\,{\sc ii}] and CO peak separation, as well. In NGC 1977, the dust optical depth indicates a density of $n\simeq 2\text{-}5\times 10^3\,\mathrm{cm^{-3}}$, {using a line-of-sight length of 1\,pc}. The radiation field in the Orion Nebula is estimated from the FIR-distance relation, that is $G_0\simeq 500 (1\,\mathrm{pc}/d)^2$ (cf. Fig \ref{Fig.FIR-d}). The FIR emission toward the molecular background of M43 suggests $G_0\simeq 2000$. In case of NGC 1977, we have used the stellar parameters of 42 Orionis to estimate the radiation field. In most of the shell, $G_0\simeq 100\text{-}200$, in concordance with \cite{Kim2016}. The radiation field in OMC3 may be slightly higher, $G_0\sim 400\text{-}1000$; \cite{Howe1991} estimate $G_0\simeq 360$.

The depth of the edge-on PDR along the line of sight has been estimated from the dust optical depth. In the Eastern Rim, $\tau_{160}\simeq 1\times 10^{-2}$, which corresponds to $A_{\mathrm{V,los}}\simeq 30$ using $A_{\mathrm{V}}\simeq N_{\mathrm{H}}/2\times 10^{21}\,\mathrm{cm}^{-2}$. In the southern part of the Eastern Rim, $\tau_{160}\simeq 5\times 10^{-3}$, hence $A_{\mathrm{V,los}}\simeq 15$. In the Southern Shell, we find $\tau_{160}\simeq 1\times 10^{-3}$, that is $A_{\mathrm{V,los}}\simeq 3$. In the east of the M43 shell, $\tau_{160}\simeq 1\times 10^{-2}$, and in the western NGC 1977 shell $\tau_{160}\simeq 3\times 10^{-3}$, that is $A_{\mathrm{V,los}}\simeq 10$. We produce PDR models for the parameters estimated here, but let them vary around those values. This way we are able to reject some combinations of input parameters. In the correlation plots, we plot the edge-on model outputs only for $x>0.07\,\mathrm{pc}$, which corresponds to the spatial resolution ($36\arcsec$) of the data.

We observe that the edge-on PDR models with the estimated parameters generally match the observations very well. As suggested by the geometry of an edge-on shell, the models require a larger column density than face-on models would provide. Regions in the Veil Shell with $d<1.0\,\mathrm{pc}$ tend to emit stronger in [C\,{\sc ii}] than accounted for by the models with $G_0=500\text{-}700$, $n= 1\times 10^4\,\mathrm{cm^{-3}}$, and $A_{\mathrm{V,los}}=30$. In the region of OMC4 and the ISF in the background of the Orion Nebula, the deviation of the expected [C\,{\sc ii}] emission from face-on models from the observations is even larger. While the face-on PDR models do not exceed an intensity of $1\times 10^{-3}\,\mathrm{erg\,s^{-1}\,cm^{-2}\,sr^{-1}}$, the observations emit as much as $4\times 10^{-3}\,\mathrm{erg\,s^{-1}\,cm^{-2}\,sr^{-1}}$ outside the Huygens Region/OMC1 and up to $6\times 10^{-3}\,\mathrm{erg\,s^{-1}\,cm^{-2}\,sr^{-1}}$ toward the background PDR in OMC1. Perhaps in OMC1 and OMC4, edge-on models would be more appropriate. In summary, the edge-on models agree well with the observations if we allow the edge-on column to vary between 3 and 30 times that of the face-on column.

The [C\,{\sc ii}] emission toward the center of M43 consists of three components: the blue-shifted expanding shell, the H\,{\sc ii} region, and the PDR on the surface of the background molecular cloud.  According to the velocity profile in \cite[Fig. 21]{Pabst2020}, only 30\% of [C\,{\sc ii}] emission toward the center of M43 stem from the background cloud, with 65\% arising in the H\,{\sc ii} region and 5\% from the foreground shell. The gray points in Fig. \ref{Fig.CII-FIR}c show to the emission from the background PDR, where we have assumed 50\% of the FIR emission to stem from the background PDR. This emission from the M43 PDR background can be fitted by a face-on PDR model with $n\simeq 1\times 10^4\,\mathrm{cm^{-3}}$.

The shell of M43 emits slightly more [C\,{\sc ii}] intensity than predicted by the edge-on models with $G_0=1000$, $n=1\times 10^4\,\mathrm{cm^{-3}}$, and $A_{\mathrm{V,los}}=10$. A larger column, that is higher $A_{\mathrm{V,los}}$, mainly increases the FIR intensity, as the [C\,{\sc ii}] line is already optically thick. The [C\,{\sc ii}] may be matched by models with lower radiation field, but larger column, that is $G_0=500$ and $A_{\mathrm{V,los}}=20\text{-}30$. The lower-$G_0$ models with higher $A_{\mathrm{V,los}}$ reproduce the [C\,{\sc ii}] versus CO(2-1) correlation in the shell more accurately.

The NGC 1977 shell seems to be characterized by $A_{\mathrm{V,los}}=3\text{-}10$, while the PDR model output is less sensitive to the exact density and radiation field. Also the correlations of the [C\,{\sc ii}] intensity with CO(2-1) intensity in the shell can be described by edge-on models with $G_0=100\text{-}200$ and $n=1\text{-}5\times 10^3\,\mathrm{cm^{-3}}$. The PDR associated with OMC3 will be analyzed in more detail by Kabanovic et al. (in prep.).

The edge-on PDR models with the same parameter as in the [C\,{\sc ii}]-FIR correlation seem to be able to explain the [C\,{\sc ii}]/FIR-CO(2-1)/FIR correlation (Fig. \ref{Fig.CII-CO-2}) in the Veil Shell. However, the behavior of the [C\,{\sc ii}]-CO(2-1) correlation (Fig. \ref{Fig.CII-CO}) reflects the presence of uncorrelated emission, as we discuss below. The points in OMC4 lie close to the face-on model with $n=1\times 10^4\,\mathrm{cm^{-3}}$ and varying radiation field ($G_0\simeq 200\text{-}2000$). Points in OMC1 continue the trend toward lower CO(2-1)/FIR ratio, but seem to deviate toward lower-density models. The neutral gas density in OMC1 is expected to be higher, however, $n\simeq 1\times 10^5\,\mathrm{cm^{-3}}$. Compared to uniform face-on models with adequate incident radiation field ($G_0\simeq 2\times 10^4$), OMC1 emits too much [C\,{\sc ii}] intensity. Invoking a clumpy PDR structure generally increases the amount of [C\,{\sc ii}] emission from the models \citep[e.g.,][]{Cubick2008}. For M43, all edge-on models produce similar [C\,{\sc ii}]/FIR ratios at similar CO(2-1)/FIR ratios, while the background PDR intensities scatter around the face-on model prediction, indicating either varying density or varying geometry. Points in NGC 1977 neatly follow the edge-on model predictions.

\begin{figure}[tb]
\includegraphics[width=0.5\textwidth, height=0.33\textwidth]{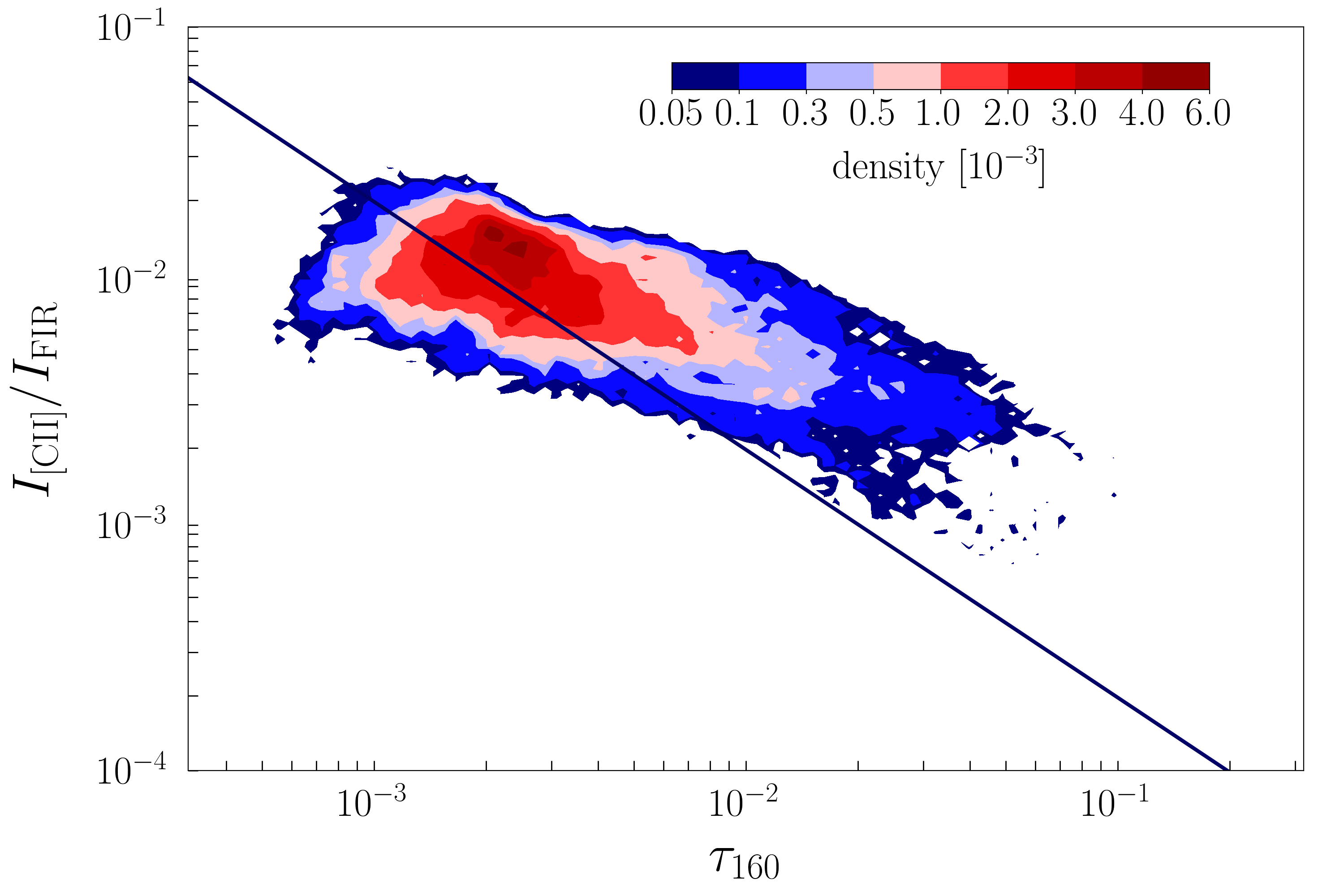}
\caption{Correlation plot of [C\,{\sc ii}]/FIR with dust optical depth $\tau_{160}$ as histogram. Colors indicate the density of data points in each bin. The line indicates the inversely linear relation expected for a cloud slab of constant dust temperature \citep{Goicoechea2015}, running through the median values of [C\,{\sc ii}]/FIR and $\tau_{160}$.}
\label{Fig.CII-tau}
\end{figure}

Figure \ref{Fig.CII-tau} shows the global dependence of the [C\,{\sc ii}]/FIR ratio on the dust optical depth $\tau_{160}$. As opposed to points lying in OMC1, studied in \cite[Fig. 15]{Goicoechea2015}, we do not observe a linear reciprocal trend in the [C\,{\sc ii}]/FIR ratio with $\tau_{160}$ on larger scales. \cite{Goicoechea2015} expect $\log_{10} I_{\mathrm{[C\,\textsc{ii}]}}/I_{\mathrm{FIR}} \simeq -\log_{10}\tau_{160}+C$ for a cloud slab of constant dust temperature viewed in face-on geometry. The fact that we do not observe this behavior globally indicates that, unlike for OMC1, the temperature varies on large scales. However, there seems to be a decreasing trend of [C\,{\sc ii}]/FIR with increasing $\tau_{160}$. Low [C\,{\sc ii}]/FIR ratios are found toward the OMC4/ISF region, where $\tau_{160}$ is large due to a large column of cooler dust. This cool dust likely emanates from deeper layers in the cloud that are not exposed to FUV radiation.

\begin{figure}[tb]
\includegraphics[width=0.5\textwidth, height=0.5\textwidth]{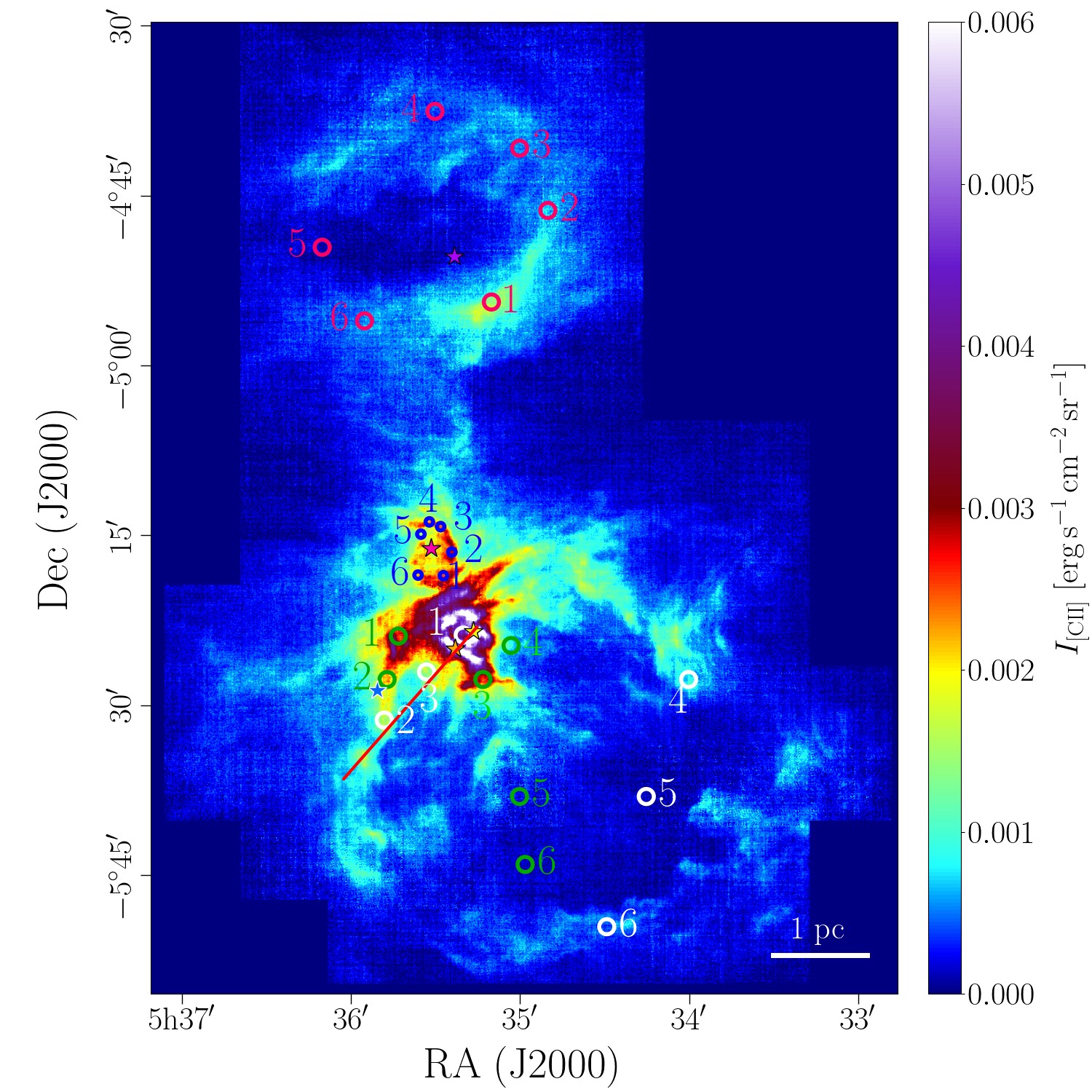}
\caption{Positions of spectra (green and white circles: spectra in M42, blue: M43, red: NGC 1977) and line cut in Fig. \ref{Fig.line-cut-behind_bar} (red line) on [C\,{\sc ii}] line-integrated emission at its native resolution. The stars indicate the most massive stars within each region: $\theta^1$ Ori C (yellow) in M42, NU Ori (pink) in M43, and 42 Orionis (purple) in NGC 1977.}
\label{Fig.map_spectra}
\end{figure}

\begin{figure*}[tb]
\begin{minipage}[t]{0.49\textwidth}
\includegraphics[width=\textwidth, height=0.67\textwidth]{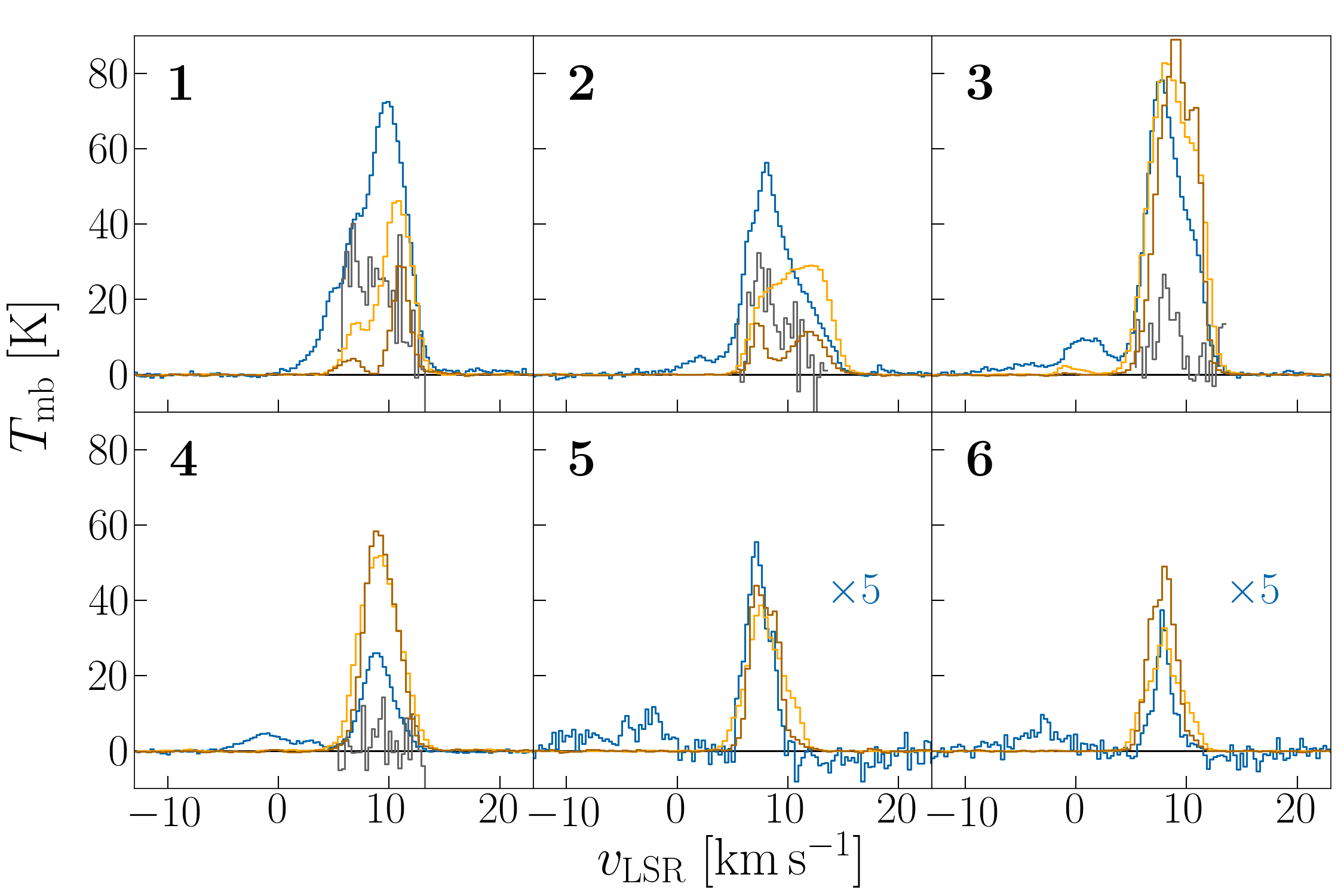}
\caption{Spectra toward high-intensity regions in M42, indicated by green circles with radius $40\arcsec$ in Fig. \ref{Fig.map_spectra}. Blue: [C\,{\sc ii}], yellow: ${}^{12}$CO(2-1), brown: ${}^{13}$CO(2-1) (multiplied by 3). Additionally, the [${}^{13}$C\,{\sc ii}] $F=2\text{-}1$ hyperfine component is plotted (gray) in its systemic velocity (multiplied by 20) in spectra 1-4.}
\label{Fig.spectra-high-CII}
\end{minipage}
\hspace{0.015\textwidth}
\begin{minipage}[t]{0.49\textwidth}
\includegraphics[width=\textwidth, height=0.67\textwidth]{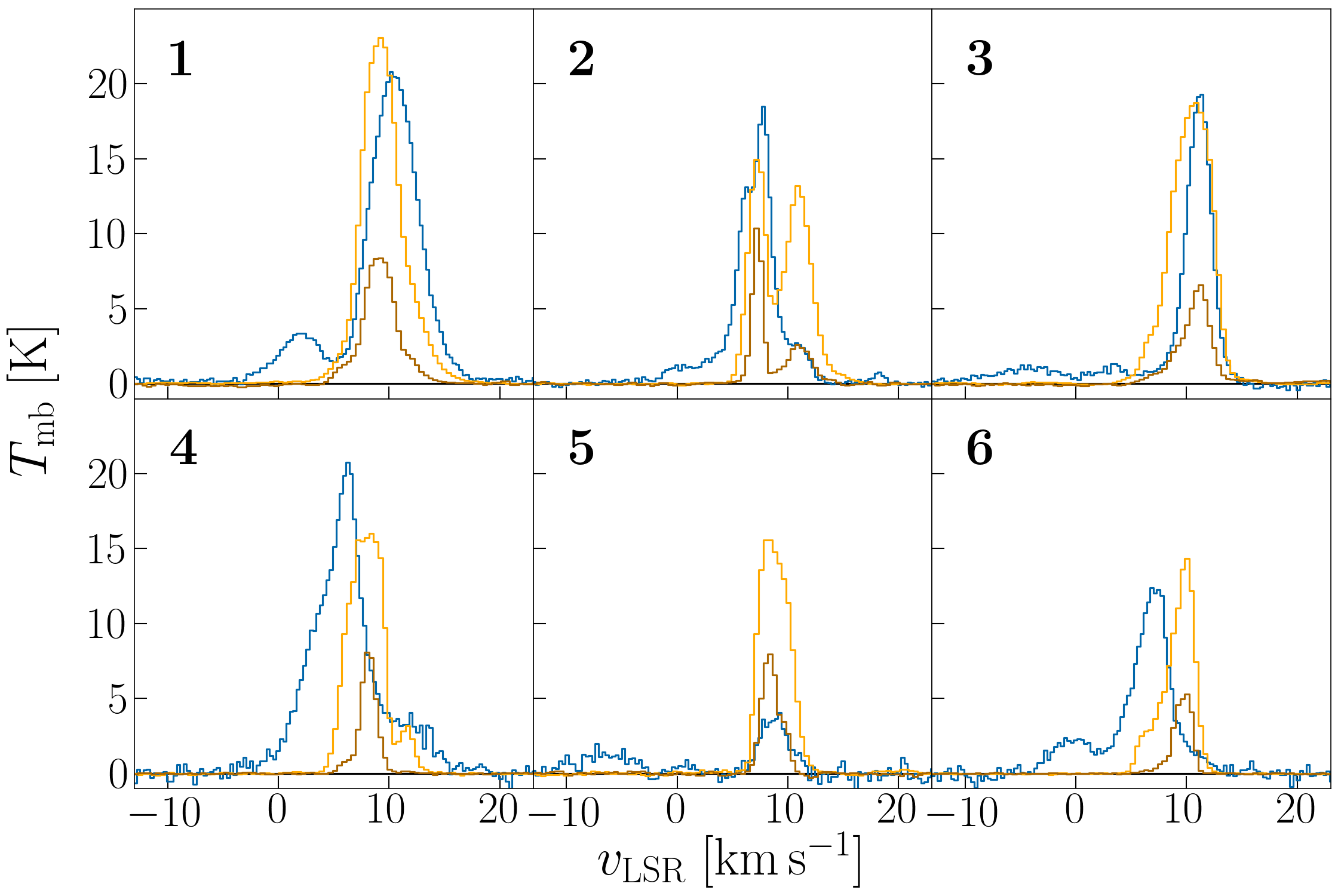}
\caption{Spectra toward six regions of different geometry, indicated by white circles with radius $40\arcsec$ in Fig. \ref{Fig.map_spectra} (OMC1, behind the Orion Bar, Eastern Rim, Western Edge, M42 cloud background, southern Veil Shell). Blue: [C\,{\sc ii}], Yellow: ${}^{12}$CO(2-1), Brown: ${}^{13}$CO(2-1) (not multiplied by 3).}
\label{Fig.spectra-turbulence}
\end{minipage}

\begin{minipage}[t]{0.49\textwidth}
\includegraphics[width=\textwidth, height=0.67\textwidth]{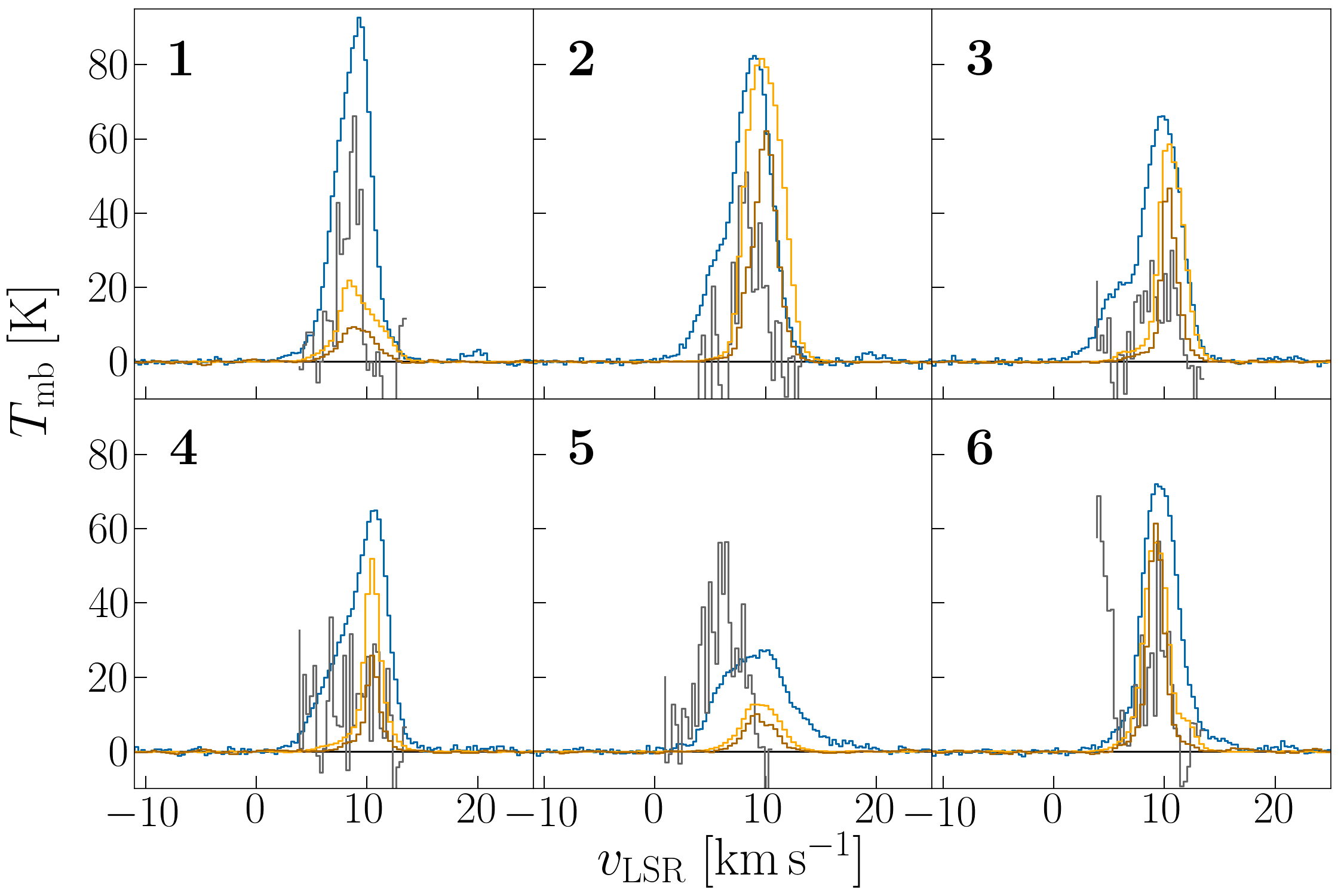}
\caption{Spectra toward the M43 shell, indicated by blue circles with radius $20\arcsec$ in Fig. \ref{Fig.map_spectra}. Blue: [C\,{\sc ii}], yellow: ${}^{12}$CO(2-1), brown: ${}^{13}$CO(2-1) (multiplied by 3). Additionally, the [${}^{13}$C\,{\sc ii}] $F=2\text{-}1$ hyperfine component is plotted (gray) in its systemic velocity (multiplied by 20).}
\label{Fig.spectra-M43-shell}
\end{minipage}
\hspace{0.015\textwidth}
\begin{minipage}[t]{0.49\textwidth}
\includegraphics[width=\textwidth, height=0.67\textwidth]{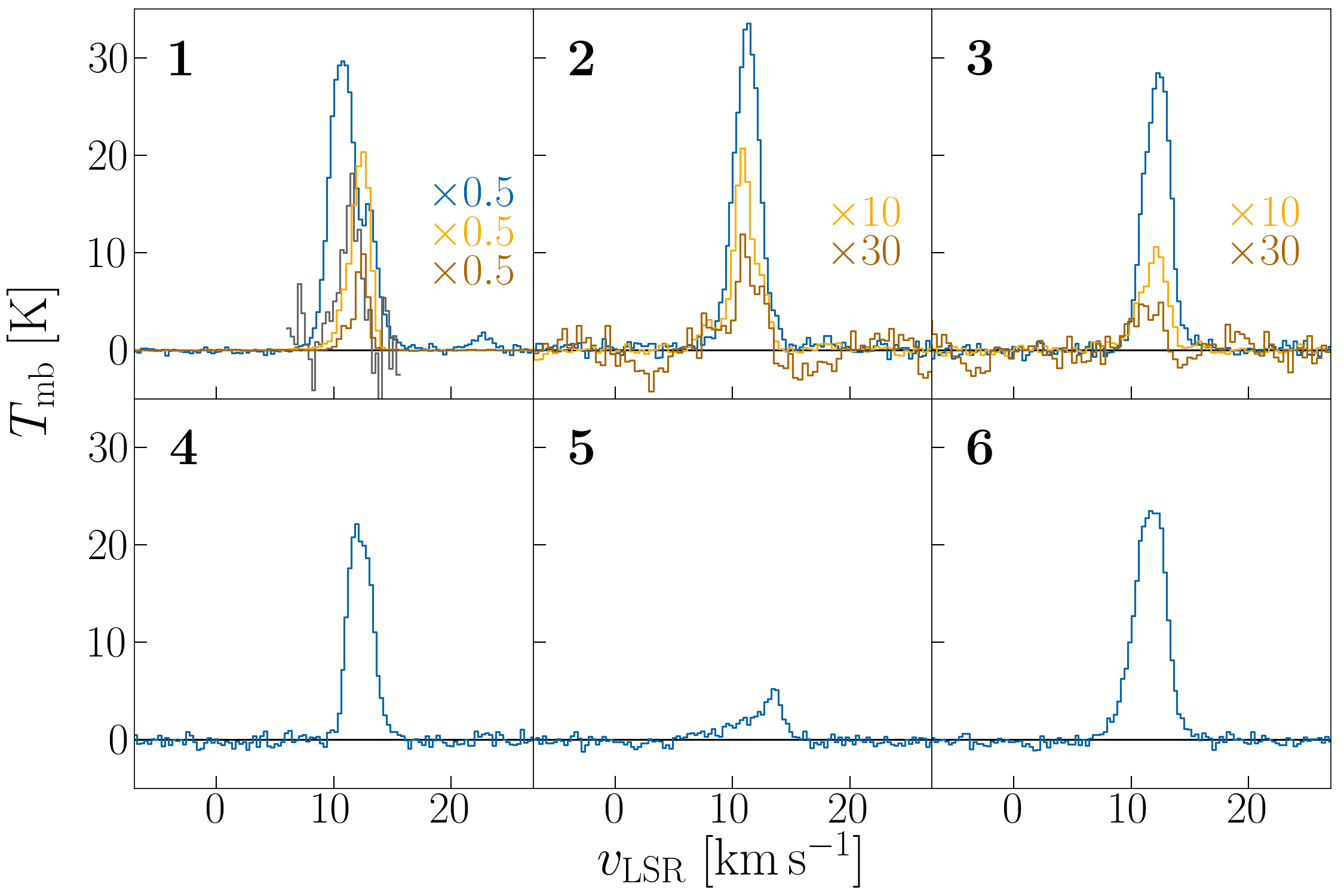}
\caption{Spectra toward the NGC 1977 shell, indicated by red circles with radius $40\arcsec$ in Fig. \ref{Fig.map_spectra}. Blue: [C\,{\sc ii}], yellow: ${}^{12}$CO(2-1), brown: ${}^{13}$CO(2-1) (multiplied by the factor indicated in the respective panel). Additionally, the [${}^{13}$C\,{\sc ii}] $F=2\text{-}1$ hyperfine component is plotted (gray) in its systemic velocity (multiplied by 20) in spectrum 1.}
\label{Fig.spectra-NGC1977-shell}
\end{minipage}
\end{figure*}

We follow up this discussion with a closer look at the spectra in these high-[C\,{\sc ii}] intensity regions. Fig. \ref{Fig.map_spectra} shows a map of the positions toward which the spectra are taken are indicated. The [C\,{\sc ii}] spectra along with CO(2-1) spectra are shown in Fig. \ref{Fig.spectra-high-CII}. We conclude that all but one spectra exhibit multiple components, which may explain the enhanced [C\,{\sc ii}] emission from these high-intensity regions. toward OMC1w (spectrum 4 in Fig. \ref{Fig.spectra-high-CII}), that has relatively less [C\,{\sc ii}] intensity, we observe a single line component. A detailed discussion of these spectra can be found in Appendix \ref{Sec.spectra}. We remark here that the analysis of the [${}^{13}$C\,{\sc ii}] $F=2\text{-}1$ line, where possible, indicates that only about 20-35\% of the gas column estimated from the dust optical depth is traced by [C\,{\sc ii}] (a more thorough analysis of the [${}^{13}$C\,{\sc ii}] emission from Orion A will be done by Kabanovic et al., in prep.) and that the power-law behavior of the [C\,{\sc ii}]-FIR and [C\,{\sc ii}]-$8\,\mu\mathrm{m}$ correlations does not seem to be caused by increasing optical depth with increasing intensity. In addition, in some regions we see the effects of local sources on the gas kinematics.

Figure \ref{Fig.spectra-turbulence} shows [C\,{\sc ii}] and CO(2-1) spectra toward regions of different geometry. In OMC1, [C\,{\sc ii}] and CO(2-1) are slightly shifted from each other. Spectra 3, 4 and 6 show the velocity shift between [C\,{\sc ii}] and CO(2-1) emission toward regions where [C\,{\sc ii}] emission is arising in an edge-on geometry. This indicates that the [C\,{\sc ii}] and CO(2-1) emission arise in different regions of the cloud. In contrast, spectra taken toward the molecular background in face-on geometry, such as spectra 2 and 5, exhibit little systematic velocity shift. However, even spectra 2 and 5 show substructure that deviates from a Gaussian spectral form. Comparing the line widths of the [C\,{\sc ii}] and CO spectra would yield insight into the turbulent motions of the respective gas layers, but this is outside the scope of this study and will be subject of a follow-up study.

In general, the PDR models also reproduce the [C\,{\sc ii}]-CO relationship, but the scatter in those correlations is generally large. From examination of the spectra, the relation of [C\,{\sc ii}] with CO(2-1) emission is not immediately clear. In most cases we find a velocity offset between the [C\,{\sc ii}] and the CO(2-1) lines. Also the global morphology is very different. While [C\,{\sc ii}] emission traces the edge-on shells and illuminated face-on PDR surfaces, CO(2-1) emission traces the deeper molecular layers of a PDR and the background molecular cloud, that might not be co-located with the shell. In case of face-on PDRs, low-J CO emission arises predominantly in deeper layers of the PDR, and the velocity difference is generally less than toward edge-on structures. 

In M43, models with $G_0=2000$, $n= 1\times 10^4\,\mathrm{cm^{-3}}$, and $A_{\mathrm{V,los}}=10$ clearly produce too much FIR emission. The observations agree better with models with $G_0=700\text{-}1000$, however [C\,{\sc ii}] emission is somewhat enhanced compared to the model predictions. As in the shell of M42, spectra toward the edge of M43 consists of at least two components, shown in Fig. \ref{Fig.spectra-M43-shell}. Again, a detailed description and discussion of the spectra is provided in Appendix \ref{Sec.spectra}. The [C\,{\sc ii}] column, computed from the [${}^{13}$C\,{\sc ii}] $F=2\text{-}1$ line, where possible, traces only a fraction (30-60\%) of the total gas column estimated from the dust optical depth. It is evident, however, that the C$^+$ column (with $A_{\mathrm{V}}\simeq 15$) is substantially larger than expected for a face-on PDR with similar conditions. The edge-on geometry of the M43 rim becomes very pronounced. The face-on model with $n= 1\times 10^4\,\mathrm{cm^{-3}}$ also reproduces the [C\,{\sc ii}] and FIR emission from the background molecular cloud, that is after subtraction of the contribution from the foreground shell (gray points).

The [C\,{\sc ii}] and CO(2-1) spectra toward OMC3 and the NGC 1977 shell are shown in Fig. \ref{Fig.spectra-NGC1977-shell}. FIR line emission toward OMC3 at the edge of NGC 1977 has been subject to modeling efforts previously. \cite{Howe1991} conclude that the spatial distribution of [C\,{\sc ii}] emission can be explained by a clumpy medium. Spectrum 1 toward OMC3 can be fitted better by an emitting and a (strong) absorbing component than by two emitting components. The [${}^{13}$C\,{\sc ii}] line is slightly asymmetric, as well, but this is hard to quantify due to noise. It seems to peak between the two emission peaks of the [${}^{12}$C\,{\sc ii}] component, which signifies self-absorption (Kabanovic et al. in prep.). In a clumpy medium, the strong emission and absorption may be easily explained by temperature gradients between the clump and inter-clump medium. While the [C\,{\sc ii}] spectrum may be fitted by one emitting and one absorbing component, both the ${}^{12}$CO and the ${}^{13}$CO spectrum comprise two emitting components. It is unclear how the CO components are related to the [C\,{\sc ii}] components. As in other spectra, we find a significant velocity offset.

While CO emission in OMC3 is strong, in the rest of the shell that is covered by the CO observations it is very weak. ${}^{13}$CO is detected, but at low temperatures the ${}^{12}$CO line in FUV-illuminated gas can still be optically thin with enhanced ${}^{13}$CO emission due to chemical fractionation \citep[e.g.,][]{Goicoechea2020}. In the case of optically thick ${}^{12}$CO emission, the excitation temperature is very low ($T_{\mathrm{ex}}\simeq 4\text{-}5\,\mathrm{K}$). Possibly CO emission stems from a molecular background, that is not associated with the [C\,{\sc ii}]-emitting shell, and that is only faintly illuminated. Another possibility is that the CO-emitting gas originates from two slightly velocity-shifted parts of the expanding gas around the NGC 1977 H\,{\sc ii} region.

Summarizing, the spectra shed light on why [C\,{\sc ii}] emission may be underestimated by the standard PDR models we employ. First of all, virtually all spectra consist of more than one component. One needs geometrically more accurate and multi-component models to understand the precise emission characteristics of both edge-on and face-on geometries. Edge-on models should incorporate temperature gradients, that can lead to self-absorption (Kabanovic et al. in prep.). In some cases, clumpy PDR models might be a good choice. Also dynamics of the gas should be taken into account. Those can lead to pressure variations within the gas by shear or compression. Furthermore, as we see velocity shifts in the [C\,{\sc ii}] and CO(2-1) line emission, photoevaporation may be important and this can alter the emission from a PDR \citep{Bertoldi1989,StoerzerHollenbach1998,Bron2018}. Most of our discussed lines can be fitted by Gaussians, but the spatial relationship of the emission components is often unclear. We notice that in regions where we attribute the emission to edge-on PDRs, the velocity of the CO-emitting gas is generally offset from the velocity of the [C\,{\sc ii}]-emitting gas. Either, this means that the gas is subject to shear, or that [C\,{\sc ii}] and CO trace different, possibly unrelated, layers of gas. In edge-on (half-)shells such as the Veil Shell, CO(2-1) emission arises in the molecular background, while [C\,{\sc ii}] emission stems from the edge-on shell. Compared to M42 and M43, NGC 1977 is evolving rather slowly at its advanced age and tends to emit narrower lines. This might reflect the mild stellar feedback it is experiencing at this stage. At the same time, M42 and M43 are still heavily affected by driving stellar feedback and lines tend to be broader, hinting at increased turbulence.

\subsection{The $G_0$-$p_{\mathrm{th}}$ relationship}
\label{Sec.G0-n}

\begin{figure}[tb]
\includegraphics[width=0.5\textwidth, height=0.33\textwidth]{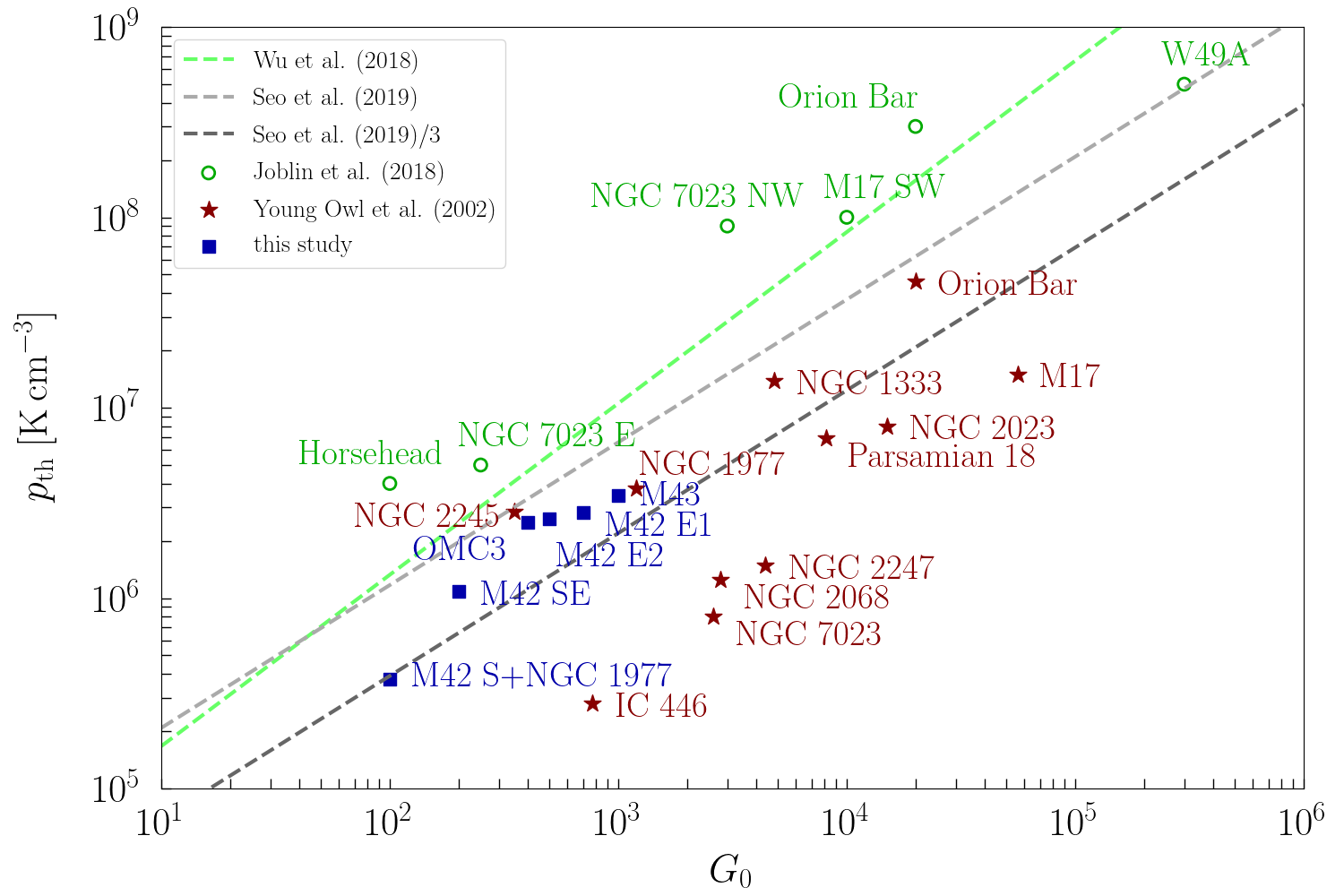}
\caption{Thermal pressure versus incident radiation field in several PDRs. Red points are from \cite{YoungOwl2002} (based on FIR line analysis), green points are from \cite{Joblin2018} (based on high-J CO line analysis), dark blue points are from this study (M42 S is the southern Veil Shell, M42 SE is the southeastern Veil Shell, M42 E1 and M42 E2 lie in the bright eastern arm of the Veil Shell (cf. green circles 1 and 2 in Fig. \ref{Fig.map_spectra})). The green dashed line is the fit by \cite{Wu2018} (based on high-J CO line analysis). The dashed gray line is the analytic solution of \cite{Seo2019}, once divided by 3 to account for pressure equipartition.}
\label{Fig.G0-pth}
\end{figure}

We estimate the thermal pressure $p_{\mathrm{th}}=nT$ in the PDR surfaces from the gas density estimated above and the temperature predicted by the respective PDR model at the estimated $G_0$. Fig. \ref{Fig.G0-pth} shows the thermal pressure versus incident radiation field in a number of regions from our study and from the literature. Our data show a clear trend of increasing thermal pressure with increasing strength of the radiation field. A similar trend is apparent in studies of fine-structure line emission from PDRs \citep{YoungOwl2002} as well as in studies of the CO ladder in PDRs \citep{Joblin2018}, albeit that those relations are displaced somewhat. For the study of \cite{YoungOwl2002}, this displacement reflects an overestimate of the incident FUV field in several of the sources. Specifically, they adopted a $G_0$ of 5000 for NGC 1977S (the bright region in the southwest of NGC 1977), while we derive a value of 1100. After correction, the data for this particular source agrees, not surprisingly as both analysis are based upon fine-structure line studies. Likewise, the $G_0$ value adopted for the Orion Bar by \cite{YoungOwl2002} is somewhat higher than current estimates \citep{Salgado2016, Joblin2018}. The results of the studies by \cite{Joblin2018} and \cite{Wu2018} are displaced toward higher pressures. Both these studies rely on high-J CO transitions with high critical densities. Consequently, clumps or gas density enhancements dominate the emission in these lines, whereas the fine-structure lines measure the interclump gas. The offset of these studies might be due to the importance of self-gravity in the clumps, which will dominate the pressure distribution. Hence, the high-J CO lines may not be good measures of the pressure in the [C\,{\sc ii}]-emitting PDR layers. As a corollary, we do not expect that constant-pressure PDR models will be applicable to these environments. \cite{Joblin2018} note that specifically the [C\,{\sc ii}] emission in their sources (bright PDRs associated with massive star formation) predominantly comes from lower-density interclump gas, that fills the beam.

Following \cite{YoungOwl2002}, using the model of an H\,{\sc ii} region bordering on a PDR in pressure equilibrium, \cite{Seo2019} derive an analytic expression for the correlation between $p_{\mathrm{th}}$ and $G_0$ in the PDR:
\begin{align}
p_{\mathrm{th}} \simeq 4.6\times 10^4 f^{-1/2}(\frac{\Phi_{\mathrm{FUV}}}{10^{51}\,\mathrm{s^{-1}}})^{-1/4} G_0^{3/4} \,\mathrm{K\,cm^{-3}}\label{eq.pth-G0},
\end{align}
where $f=\Phi_{\mathrm{FUV}}/\Phi_{\mathrm{EUV}}$ is the ratio of the FUV flux to the H-ionizing flux. For relevant spectral types, this ratio is constant to within a factor 1.5 \citep{YoungOwl2002}. This relation is also shown in Fig. \ref{Fig.G0-pth}. The theoretical relation is well above the empirical relation derived from the fine-structure lines but below the data derived from the CO analysis.

We note that all our analysis and that of \cite{YoungOwl2002} only measures the thermal pressure in the PDR. The total pressure includes contributions from turbulence and magnetic fields and should be balanced by the thermal pressure from the ionized gas and radiation pressure. For the three sources in Orion, we have compared these and concluded that there is equipartition between thermal, turbulent and magnetic pressure in the PDR and that radiation pressure is typically not very important \citep{Pabst2020}. A similar conclusion was reached by \citet{Pellegrini2009} in their detailed analysis of the PDR associated with the Orion Bar. Moreover, while there is approximate pressure equilibrium between the plasma and the ionized gas within the Veil Shell, there is a strong pressure gradient from the background molecular cloud toward us, which drives the rapid expansion of the shell. In the PDRs of M43 and NGC 1977, the pressure of the ionized gas exceeds the PDR pressure, indicating that magnetic or turbulent pressure may be quite important in these regions.

In a linear fit, we find $p_{\mathrm{th}}/G_0\simeq 5\times 10^3\,\mathrm{K\,cm^{-3}}$ for our data points. The data points of \cite{Joblin2018} agree well with the empirical relation of \cite{Wu2018}. We note that the error bars of the data points are usually large, which is mostly due to the uncertainty in the estimated gas density ($\sigma\sim 0.5\,\mathrm{dex}$). As discussed before, the detailed physics of the models used to infer physical conditions strongly affect the resulting estimates. This applies to our data points, as well, since we rely on the standard PDR structure and dust properties to estimate the gas density in M42, M43, and NGC 1977. Besides the pressure of the H\,{\sc ii} region acting on the PDR, the correlation between thermal pressure or gas density and the incident radiation field may also reflect the star-formation history of a region, as \cite{YoungOwl2002} comment.

\cite{Howe1991} demonstrated that [C\,{\sc ii}] emission from OMC3 can be modelled successfully employing a clumpy PDR model. Considering the geometry of a source is certainly important in modeling efforts. A slight deviation from pure edge-on or face-on geometry results in differing estimates of the physical conditions. Considering the detailed line shape of the observed lines is equally important. Often emission lines reveal several line components mixed within the line of sight or foreground components by self-absorption. Complex velocity profiles may also hint at the presence of a clumpy medium, hydrodynamic instabilities and/or colliding filaments and gas flows. These processes imply a less homogeneous ISM than usually accounted for in PDR models.

While we recognize that real PDRs might contain clumps and gas-density enhancements \citep{Joblin2018, Goicoechea2016, Cubick2008} and that this may influence the analysis, we want to emphasize that the physical properties of PDRs derived from the analysis of fine-structure lines are in good agreement with the simple theoretical model based upon the Str\"omgren relation in combination with pressure equilibrium between the ionized and PDR gas and taking equipartition of thermal, turbulent, and magnetic pressures into account.

\begin{table*}[tb]
\centering
\addtolength{\tabcolsep}{-1.5pt}
\def\arraystretch{1.2}
\caption{Luminosities and masses of M42, M43 and NGC 1977 per subregion.}
\begin{tabular}{llccccccccc}
\hline\hline
Region & & $A$ & $L_{\mathrm{FIR}}$ & $L_{\mathrm{[C\,\textsc{ii}]}}$ & $L_{\mathrm{[C\,\textsc{ii}]}}/L_{\mathrm{FIR}}$ & $M_{\mathrm{gas}}$\tablefootmark{a} & $L_{\mathrm{CO(2\text{-}1)}}$\tablefootmark{b} & $M_{\mathrm{CO}}$\tablefootmark{c} & $L_{\mathrm{24\mu m}}$\tablefootmark{d} & $L_{\mathrm{H\alpha}}$\tablefootmark{e} \\
&  &  $\mathrm{[pc^2]}$ & $\mathrm{[L_{\sun}]}$ & $\mathrm{[L_{\sun}]}$ & & $\mathrm{[M_{\sun}]}$ & $\mathrm{[L_{\sun}]}$ & $\mathrm{[M_{\sun}]}$ & $\mathrm{[L_{\sun}]}$ & $\mathrm{[L_{\sun}]}$ \\ \hline
M42 & all & 45 & $1.8\times 10^5$ & 510 & $2.8\times 10^{-3}$ & 10900 & 1.1 & 8800 & 4700 & 3500 \\
 & OMC1\tablefootmark{f} & 0.31 & $8.2\times 10^4$ & 53 & $6.5\times 10^{-4}$ & 1500 & $8.9\times 10^{-2}$ & 720 & -- & 1500 \\
 & Bright Ridge\tablefootmark{f} & 0.07 & $3.7\times 10^4$ & 11 & $2.9\times 10^{-4}$ & 870 & $2.8\times 10^{-2}$ & 220 & -- & 470 \\
 & Eastern Rim & 1.3 & $1.7\times 10^4$ & 59 & $3.5\times 10^{-3}$ & 510 & $9.2\times 10^{-2}$ & 740 & 810 & 260 \\
 & OMC4 & 2.5 & $1.6\times 10^4$ & 45 & $2.8\times 10^{-3}$ & 2100 & $1.5\times 10^{-1}$ & 1200 & 640 & 470 \\
 & Veil Shell & 8.7 & $3.5\times 10^4$ & 150 & $4.3\times 10^{-3}$ & 2500 & $2.3\times 10^{-1}$ & 1800 & 1200 & 590 \\

M43 & all & 0.74 & $1.4\times 10^4$ & 38 & $2.7\times 10^{-3}$ & 530 & $3.0\times 10^{-2}$ & 240 & 740 & 190 \\
 & background, H\,{\sc ii} & 0.13 & $3.1\times 10^3$ & 8.9\tablefootmark{g} & $2.8\times 10^{-3}$ & 53 & $4.6\times 10^{-3}$ & 37 & 100 & 99 \\
 & {\footnotesize faint CO, faint [C\,{\sc ii}]} & 0.14 & $8.8\times 10^2$ & 4.1 & $4.7\times 10^{-3}$ & 29 & $2.5\times 10^{-3}$ & 20 & 49 & 12 \\
 & {\footnotesize bright CO, faint [C\,{\sc ii}]} & 0.16 & $2.0\times 10^3$ & 6.0 & $3.0\times 10^{-3}$ & 200 & $9.2\times 10^{-3}$ & 74 & 87 & 18 \\
 & shell & 0.31 & $8.2\times 10^3$ & 19 & $2.3\times 10^{-3}$ & 250 & $1.4\times 10^{-2}$ & 110 & 410 & 63 \\

NGC & all & 17 & $1.8\times 10^4$ & 160 & $8.9\times 10^{-3}$ & 2600 & $1.2\times 10^{-1}$ & 950 & 1000 & 63 \\
 1977 & H\,{\sc ii} & 2.9 & $4.7\times 10^3$ & 37\tablefootmark{h} & $7.9\times 10^{-3}$ & 310 & $1.1\times 10^{-2}$ & 87 & 300 & 17 \\
 & OMC3 & 2.6 & $5.8\times 10^3$ & 34 & $5.9\times 10^{-3}$ & 1200 & $8.9\times 10^{-2}$ & 720 & 190 & 8.6 \\
 & faint H$\alpha$ & 6.9 & $3.2\times 10^3$ & 33 & $9.7\times 10^{-3}$ & 730 & $3.0\times 10^{-2}$ & 240 &280 & 19 \\
 & shell & 6.2 & $5.4\times 10^3$ & 69 & $1.3\times 10^{-2}$ & 670 & $1.6\times 10^{-2}$ & 130 & 100 & 7.9 \\ \hline
\end{tabular}
\tablefoot{\tablefoottext{a}{Gas mass derived from dust mass, using $N_H\simeq 6\times 10^{24}\,\mathrm{cm^{-2}}\,\tau_{160}$ \citep{Weingartner2001}.}
\tablefoottext{b}{ The CO(2-1) map does not cover the entire NGC 1977 region (cf. Fig. 2 in Paper I).}
\tablefoottext{c}{Gas mass from CO(2-1) intensity, using $X(\mathrm{CO})\simeq 2\times 10^{20}\,\mathrm{cm^{-2}}\,(\mathrm{K\,km\,s^{-1}})^{-1}$ \citep{Bolatto2013}.}
\tablefoottext{d}{The MIPS $24\,\mu\mathrm{m}$ image is saturated toward the Huygens/OMC1 region.}
\tablefoottext{e}{The H$\alpha$ surface brightness in the EON is largely due to scattered light from the bright Huygens Region \citep{ODell2010}. Also H$\alpha$ emission in M43 has to be corrected for a contribution of scattered light from the Huygens Region \citep{Simon-Diaz2011}.}
\tablefoottext{f}{The Bright Ridge contains the BN/KL region and Orion S and is part of OMC1. Values for OMC1 include the Bright Ridge, where we have estimated the FIR luminosity from an SED fit to the three PACS bands, that are only saturated at the core of BN/KL.}
\tablefoottext{g}{$2.4\,L_{\sun}$ from background PDR, $6.0\,L_{\sun}$ from ionized gas, $0.5\,L_{\sun}$ from foreground neutral shell.}
\tablefoottext{h}{$6.2\,L_{\sun}$ from ionized gas, $31\,L_{\sun}$ from neutral shell.}}
\label{Tab.Luminosities}
\end{table*}

\subsection{The origin of [C\,{\sc ii}] emission and its relationship to CO-dark gas}

In this section, we discuss the morphology of the emission and the implications thereof. This was done briefly in Paper I, but we continue with a more detailed discussion in this paper. We explore the origin of the [C\,{\sc ii}] emission and the amount of CO-dark gas associated with it in the following. Table \ref{Tab.Luminosities} compiles the luminosities and masses toward the regions we have defined in our analysis and summarizes the results.

As we have stated in Paper I, most of the [C\,{\sc ii}] emission (about 50\%) comes from the limb-brightened shell edges in M42, M43, and NGC 1977, that is moderately irradiated PDR surfaces\footnote{From Table \ref{Tab.Luminosities} we add the [C\,{\sc ii}] luminosities of the shells in M42, M43, and NGC 1977 plus the Eastern Rim in M42. This adds up to $328\,L_{\sun}$, which is 46\% of $708\,L_{\sun}$, the total [C\,{\sc ii}] luminosity.}, but a significant contribution (25\%) also comes from the more diffuse emission from weakly irradiated surfaces\footnote{i.e., all emission that is not captured in Table \ref{Tab.Luminosities} in the defined regions}. Although the heavily irradiated PDRs close to the central stars emit very bright [C\,{\sc ii}] emission, their contribution to the total [C\,{\sc ii}] emission is minor (10-20\%) compared to the large shells.

The massive shells comprise a significant amount of CO-dark gas, with only a few solar masses contained in dense CO-emitting globules found in the Veil Shell \citep{Goicoechea2020}. If we rely on the gas mass estimated from the dust opacity, we find that 25\% of the gas mass in the Orion Nebula is contained in the expanding shell. However, the shell mass estimate might be contaminated by the molecular background. In a conservative estimate, we attribute only 10\% of the total gas mass (i.e., $1000\,M_{\sun}$) to the shell. The mass in the dense molecular core, OMC1 with BN/KL and Orion S, is also 15\% of the total gas mass within the mapped area. Another 20\% are comprised in the dense filament, the ISF. Both fractions agree with the mass fractions traced by CO emission (8\% and 14\%, respectively). The CO emission in the line of sight toward the expanding shell likely stems from the background gas rather than the shell itself (as judged from the velocity structure of the spectra). While most of the CO emission in the region of the Orion Nebula originates from the extended background cloud, H$\alpha$ emission is concentrated around OMC1, almost 50\% stemming from the small OMC1 region.

Also 30\% of the FIR emission stem from the OMC1 region, including the FIR emission of the Bright Ridge comprising BN/KL and Orion S in the three PACS bands, that accounts for 15\%, that is half of the FIR emission from the OMC1 region (see Paper I for a discussion of the ``[C\,{\sc ii}] deficit'' or ``FIR excess''). About 25\% stem from the much larger shell. About 10\% stem from the region of the OMC4, that is not bright in surface tracers, but contributes by its large area. The same amount arises from the bright Eastern Rim, that has half the area of the OMC4 region. While FIR emission from the OMC1 region contributes significantly to the total FIR emission from the Orion Nebula, only 10\% of the total [C\,{\sc ii}] emission originate in OMC1. About 30\% of the [C\,{\sc ii}] emission stem from the shell. Also OMC4 is a minor contributor with 10\%, where we note that most of the emission associated with OMC4 actually stems from the regions close to OMC1, which can therefore be attributed to relatively warm, heavily irradiated gas. The bright Eastern Rim, with its four times larger area compared to OMC1, contributes with 10\% as well. Adding the contributions from the defined areas, leaves 40\% of [C\,{\sc ii}] emission that stems from less pronounced features, that is the extended background and structures lying outside the shell, which may be excited by local sources or bubbles spawning off the large bubble blown by $\theta^1$ Ori C.

We now turn to the other two regions within the Orion Nebula complex. In M43, 50\% of [C\,{\sc ii}] emission arise in the region we have defined as the shell. However, this emission is likely contributed to by a significant emitting background. About 25\% stem from the inner region. Of this emission, a spectral analysis shows that only 25\% arise from the background, while 5\% stem from the expanding shell and 70\% from the H\,{\sc ii} gas filling the bubble. Faint [C\,{\sc ii}] emission can be observed outwards of the shell, accounting for 25\% of the total emission. The percentages in the FIR emission are similar, slightly more (60\%) stemming from the shell region. H$\alpha$ emission is concentrated in the inner H\,{\sc ii} region. Emission at 24\,$\mu$m is generally connected to either temperature fluctuations in very small grains or to dust in the ionized gas heated by, for instance, trapped Lyman $\alpha$ radiation \citep{Desert1986,Churchwell2006,Salgado2016}. Also CO emission is dominated by the shell region and follows its morphology. The face-on shell seen toward the H\,{\sc ii} region does not exhibit CO emission, hence we conclude that the expanding shell is not very massive and that the edge-on shell structures are sculpted within the molecular background. In \cite{Pabst2020} we estimated an expanding shell mass of only $8\,M_{\sun}$, which amounts to 1.5\% of the total gas mass in M43. This compares to 5\% of [C\,{\sc ii}] emission stemming from the face-on shell. If we adopt a shell mass of $1000\,M_{\sun}$ in M42, 10\% of the gas mass account for 25\% of the [C\,{\sc ii}] emission, which is an equal relative mass to luminosity ratio as in M43.

Most of the gas mass in NGC 1977 as traced by the dust opacity, almost 50\%, is concentrated in OMC3, a dense region where the ISF bends westward. 25\% of the gas mass is situated in the shell surrounding the H\,{\sc ii} region. However, we cannot exclude the possibility that not all the mass is part of the expanding shell. As the expansion velocity is rather low, $1.5\,\mathrm{km\,s^{-1}}$ \citep{Pabst2020}, the expansion may have stalled in denser parts of the NGC 1977 region. Also we find significant dust emission outside of the shell, accounting for 25\% of the total mass. In the inner part of the shell, the enclosed H\,{\sc ii} region, we find about 10\% of the mass, amounting to a similar mass per area ratio as in the outward regions. H$\alpha$ and $24\,\mu\mathrm{m}$ are distributed in a similar fashion, predominantly arising from the H\,{\sc ii} region. The total CO emission originates almost entirely from OMC3, but this is due to the fact that the CO observations have insufficient coverage in the less CO-bright areas. We can draw the preliminary conclusion, however, that we find a significant amount of CO emission besides OMC3, which may stem from a more extended molecular background or is associated with the shell structure. The rather extended shell in NGC 1977 dominates the total [C\,{\sc ii}] emission in this region, contributing 43\%. The dense and bright OMC3 contributes only about 20\%. 25\% originate in the [C\,{\sc ii}]-emitting shell toward the H\,{\sc ii} region, with a minor contribution from the ionzed gas \citep[cf.][]{Pabst2020}. As before, a significant amount of [C\,{\sc ii}] emission stems from weakly emitting surfaces, in NGC 1977 this is about 20\% within the map coverage. Only 30\% of the total FIR emission stem from the shell. In NGC 1977, the total [C\,{\sc ii}]/FIR ratio is significantly increased compared to M42 and M43, at almost 1\%, while M42 and M43 exhibit a [C\,{\sc ii}]/FIR ratio of $\sim$0.3\%. We attribute this to the lower radiation field in NGC 1977.

In summary, when considering the individual regions (see Table \ref{Tab.Luminosities}), [C\,{\sc ii}] emission from PDRs is the main origin of [C\,{\sc ii}] emission from the Orion Nebula complex, with about 70\%. The [C\,{\sc ii}] emission from ionized gas is a minor contributor, less than 5\% in M42 and NGC 1977, but 15\% in M43. The remainder, CO-dark H$_2$ gas that is not captured in bright PDR surfaces, contributes with about 20\% (cf. Table \ref{Tab.Luminosities}). These percentages elucidate the importance of observations toward fainter extended regions, as we have seen that those can carry a significant amount of [C\,{\sc ii}] luminosity. This assessment is in agreement with the conclusion reached in Paper I when considering the overall [CII] emission observed in this survey.

The fraction of CO-dark molecular gas in these regions of Orion is much less than found elsewhere in the LMC and Milky Way. A recent study by \cite{Lebouteiller2019} concerning the origin of the [C\,{\sc ii}] line emission and pertaining to its utility as a star-formation tracer, concludes that in the giant H\,{\sc ii} region N11 in the LMC more than 90\% of the [C\,{\sc ii}] emission arises in CO-dark H$_2$ gas and that most of the molecular gas is CO-dark (above 40\%). Dwarf galaxies generally have lower metallicity and penetration of FUV photons, and therefore photodissociation of CO is enhanced. The GOT C+ survey has shown that 20\% of the Galactic [C\,{\sc ii}] emission arise from ionized gas \citep{Pineda2014}, while 30\% stem from PDRs, and only 25\% from CO-dark H$_2$ gas. \cite{Pineda2013} note that most of the [C\,{\sc ii}] emission in the Milky Way arises in moderately FUV-illuminated regions ($G_0\simeq 2\text{-}50$), that is large faint surfaces rather than small dense and bright PDRs. Our observations toward the Orion Nebula are in good agreement with this conclusion, the bright inner OMC1 region being a minor contributor to the total [C\,{\sc ii}] luminosity. The [C\,{\sc ii}] observations toward L1630, comprising the Horsehead Nebula, predominantly highlight the PDR surfaces illuminated by $\sigma$ Ori, where 95\% of the total [C\,{\sc ii}] emission in the mapped area arise. Yet, regions that are bright in [C\,{\sc ii}] trace only 8\% of the gas mass, while 85\% of the mass is associated with strong CO emission \citep{Pabst2017}. In the Orion Nebula complex, we find that about 70\% of the [C\,{\sc ii}] emission arises in PDR surfaces (i.e. massive shells and surfaces of cores as tabulated in Table \ref{Tab.Luminosities}). The [C\,{\sc ii}]-emitting regions we have defined contain 64\% of the gas mass (traced by dust). A portion of that gas mass, about 30-50\%, is likely associated with the molecular background rather than the PDR surface. Hence, bright [C\,{\sc ii}] emission, that traces the H/H$_2$ transition in a PDR, traces about 30\% of the total gas mass in the Orion Nebula complex. On a larger scale, the Orion molecular clouds and star-formation regions are embedded in the Orion-Eridanus superbubble. In the Orion-Eridanus region, an area of $\sim 400\,\mathrm{pc}\times 400\,\mathrm{pc}$, [C\,{\sc ii}] emission (observed with COBE) is dominated by the emission from extended low-density, low-UV field molecular cloud surfaces associated with the active regions of massive star formation \citep{Abdullah2020}.

\subsection{Photoelectric heating efficiency}
\label{Sec.pe-efficiency}

In Paper I we have discussed the observed [C\,{\sc ii}] deficit, that is the decrease in the [C\,{\sc ii}]/FIR intensity with increasing FIR intensity, and related it to a decreased photoelectric heating efficiency in dense PDRs, in addition to the importance of [O\,{\sc i}] cooling. Here, we examine the heating efficiency of PDR material in more detail. We estimate the photoelectric heating efficiency by the observed cooling efficiency of the gas, which we approximate by the observed [C\,{\sc ii}]/FIR ratio. At intermediate temperatures and densities, the cooling is dominated by cooling through the [C\,{\sc ii}] line. At higher temperatures and densities, like those in the Huygens Region, the [O\,{\sc i}] 63\,$\mu$m line becomes important \citep{HollenbachTielens1999}. However, in the shell surrounding the Extended Orion Nebula, gas cooling is most likely dominated by [C\,{\sc ii}], as the temperatures and densities are rather moderate. In M43 and the densest part of NGC 1977 (in OMC3), [O\,{\sc i}] emission will contribute significantly \citep{Herrmann1997,YoungOwl2002}.

\begin{figure*}[tb]
\begin{minipage}[t]{0.49\textwidth}
\includegraphics[width=\textwidth, height=0.67\textwidth]{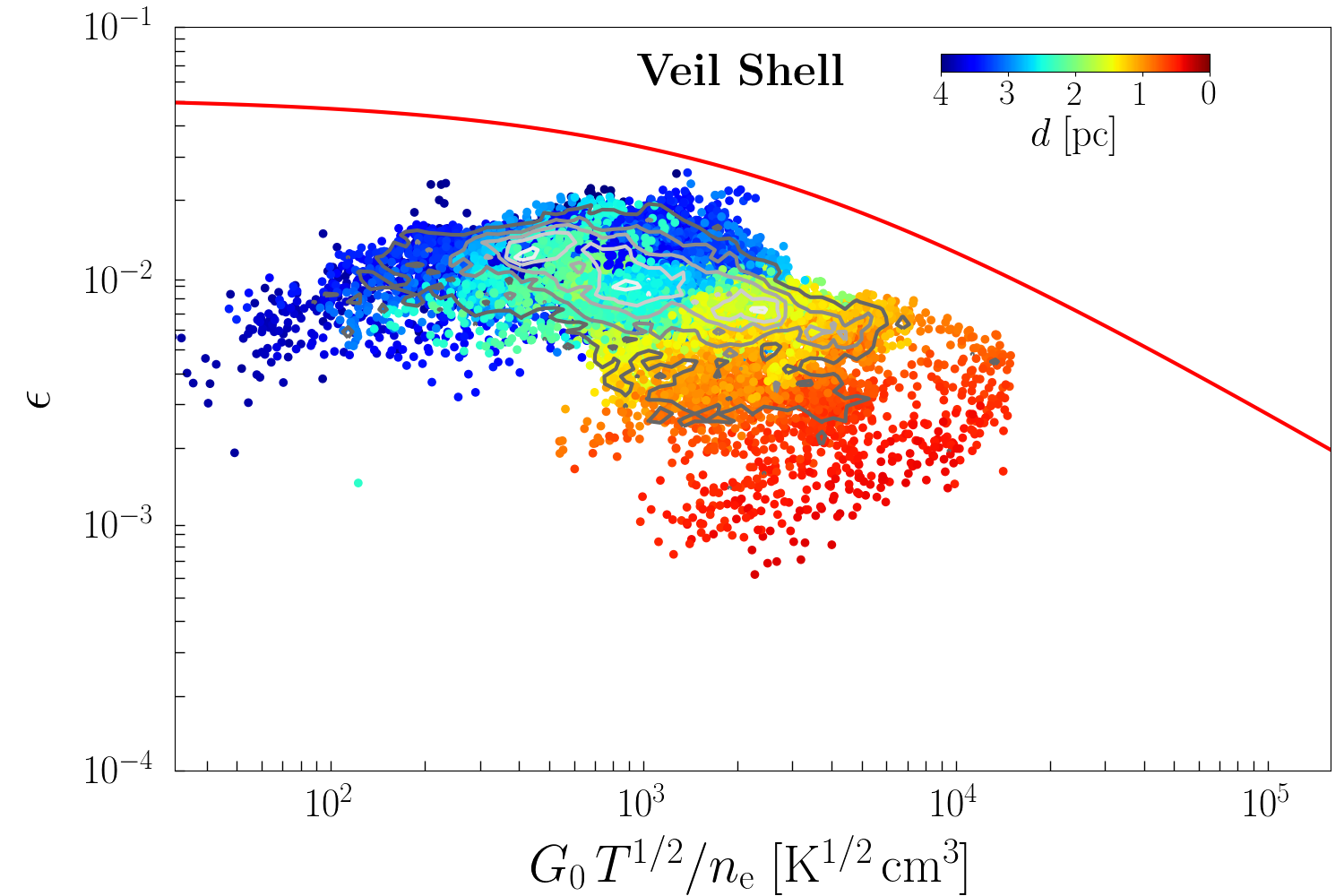}
\end{minipage}
\begin{minipage}[t]{0.49\textwidth}
\includegraphics[width=\textwidth, height=0.67\textwidth]{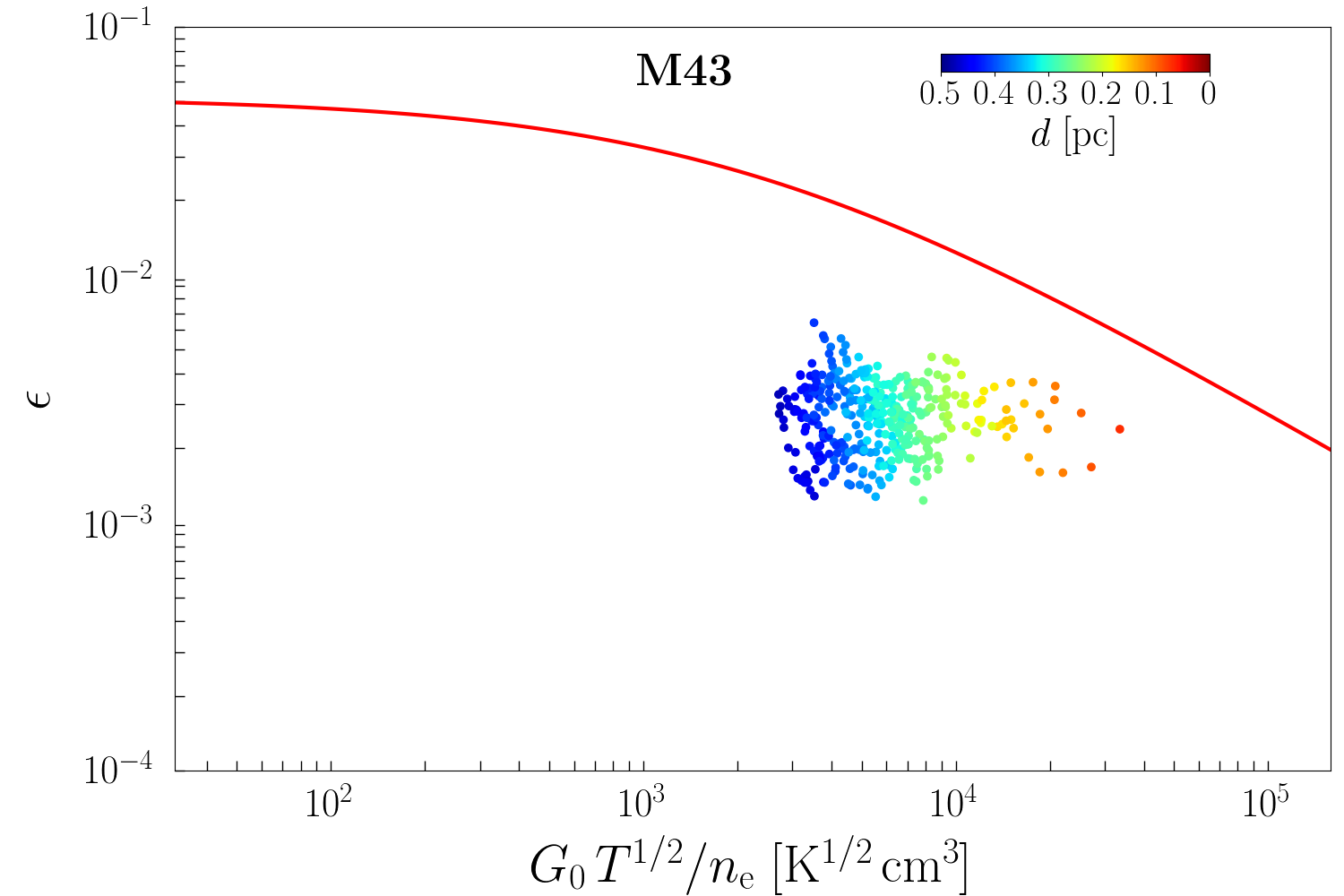}
\end{minipage}

\begin{minipage}[t]{0.49\textwidth}
\includegraphics[width=\textwidth, height=0.67\textwidth]{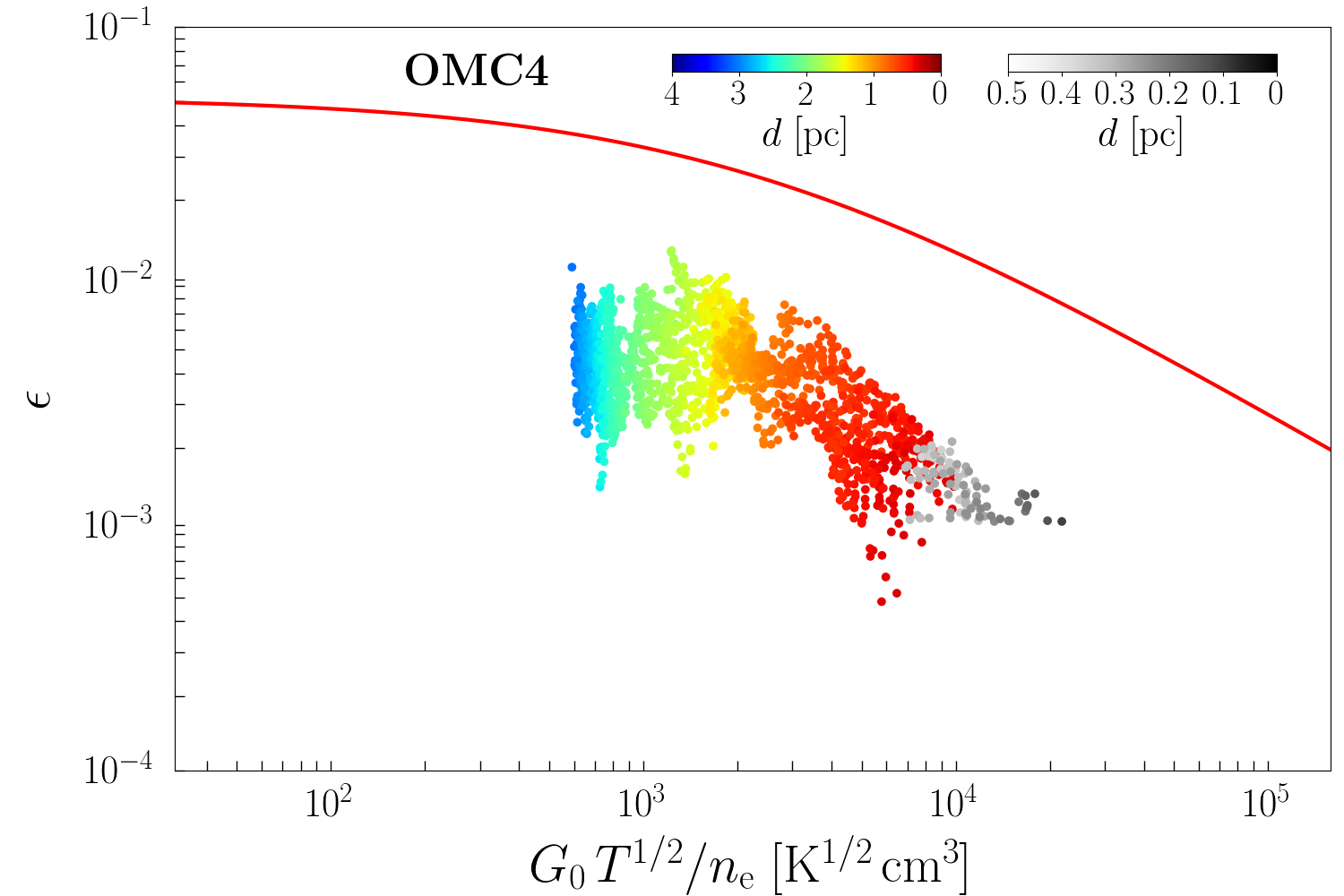}
\end{minipage}
\begin{minipage}[t]{0.49\textwidth}
\includegraphics[width=\textwidth, height=0.67\textwidth]{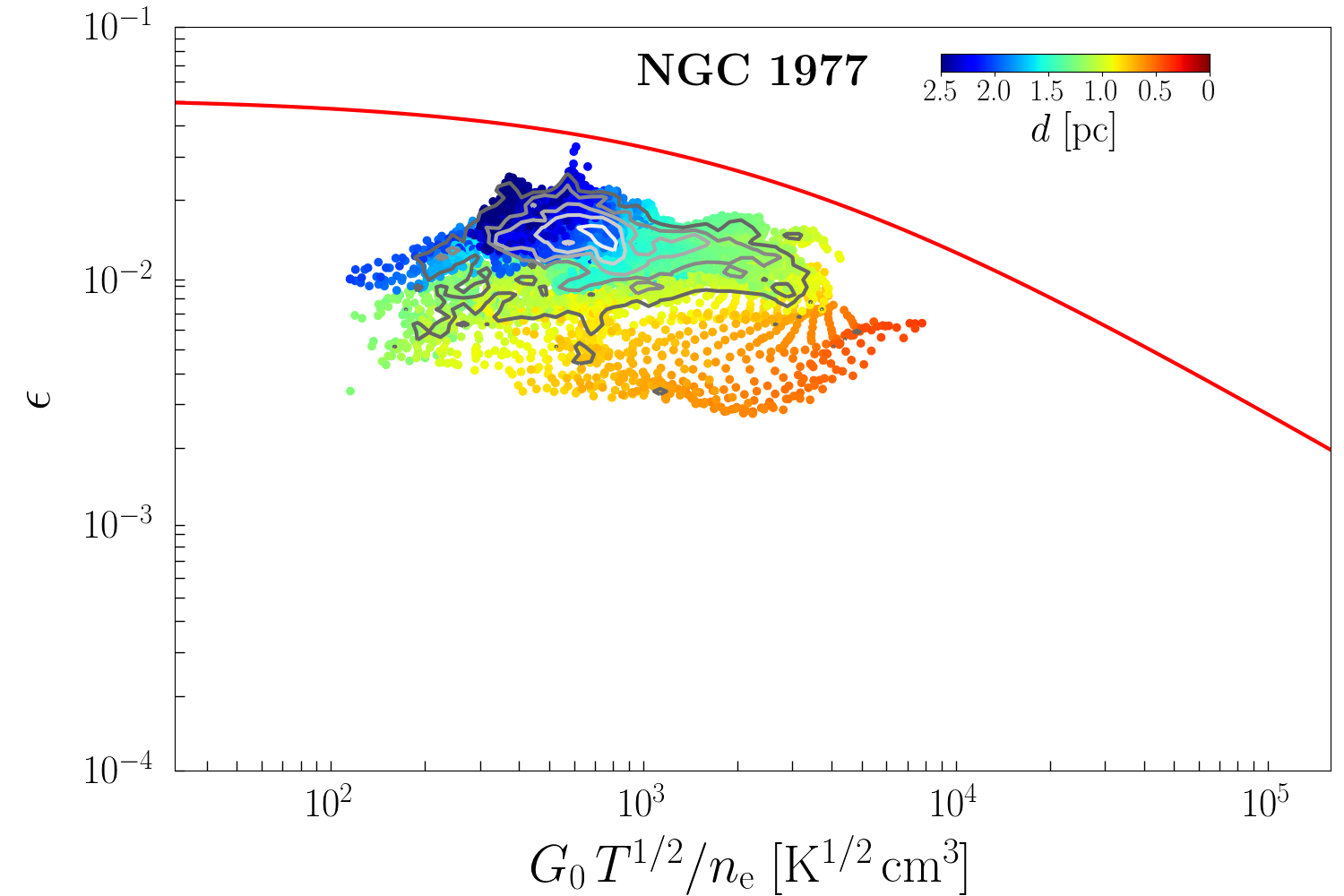}
\end{minipage}
\caption{Photoelectric heating efficiency $\epsilon$, equated as cooling efficiency [C\,{\sc ii}]/FIR, in M42 (top left: Veil Shell, bottom left: OMC4 with OMC1 (in gray scale)), the shell of M43 (top right), and the shell of NGC 1977 (bottom right) as function of the ionization parameter $\gamma=G_0 T^{1/2}/n_{\mathrm{e}}$. Colors indicate the distance from $\theta^1$ Ori C in the Orion Nebula, from NU Ori in M43, and from 42 Orionis in NGC 1977, respectively. Contours in the upper-left and lower-right panel indicate the density levels above which 90, 70, 50, 30, and 10\% of the points lie. The red curve is the theoretical prediction of Eq. 43 by \cite{BakesTielens1994}.}
\label{Fig.pe_scatter}
\end{figure*}

Theoretically, the photoelectric heating efficiency depends on the ionization parameter $\gamma=G_0 T_{\mathrm{gas}}^{1/2}/n_{\mathrm{e}}$ \citep[Eq. 43]{BakesTielens1994}. In Paper I, we estimated $\gamma$ in each pixel of the map. Here, we set the gas density in OMC4 so that it varies approximately equally with distance from OMC1 as in the Veil Shell, $n\simeq 10^4 (1\mathrm{pc}/(d+0.1\,\mathrm{pc}))$. The radiation field varies as $G_0\simeq 500 (1\mathrm{pc}/(d+0.1\,\mathrm{pc}))^2$ (cf. Paper I). The gas density in the shell of M43 is set to $n\simeq 1\times 10^4\,\mathrm{cm^{-3}}$ \citep[cf.][]{Pabst2020}. In NGC 1977, we approximate the gas density, using a column of $1\,\mathrm{pc}$, by $n\sim 6\times 10^{24} \tau_{160}\,\mathrm{cm^{-2}}/1\,\mathrm{pc}$. We use $G_0\simeq 150 (1\,\mathrm{pc}/d)^2$ in M43 and $G_0\simeq 200 (1\,\mathrm{pc}/d)^2$ in NGC 1977, approximated from the total FIR luminosity of M43 and NGC 1977, respectively. The correlations between the photoelectric heating efficiency $\epsilon$ and the ionization parameter $\gamma$ in the Orion Nebula (the Veil Shell and OMC4), the shell of M43, and the shell of NGC 1977 are shown in Fig. \ref{Fig.pe_scatter}. For M42, the data reveals a similar trend with a decreased efficiency with increased ionization parameter as the theoretical predictions of \cite{BakesTielens1994}. The cooling efficiency is generally lower in OMC4 compared to the Veil Shell, M43, and NGC 1977, due to a ``FIR excess'' from dense gas behind the [C\,{\sc ii}]-emitting layers (cf. Fig. 2f in Paper I). M43 comprises too few data points to reveal a clear trend. For NGC 1977, the observations reveal a heating efficiency that is largely independent of the ionization parameter, but this is complying with the near-constant theoretical heating efficiency for low $\gamma$. However, virtually all points lie below the theoretical heating curve. As a caveat, we note that it is hard to constrain $\gamma$ observationally, which is mainly due to the difficulty of obtaining a robust gas density estimate, but we deem it unlikely that we overestimate the density (or electron density, assuming all electrons stem from carbon ionization) systematically by an order of magnitude. We observe this same behavior when averaging over the regions and employing average values of $\gamma$.

\begin{figure}[tb]
\includegraphics[width=0.5\textwidth, height=0.33\textwidth]{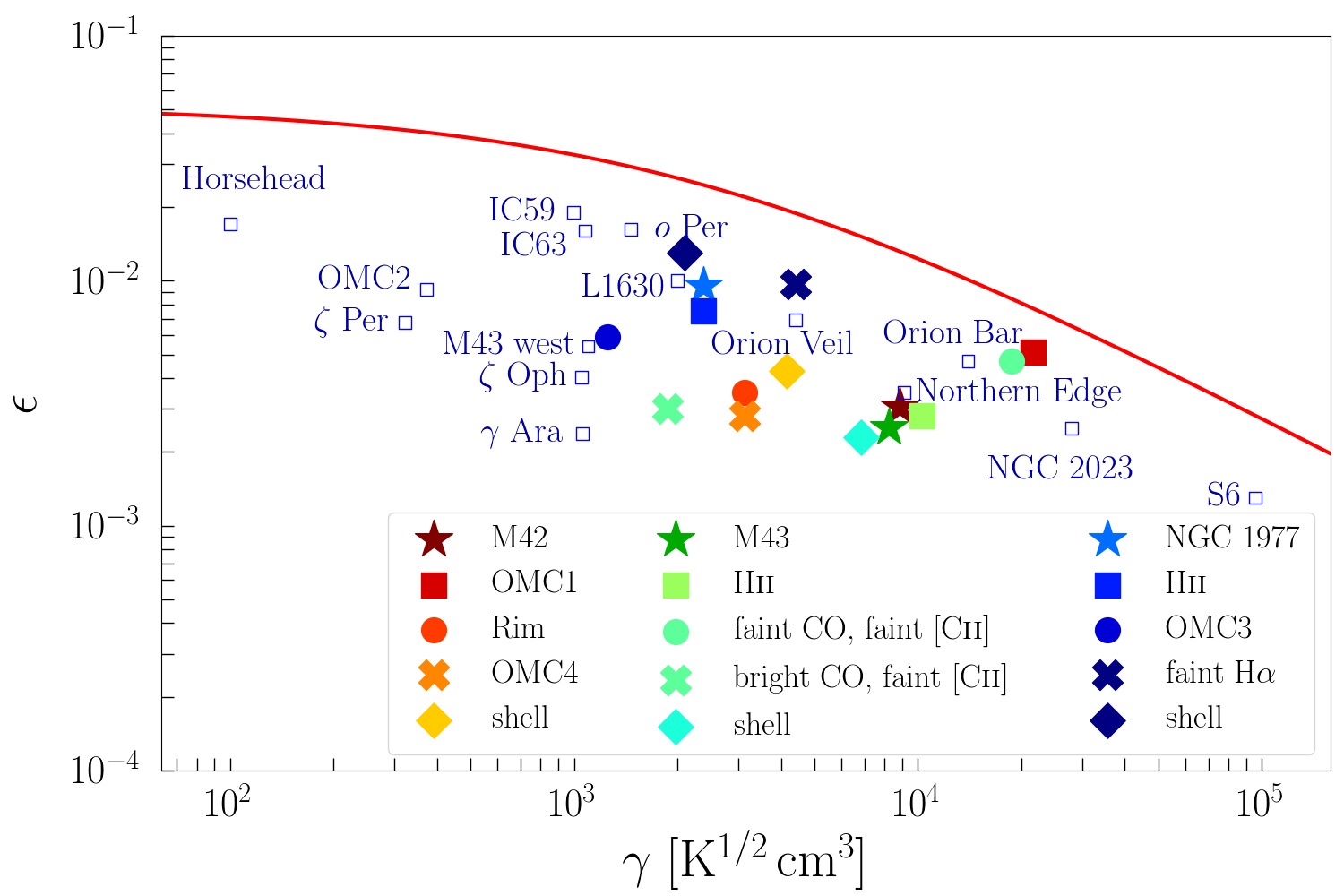}
\caption{The average photoelectric heating efficiency $\epsilon$, equated as cooling efficiency [C\,{\sc ii}]/FIR, in M42, M43, and NGC 1977, divided into the previously defined subregions, as function of the ionization parameter $\gamma=G_0 T^{1/2}/n_{\mathrm{e}}$. OMC1 includes [O\,{\sc i}] cooling, but excludes the BN/KL region and Orion S. Parameters for Orion A are given in Table \ref{Tab.pe}. The photoelectric heating efficiency in PDRs and diffuse sight lines from the literature is plotted as open blue squares (cf. Fig. 9 in Paper I). Data for the diffuse ISM are taken from \citet{Gry1992} and \citet{vanDishoeck1986}, data of NGC 2023 are from \citet{HollenbachTielens1999}, data for the Horsehead and L1630 are from \citet{Pabst2017}, data for IC 59 and IC 63 are from \citet{Andrews2018}, data for the Orion Bar, S6, M43 west, and OMC2 are from \citet{Herrmann1997}, and data for the Orion Veil are from \citet{Salas2019}. The red curve is the theoretical prediction of Eq. 43 by \cite{BakesTielens1994}.}
\label{Fig.pe_regions}
\end{figure}

\begin{table}[tb]
\addtolength{\tabcolsep}{-5.5pt}
\def\arraystretch{1.2}
\caption{Cooling efficiency and estimated physical parameters to compute $\gamma$ in Fig. \ref{Fig.pe_regions}.}
\begin{tabular}{ll|cccc}
\hline\hline
Region & & $L_{\mathrm{[C\,\textsc{ii}]}}/L_{\mathrm{FIR}}$ & $G_0$\tablefootmark{a} & $n$\tablefootmark{b} & $T$\tablefootmark{c} \\
 & &  &  & $\mathrm{[cm^{-3}]}$ & $\mathrm{[K]}$ \\ \hline
M42 & all & $3.6\times 10^{-3}$ & $1\times 10^{3}$ & $1\times 10^{4}$ & 200 \\
& OMC1\tablefootmark{d} & $6.5\times 10^{-4}$ & $2\times 10^{4}$ & $1\times 10^{5}$ & 300 \\
& Eastern Rim & $3.5\times 10^{-3}$ & 500 & $1\times 10^{4}$ & 100 \\
& OMC4 & $2.8\times 10^{-3}$ & 500 & $1\times 10^{4}$ & 100 \\
& Veil Shell & $4.3\times 10^{-3}$ & 200 & $3\times 10^{3}$ & 100 \\
M43 & all & $2.7\times 10^{-3}$ & $1\times 10^{3}$ & $1\times 10^{4}$ & 120 \\
& background, H\,{\sc ii} & $2.8\times 10^{-3}$ & $1.5\times 10^{3}$ & $1\times 10^{4}$ & 120 \\
& {\footnotesize faint CO, faint [C\,{\sc ii}]} & $4.7\times 10^{-3}$ & 300 & $1\times 10^{3}$ & 100 \\
& {\footnotesize bright CO, faint [C\,{\sc ii}]} & $3.0\times 10^{-3}$ & 300 & $1\times 10^{4}$ & 100 \\
& shell & $2.3\times 10^{-3}$ & $1\times 10^{3}$ & $1\times 10^{4}$ & 120 \\
NGC & all & $8.9\times 10^{-3}$ & 100 & $5\times 10^{3}$ & 90 \\
1977 & H\,{\sc ii} & $7.9\times 10^{-3}$ & 200 & $5\times 10^{3}$ & 90 \\
& OMC3 & $5.9\times 10^{-3}$ & 200 & $2\times 10^{4}$ & 100 \\
& faint H$\alpha$ & $9.7\times 10^{-3}$ & 100 & $1\times 10^{3}$ & 50 \\
& shell & $1.3\times 10^{-2}$ & 100 & $5\times 10^{3}$ & 70 \\ \hline
\end{tabular}
\tablefoot{\tablefoottext{a}{Estimated from average distance to respective central star with systematic uncertainty of $\pm 50\%$.}
\tablefoottext{b}{Estimated from [C\,{\sc ii}]-CO separation in Eastern Rim and M43 shell, otherwise $\tau_{160}$, with systematic uncertainty of $\pm 50\%$.}
\tablefoottext{c}{Estimated from [C\,{\sc ii}] excitation temperature (cf. Appendix A in Paper I). PDR models predict up to twice as large temperature for the [C\,{\sc ii}]-emitting layers.}
\tablefoottext{d}{[O\,{\sc i}] $63\,\mu\mathrm{m}$ cooling contributes significantly in OMC1. We find $L_{\mathrm{[O\,\textsc{i}]}} \simeq 320\,L_{\sun}$ (Higgins et al. in prep.), hence $(L_{\mathrm{[C\,\textsc{ii}]}}+L_{\mathrm{[O\,\textsc{i}]}})/L_{\mathrm{FIR}} \simeq 4.5\times 10^{-3}$. The Bright Ridge with BN/KL and Orion S accounts for $L_{\mathrm{[O\,\textsc{i}]}} \simeq 90\,L_{\sun}$, hence including this area yields a cooling efficiency of $(L_{\mathrm{[C\,\textsc{ii}]}}+L_{\mathrm{[O\,\textsc{i}]}})/L_{\mathrm{FIR}} \simeq 5.1\times 10^{-3}$ for the PDR.}}
\label{Tab.pe}
\end{table}

Figure \ref{Fig.pe_regions} shows the average cooling rate as a function of $\gamma$ averaged over the regions defined in M42, M43, and NGC 1977, together with data from the literature. The estimated parameters for M42, M43, and NGC 1977 employed in this figure are given in Table \ref{Tab.pe}. We include [O\,{\sc i}] $63\,\mu\mathrm{m}$ cooling from Higgins et al. (in prep.) in OMC1. The [O\,{\sc i}] $63\,\mu\mathrm{m}$ line is expected to be important in the Eastern Rim and M43, as well, with $I_{\mathrm{[O\,\textsc{i}]}}/I_{\mathrm{[C\,\textsc{ii}]}}\sim 3$ \citep{PoundWolfire2008, Kaufman2006}, but will be of little significance in the extended shell of NGC 1977. Observations of [O\,{\sc i}] $63\,\mu\mathrm{m}$ and $145\,\mu\mathrm{m}$ line emission in the Orion Nebula would provide the means to establish better estimates of the gas temperature and gas density, besides the fact that it substantially contributes to the gas cooling.

These average heating efficiencies show the same trend as the overall data in Fig. \ref{Fig.pe_scatter} with a decrease with increasing ionization parameter but the spread is rather large. The Orion data also agrees well with previous studies, both for bright PDRs using fine-structure line measurements as also for diffuse clouds from UV absorption line studies. This earlier data set also shows a large spread. Comparing the data to the theoretical results of \cite{BakesTielens1994}, we see a similar behavior of decreasing heating efficiency with increasing $\gamma$, consistently in Orion and earlier studies, but displaced by a factor of $\sim 5$ to lower $\epsilon$. There seems to be an additional factor, besides the ionization parameter, that affects the coupling of FUV photons to the neutral atomic gas. This factor could be a variable abundance of PAHs and VSGs in different regions within Orion (and other lines of sight). A lower abundance of PAHs and VSGs diminishes the heating efficiency systematically, while a variable composition of the PAH family or a varying ratio of PAHs to VSGs can cause variation in the heating efficiency. As shown by \citet{Berne2015} and \citet{Croiset2016} in the case of NGC 7023, the population of PAHs is very sensitive to the physical conditions ($G_0$ and $n$). Furthermore, the recombination rates of PAHs, employed in the theoretical model, are not well-constrained experimentally.

The heating efficiency describes the energy input per FUV photon. When equating the cooling efficiency [C\,{\sc ii}]/FIR with the heating efficiency, we assume that only FUV photons are re-radiated in FIR energy. We note that a fraction of energy of O stars is emitted in the EUV and redistributed into optical photons by the surrounding H\,{\sc ii} region while cooler stars emit a significant amount of their energy in the spectrum below 6\,eV. These lower-energy photons contribute to the heating of the large dust grains and increase the FIR intensity emitted by those. Using CLOUDY \citep[C17.02][]{Ferland2017} models we find an average factor of $\sim 1.5$ increase depending on the spectral type. Also the energy spectrum changes with spectral type, which induces changes in the heating efficiency as derived by \cite{BakesTielens1994} for the radiation field at $3\times 10^4\,\mathrm{K}$. For early B stars ($T_{\mathrm{eff}}\simeq 2.5\times 10^4\,\mathrm{K}$), \cite{Spaans1994} derive a $\sim$15\% lower heating efficiency. While those two corrections might bridge the gap between the observations and predictions in NGC 1977 (except OMC3), cooling efficiencies in regions in M42 and M43 are about 5 times smaller than predicted by theory at the estimated $\gamma$. Another complication in the interpretation of the observed [C\,{\sc ii}]/FIR ratio is the fact that most structures are observed in edge-on geometry. In the M42 and M43 edges, the [C\,{\sc ii}] line tends to be optically thick toward the observer. The real cooling efficiency might be higher (by a factor of $\sim 2$) than observed if the shell structures cool mainly through the (unobservable) face-on surfaces. Dust in H\,{\sc ii} regions can also produce low [C\,{\sc ii}]/FIR ratios. Only OMC1 lies close to the theoretical curve, but we note that the cooling is dominated by the [O\,{\sc i}] $63\,\mu\mathrm{m}$ line\footnote{The observed [C\,{\sc ii}] intensity in OMC1 is higher than predicted by standard face-on PDR models \citep{PoundWolfire2008, Kaufman2006}, but the [O\,{\sc i}] $63\,\mu\mathrm{m}$ intensity is predicted accurately.}.

\begin{figure}[tb]
\includegraphics[width=0.5\textwidth, height=0.33\textwidth]{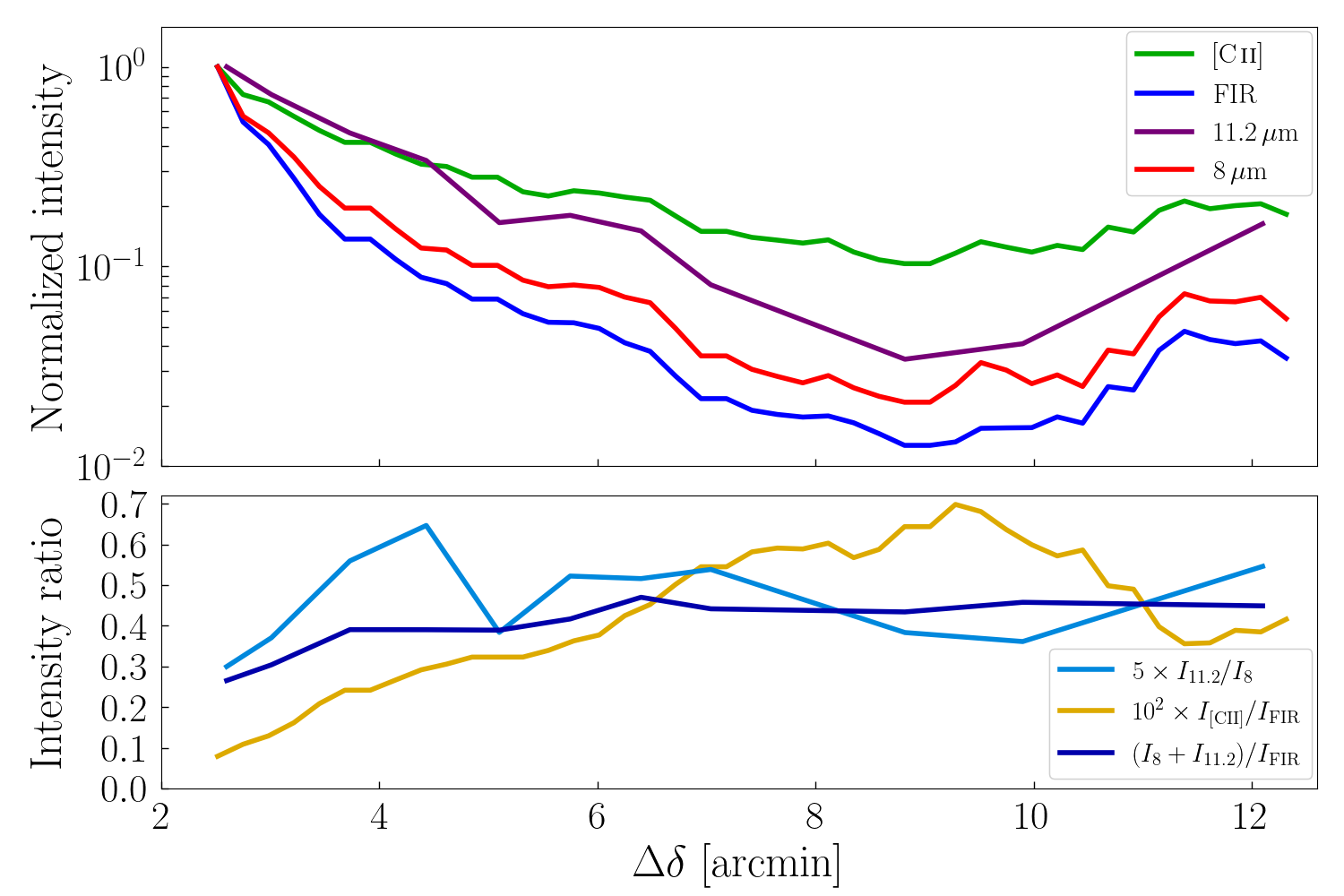}
\caption{Line cut from the Orion Bar to the southwest, reaching the Eastern Rim at $12\arcmin$, indicated in Fig. \ref{Fig.map_spectra}. The offset is given with respect to $\theta^1$ Ori C. The IRAC 8\,$\mu$m filter measures predominantly the emission in the 7.7\,$\mu$m PAH band. IRS has measured the strength of the 11.2\,$\mu$m PAH feature. The $11.2\,\mu\mathrm{m}$ intensities are taken from Table 1 in \cite{Boersma2012}.}
\label{Fig.line-cut-behind_bar}
\end{figure}

The Orion [C\,{\sc ii}] survey also provides a way to link the observed heating efficiency directly to the characteristics of the local PAH family. \cite{Rubin2011} and \cite{Boersma2012} have analyzed a set of {\it Spitzer}/IRS spectra that systematically explored the behavior with distance of PAHs in the PDR associated with the Veil Shell. As Fig. \ref{Fig.line-cut-behind_bar} shows, the intensity of the [C\,{\sc ii}], FIR, and PAH bands all smoothly decrease with distance from the Orion Bar (and hence the illuminating star). The gas heating efficiency, however, shows a clear peak at about $10\arcmin$ from the Bar. In contrast, the ratio of the IRAC 8\,$\mu$m band -- dominated by the 7.7\,$\mu$m PAH feature -- to the IRS 11.2\,$\mu$m band -- a blend of the PAH 11.0\,$\mu$m and 11.25\,$\mu$m features -- is largely constant over the full distance range probed. This band ratio is generally thought to probe the degree of ionization of the emitting PAHs as the former band traces emission by ionized PAHs while the latter is dominated by neutral PAHs \citep{Peeters2002, Boersma2012}. Given the general relation between $G_0$ and $n$ discussed in Sec. \ref{Sec.G0-n}, we expect little variation in the ionization parameter with distance and that is borne out by the constant $8\,\mu\mathrm{m}$/$11.2\,\mu\mathrm{m}$ band ratio. As a corollary, this again suggests that there is another factor that affects the gas heating efficiency besides ionization. That could be abundance of PAHs and VSGs. However, the intensity ratio ($I_{8\,\mu\mathrm{m}}+I_{11.2\,\mu\mathrm{m}})/I_{\mathrm{FIR}}$ -- a good proxy for the PAH abundance \citep{Tielens2008}-- is constant along the line cut, which hints that VSGs play an important role.

Most of the observed [C\,{\sc ii}] and FIR emission will stem from the background PDR. The [C\,{\sc ii}]/FIR increases with distance from the origin of the line cut, until it hits the Eastern Rim, where it gradually decreases as the Rim piles up along the line of sight. The decrease of the cooling efficiency in the Rim may be due to increased $G_0$ (as a result of the geometry). Alternatively, the edge-on geometry of the Rim increases the optical depth of the [C\,{\sc ii}] line compared to the FIR intensity, which leads to a diminished observed [C\,{\sc ii}]/FIR ratio. As we expect the [C\,{\sc ii}] line to be optically thick in the Rim, the latter will certainly be an important effect. Yet, we cannot exclude that $G_0$ in the Rim is higher than in the molecular background, as reflected in the higher dust temperature in the Rim (cf. Fig. \ref{Fig.T_tau}).

\begin{figure}[tb]
\includegraphics[width=0.5\textwidth, height=0.33\textwidth]{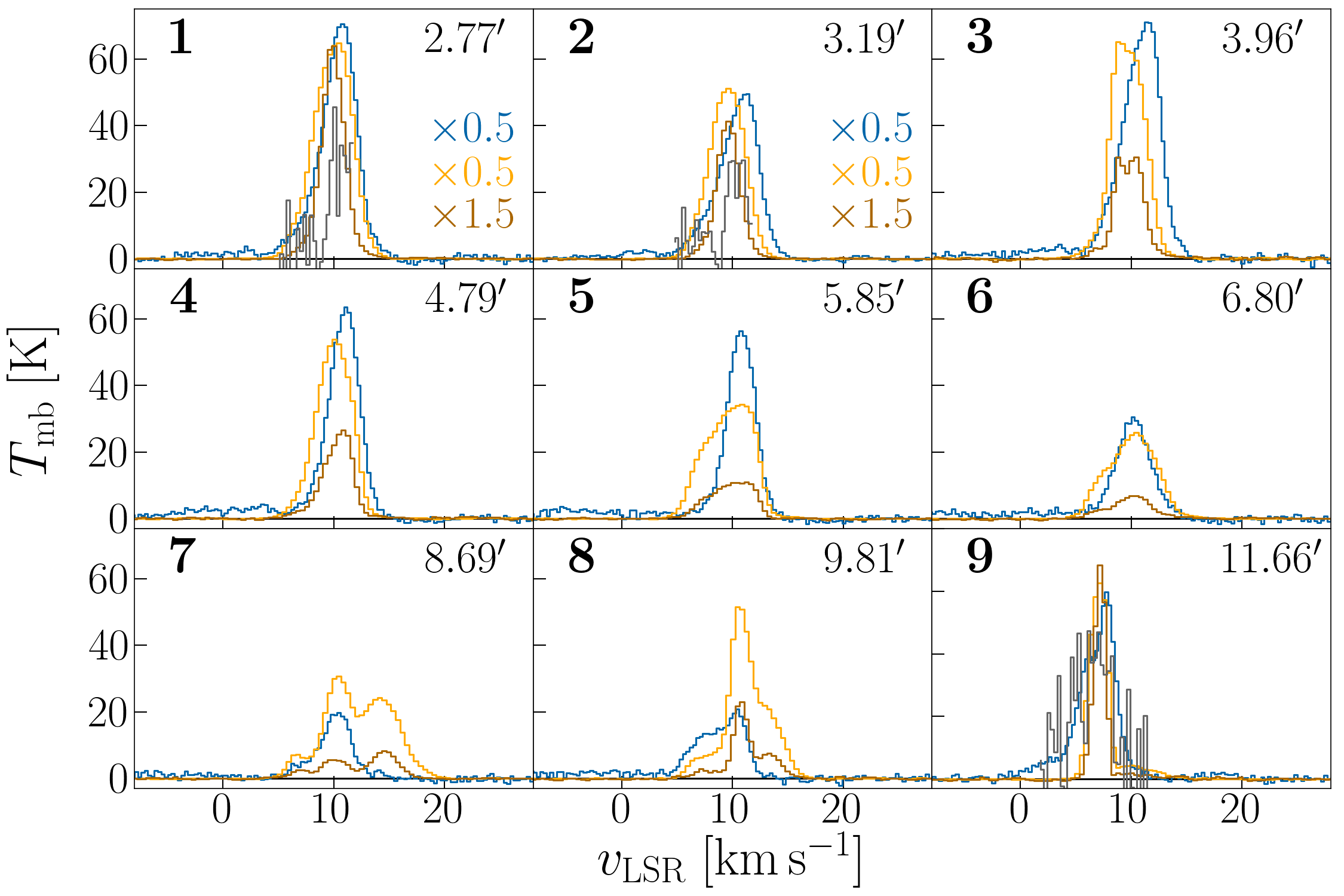}
\caption{[C\,{\sc ii}] (blue), ${}^{12}$CO(2-1) (yellow), and ${}^{13}$CO (brown) spectra along the line cut behind the Orion Bar in Fig. \ref{Fig.line-cut-behind_bar}. The offset with respect to $\theta^1$ Ori C is given in arcminutes. In addition, the [$^{13}$C\,{\sc ii}] $F=2\text{-}1$ component is plotted (gray) in its systemic velocity and multiplied by 20 in spectra 1, 2, and 9.}
\label{Fig.spectra-behind_bar}
\end{figure}

Spectra along the line cut, shown in Fig. \ref{Fig.spectra-behind_bar}, reveal complex dynamics of the gas, and potentially foreground or self-absorption of the [C\,{\sc ii}] line, that can effect the cooling and heating efficiency in this region. The CO(2-1) intensity peaks strongly between 8 and 10 arcmin. Those are the points (positions V1 and V2) that show deviating characteristics in the study of \cite{Boersma2012}, that is a shift in peak position and line width. Positions V1 and V2 positions also exhibit a higher fullerene-to-PAH ratio than other positions. CO pv diagrams reveal dynamic structures in this region. While those structures are more manifest in CO, spectra toward selected regions reveal weaker corresponding [C\,{\sc ii}] components (spectra 7 and 8 in Fig. \ref{Fig.spectra-behind_bar}). FIR emission seems to be dominated by re-radiation of FUV photons from $\theta^1$ Ori C and/or $\theta^2$ Ori A, as concluded from the persistent decline in FIR intensity.

\section{Conclusion}

In Paper I, we have studied the correlations of the [C\,{\sc ii}] intensity with the dust continuum intensity at 70\,$\mu$m, the FIR intensity (40-500\,$\mu$m), PAH 8\,$\mu$m intensity, and CO(2-1) intensity in the Orion Nebula complex. Here, we have focused on these same correlations divided into several subregions (Veil Shell, OMC4, M43, and NGC 1977, further divided up into more subregions). Typically, the observed correlations show only minor variations between these regions. However, the [C\,{\sc ii}]-CO(2-1) correlations are very different between the regions as they are very sensitive to the geometry. The geometric complexity of these regions is quite apparent when examining individual spectra toward several representative regions. We have compared the observations with constant-density PDR models and concluded that, in general, the edge-on PDR models are able to reproduce the observed intensities, while face-on PDR models cannot explain the observations. The estimated physical parameters, that serve as model inputs, correspond to a constant $p_{\mathrm{th}}/G_0$ ratio, which has been observed toward several other PDRs, as well. We have explicated the origin of the [C\,{\sc ii}] emission within the mapped area in greater detail than in Paper I, focusing on the detailed PDR processes and source properties. While the bulk of [C\,{\sc ii}] emission from the mapped area originates in PDR surfaces (only 7.5\% from the bright OMC1), about 5\% each in the Veil Shell and NGC 1977 and 15\% in M43 stem from H\,{\sc ii} gas. Furthermore, we have compared the properties of PAHs in the southeast of the Orion Bar (as measured by {\it Spitzer}) to the [C\,{\sc ii}] cooling efficiency, but do not find a correlation between the two: While the [C\,{\sc ii}] cooling efficiency varies with distance from the Orion Bar, the PAH charge does not do so. Future studies may include the [O\,{\sc i}] 63\,$\mu$m and 145\,$\mu$m cooling lines and alter the picture obtained by the study of the [C\,{\sc ii}] 158\,$\mu$m line alone. Moreover, the James Webb Space Telescope (JWST) will be able to measure ISM/PAH properties with higher accuracy, and these results may shine further light on the observed variations in the cooling efficiency.

\begin{acknowledgements}
We thank the anonymous referee for valuable comments that helped improve the manuscript.

This work is based on observations made with the NASA/DLR Stratospheric Observatory for Infrared Astronomy (SOFIA). SOFIA is jointly operated by the Universities Space Research Association, Inc. (USRA), under NASA contract NNA17BF53C, and the Deutsches SOFIA Institut (DSI) under DLR contract 50 OK 0901 to the University of Stuttgart.

JRG thanks the Spanish MCIU for funding support under grant AYA2017-85111-P and PID2019-106110GB-I00. This project has received funding from the European Research Council (ERC) under the European Union's Horizon 2020 research and innovation programme (Grant agreement No. 851435). Research on the interstellar medium at Leiden Observatory is supported through a Spinoza award of the Dutch Science Organisation (NWO).

We thank C.~R. O'Dell for constructive discussions and S.~T. Megeath for helpful advice.

This work made use of the TOPCAT software \citep{Taylor2005}.
\end{acknowledgements}

\bibliographystyle{aa} 
\bibliography{article3} 

\appendix

\section{Comparison with Lombardi SED fits}
\label{Sec.SED_Lombardi}

\begin{figure*}[p]
\includegraphics[width=\textwidth, height=0.515\textwidth]{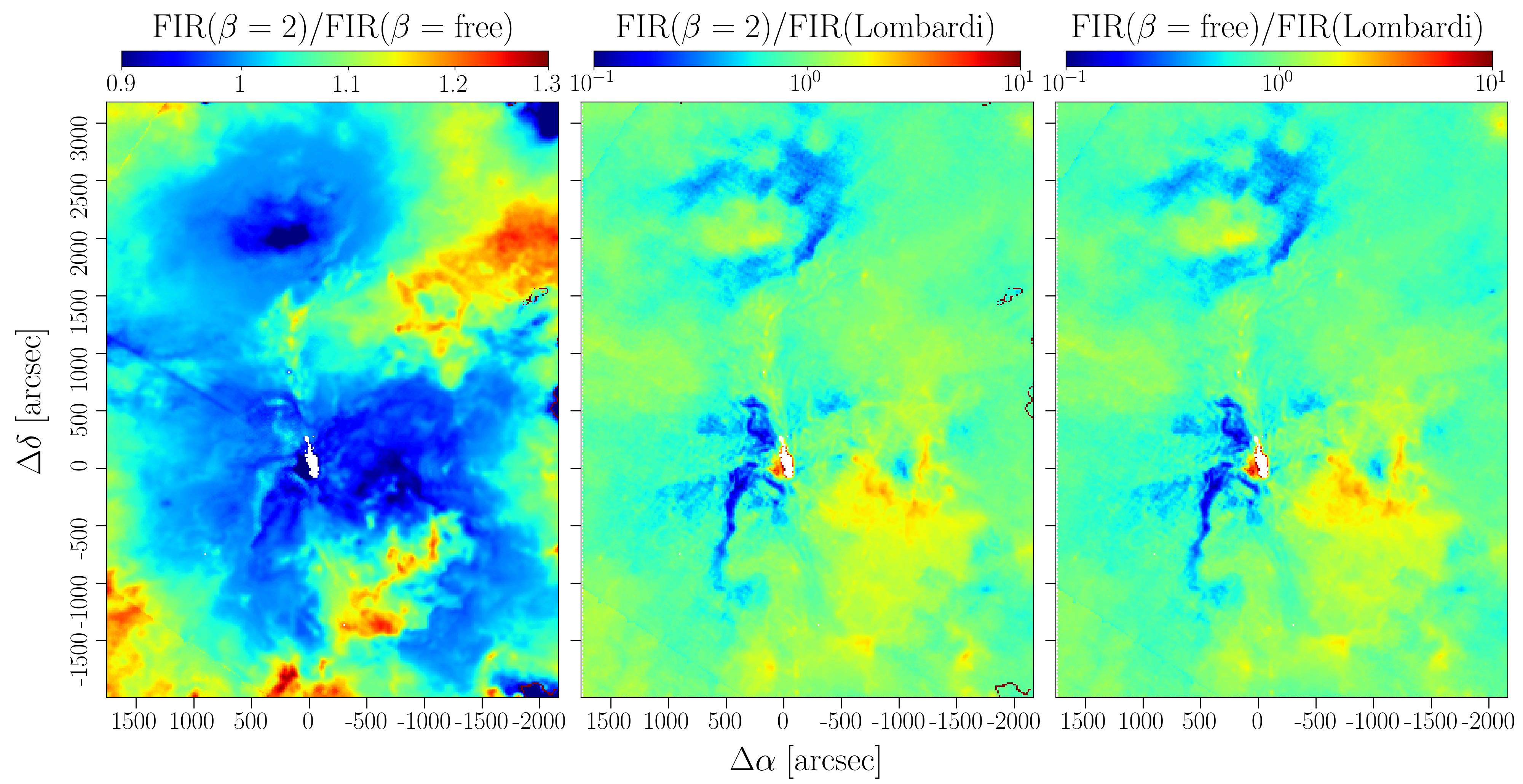}
\caption{Ratios of FIR intensity from different SED fits ($\beta=2$, $\beta$ free and the \citet{Lombardi2014} result).}
\label{Fig.sed_fir_ratios}
\end{figure*}

\begin{figure*}[p]
\includegraphics[width=\textwidth, height=0.515\textwidth]{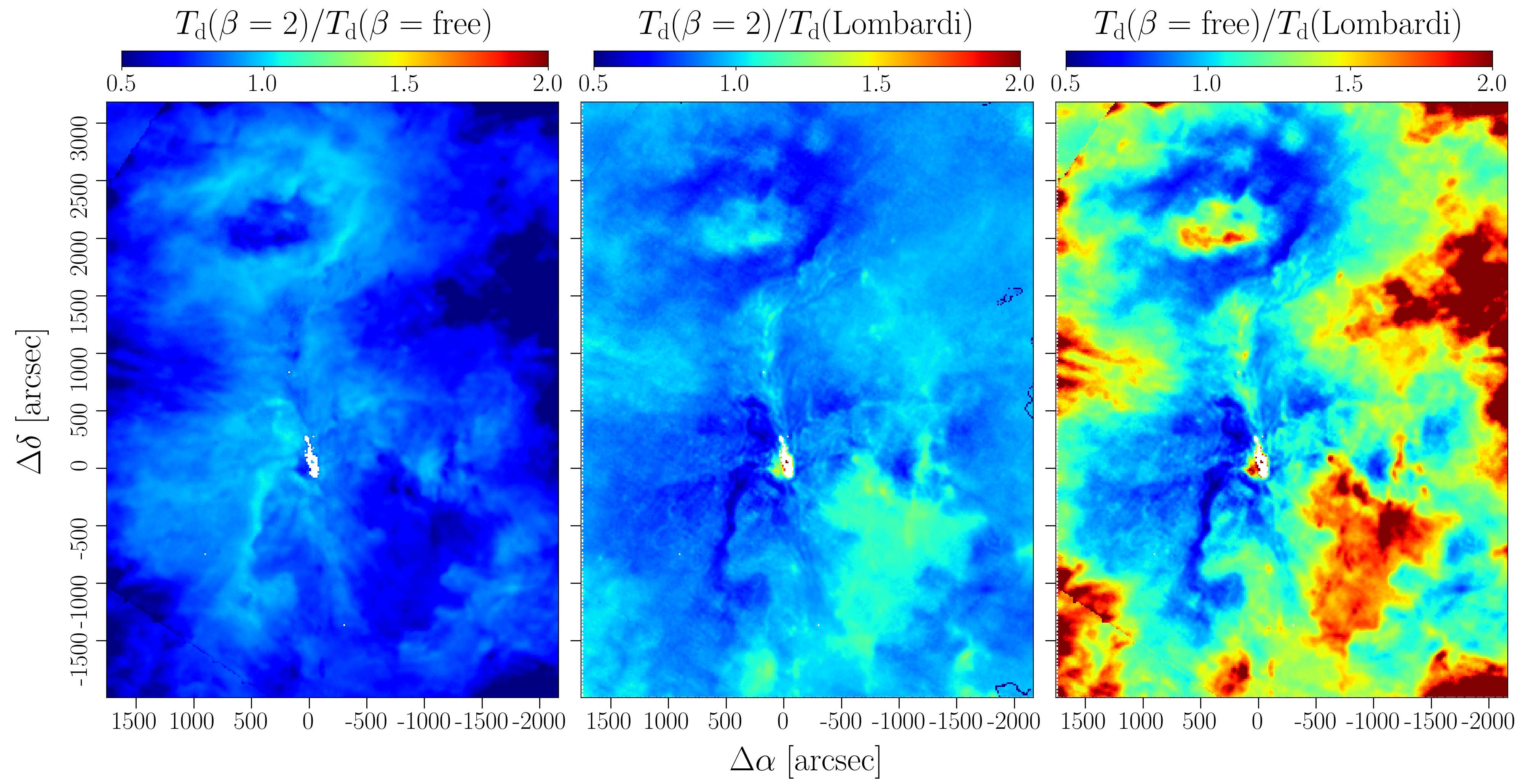}
\caption{Ratios of the dust temperature $T_{\mathrm{d}}$ from different SED fits ($\beta=2$, $\beta$ free and the \citet{Lombardi2014} result).}
\label{Fig.sed_Td_ratios}
\end{figure*}

As opposed to \cite{Lombardi2014}, we include the shorter-wavelength PACS $70\,\mu\mathrm{m}$ and $100\,\mu\mathrm{m}$ bands, since we do not expect them to be opaque in the extended structures we are mainly interested in. In fact, the extended shell of the Orion Nebula has $\tau_{160}<10^{-2}$, which translates into $\tau_{70}<3\times 10^{-2}$; even the densest cores along the ISF are described by $\tau_{160}\lesssim 10^{-1}$. Comparing our derived FIR intensity, we note that it agrees with the FIR intensity computed from the SED maps of \cite{Lombardi2014}, except in the bright eastern arm, the Rim, and the bubble edge of NGC 1973, 1975, and 1977. Here, we obtain significantly lower FIR intensity than \cite{Lombardi2014} (cf. Fig. \ref{Fig.sed_fir_ratios}). These filaments contain warmer dust that is only captured by the shorter-wavelength bands, leading \cite{Lombardi2014} to derive too high a dust temperature and too low a dust optical depth. \cite{Lombardi2014} obtain higher dust temperatures in the Eastern Rim, yet significantly lower dust optical depth.

Figs. \ref{Fig.sed_fir_ratios} and \ref{Fig.sed_Td_ratios} show the ratios of the results of different SED fits performed on the region of the EON. Fig. \ref{Fig.sed_fir_ratios} shows the ratios of the FIR intensity of our $\beta=2$ SED fit, an SED fit where we let $\beta$ a free parameter, and the \cite{Lombardi2014} result. Fig. \ref{Fig.sed_Td_ratios} shows the ratios of the dust temperatures from these same fits. From the right-hand panel it becomes obvious that the fit of \cite{Lombardi2014} yields significantly higher FIR intensity and dust temperature in the brightest parts of the EON, especially the eastern Rim, the OMC1 region and M43. This might be due to the fact that \cite{Lombardi2014} uses only the longer wavelength bands, that do not capture the peak of the SED of the warmer dust in these bright regions. This apparently results in an overestimation of the dust temperature and consequentially of the FIR intensity.

\section{Correlation plots from free-$\beta$ SED fit}
\label{Sec.SED_beta_free}

\begin{figure*}[p]
\includegraphics[width=\textwidth, height=0.515\textwidth]{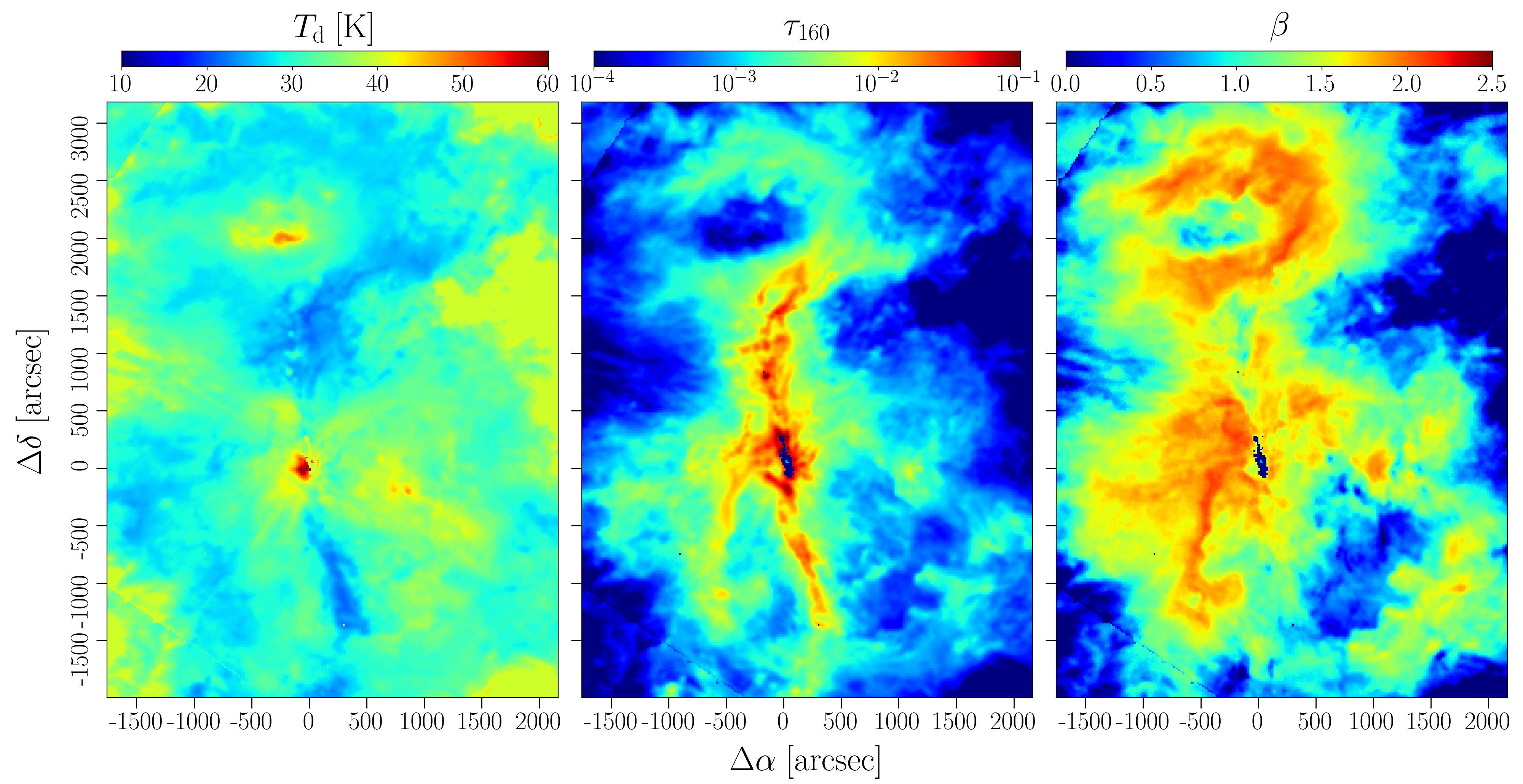}
\caption{Maps of the three SED fit output parameters $T_{\mathrm{d}}$, $\tau_{160}$ and $\beta$.}
\label{Fig.SED_results_beta_free}
\end{figure*}

In this section, we present the correlation plots of results of an SED fit where we let the grain emissivity index $\beta$ be a free parameter. Fig. \ref{Fig.SED_results_beta_free} shows the resulting maps of the three SED output parameters $T_{\mathrm{d}}$, $\tau_{160}$ and $\beta$. From this figure, we notice that there seems to be a rather tight correlation between the resulting $\beta$ and $\tau_{160}$, which we quantify in Fig. \ref{Fig.corr_tau_beta_free}.

Most of the pixels seem to be consistent with a $\beta$ of $\simeq 1.5$, as suggested by \cite{Lombardi2014}. The latter use low-resolution {\it Planck} images to constrain $\beta$, which amounts to $\beta\simeq 1.6$ in the Orion Nebula complex. The median of all pixels in our SED fit is $\beta$ of $\simeq 1.2$, but when including only pixels with $\tau_{160}>5\times 10^{-4}$, which comprise most of the visible structure, the median shifts to $\beta$ of $\simeq 1.4$. The peak values, around which the $\beta$ values are clustered, is $\beta\simeq 1.6$. Standard models, however, suggest $\beta\simeq 2$ \citep{Weingartner2001}. The study of \cite{Extaluze2013} suggests varying $\beta$ according to the environment (radiation field). We note that $\beta$ can be affected by the presence of an additional warm dust population in the hot gas and we determine low $\beta$ in the EON cavity. $\beta$ may vary with the grain composition, grain-size distribution, and with the presence of ices on the grains \citep[e.g.,][]{Ossenkopf1994, Liseau2015}. Extragalactic studies show systematic variations of $\beta$ with galactocentric radius, which may be linked to metallicity gradients \citep[e.g.,][]{Kramer2010, Tabatabaei2014}.

In ratio plots of the dust temperature from the three different SED fits, shown in Fig. \ref{Fig.sed_Td_ratios}, shell structures and interior show distinct deviations from unity, seemingly confirming the assumption that dust properties and, hence, $\beta$ indeed vary with environment \citep[cf.][]{Extaluze2013}. Denser, warmer structures close to the Trapezium stars seem to be characterized by $\beta \gtrsim 2$, while the extended shell has $\beta\simeq 1\text{-}2$ and the dilute shell interior $\beta \simeq 0\text{-}1$.

\begin{figure*}[p]
\includegraphics[width=\textwidth, height=0.4\textwidth]{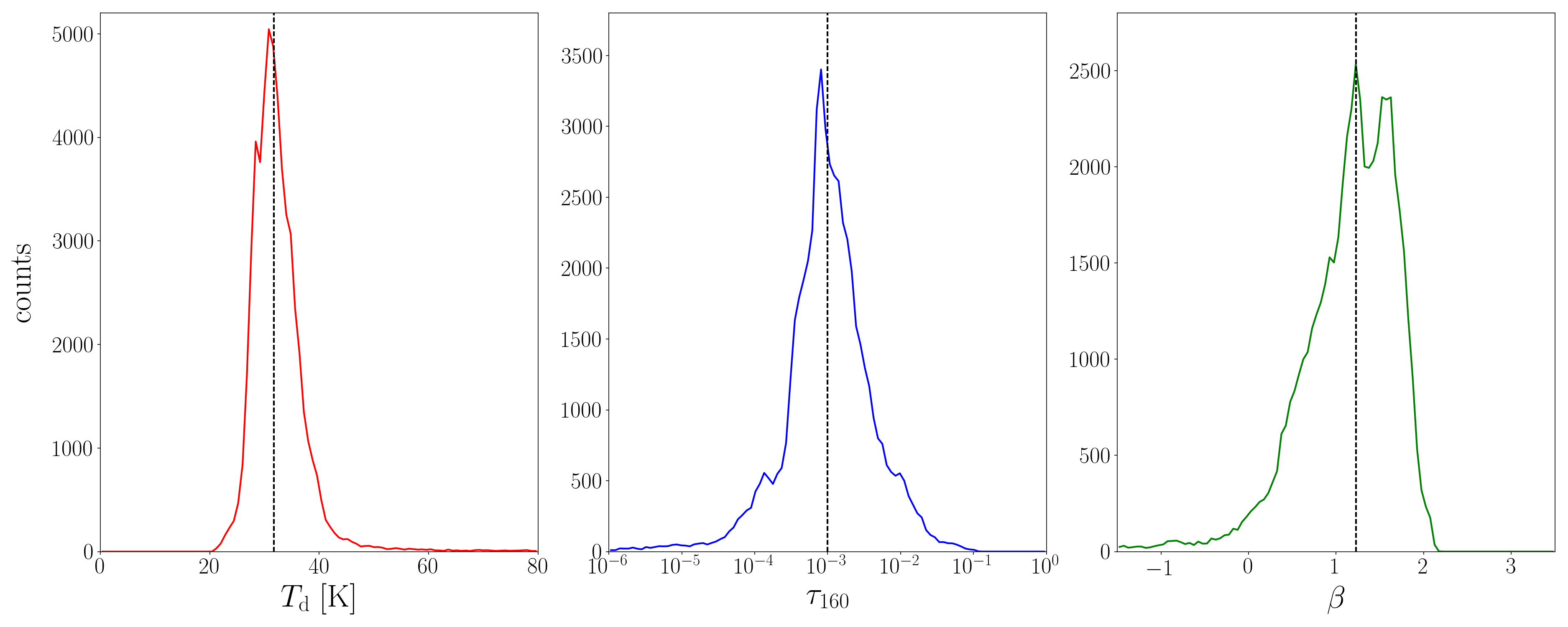}
\caption{Histograms of the three SED fit output parameters $T_{\mathrm{d}}$, $\tau_{160}$ and $\beta$. Dashed black lines indicate the median values ($\hat T_{\mathrm{d}}\simeq 31.7\,\mathrm{K}$, $\hat \tau_{160}\simeq 1.0\times 10^{-3}$, $\hat \beta \simeq 1.23$).}
\label{Fig.SED_histrogram_3}
\end{figure*}

In histograms of the SED output parameters (Fig. \ref{Fig.SED_histrogram_3}), the dust temperature lies mostly in the range of $T_{\mathrm{d}}\simeq 20 \text{-} 50\,\mathrm{K}$, dust opacity can be mostly found in $\tau_{160}\simeq 10^{-4} \text{-} 10^{-2}$, and the grain emissivity index varies mostly between $\beta\simeq 0\text{-} 2$, yet with a significant tail of negative values. The latter correspond to areas with little FIR emission and thus poor fit results.

\begin{figure*}[p]
\begin{minipage}[t]{0.49\textwidth}
\includegraphics[width=\textwidth, height=0.67\textwidth]{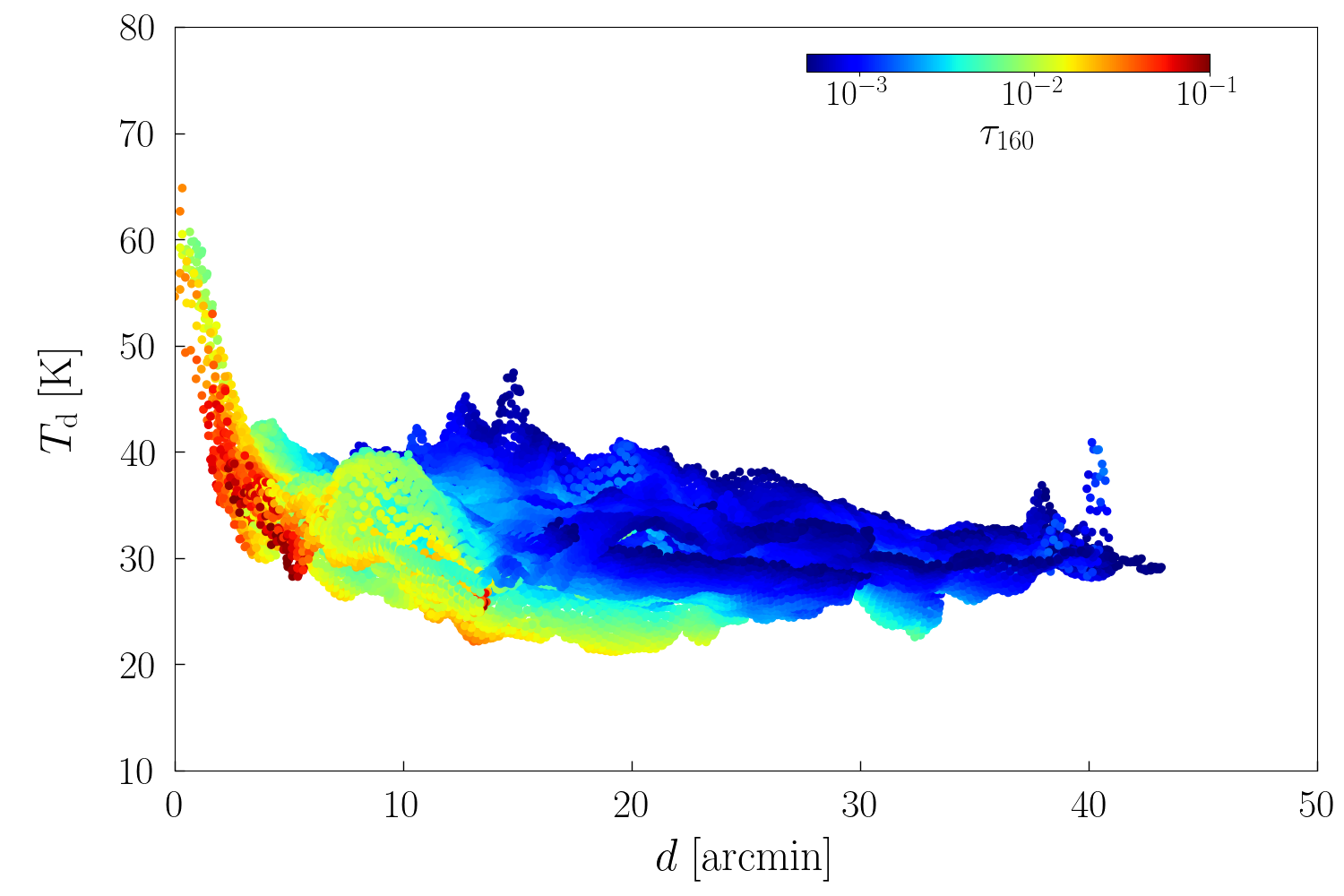}
\caption{Dust temperature $T_{\mathrm{d}}$ from an SED fit with $\beta$ a free parameter versus distance from $\theta^1$ Ori C. Colors code the dust optical depth for $\tau_{160}>5\times 10^{-4}$.}
\label{Fig.corr_Td_d_free}
\end{minipage}
\hspace{0.015\textwidth}
\begin{minipage}[t]{0.49\textwidth}
\includegraphics[width=\textwidth, height=0.67\textwidth]{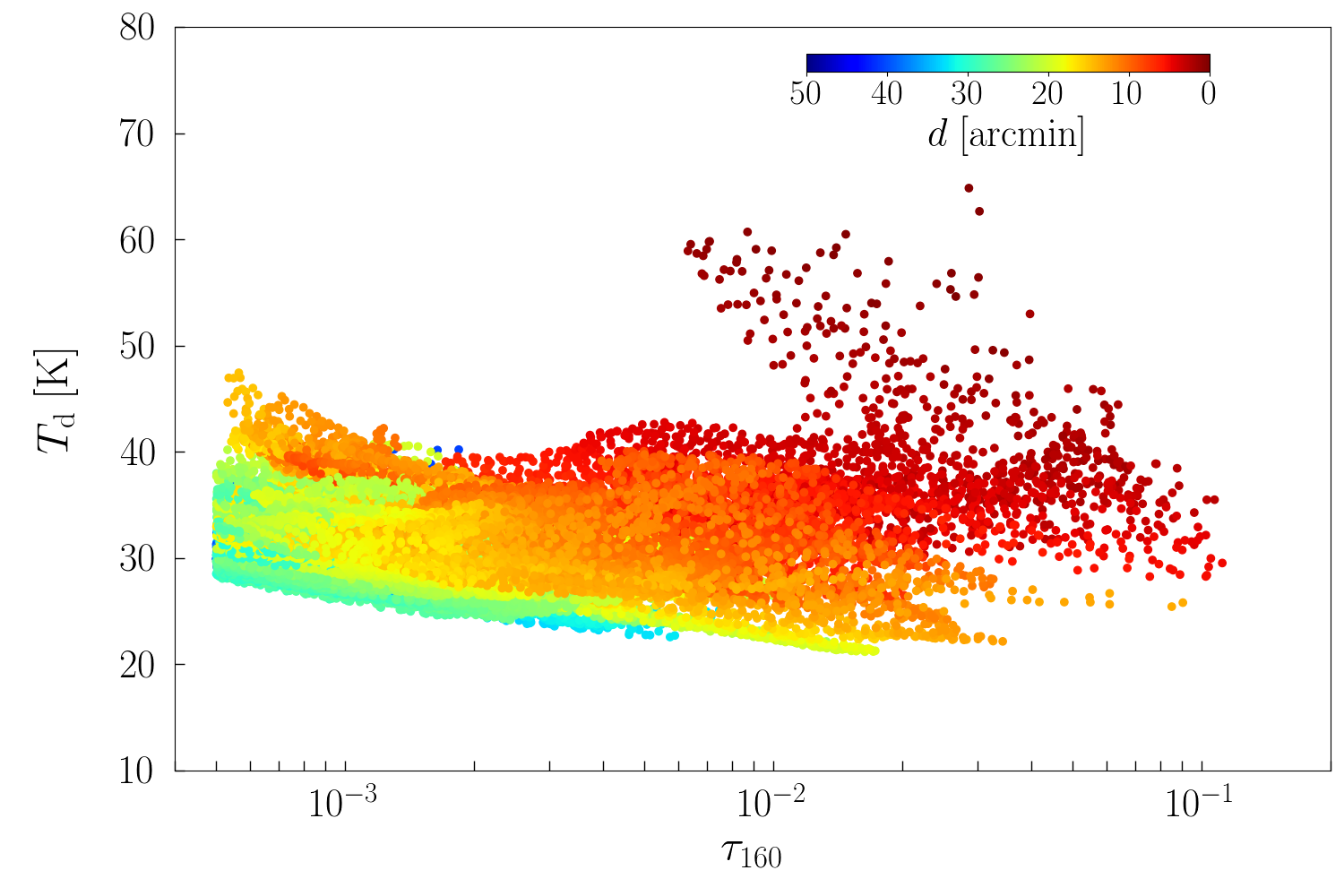}
\caption{Dust temperature $T_{\mathrm{d}}$ versus dust optical depth $\tau_{160}(>5\times 10^{-4})$ in a SED fit with $\beta$ a free parameter ($\rho\simeq 0.11$). Colors code the distance from $\theta^1$ Ori C.}
\label{Fig.corr_Td_tau_free}
\end{minipage}

\begin{minipage}[t]{0.49\textwidth}
\includegraphics[width=\textwidth, height=0.67\textwidth]{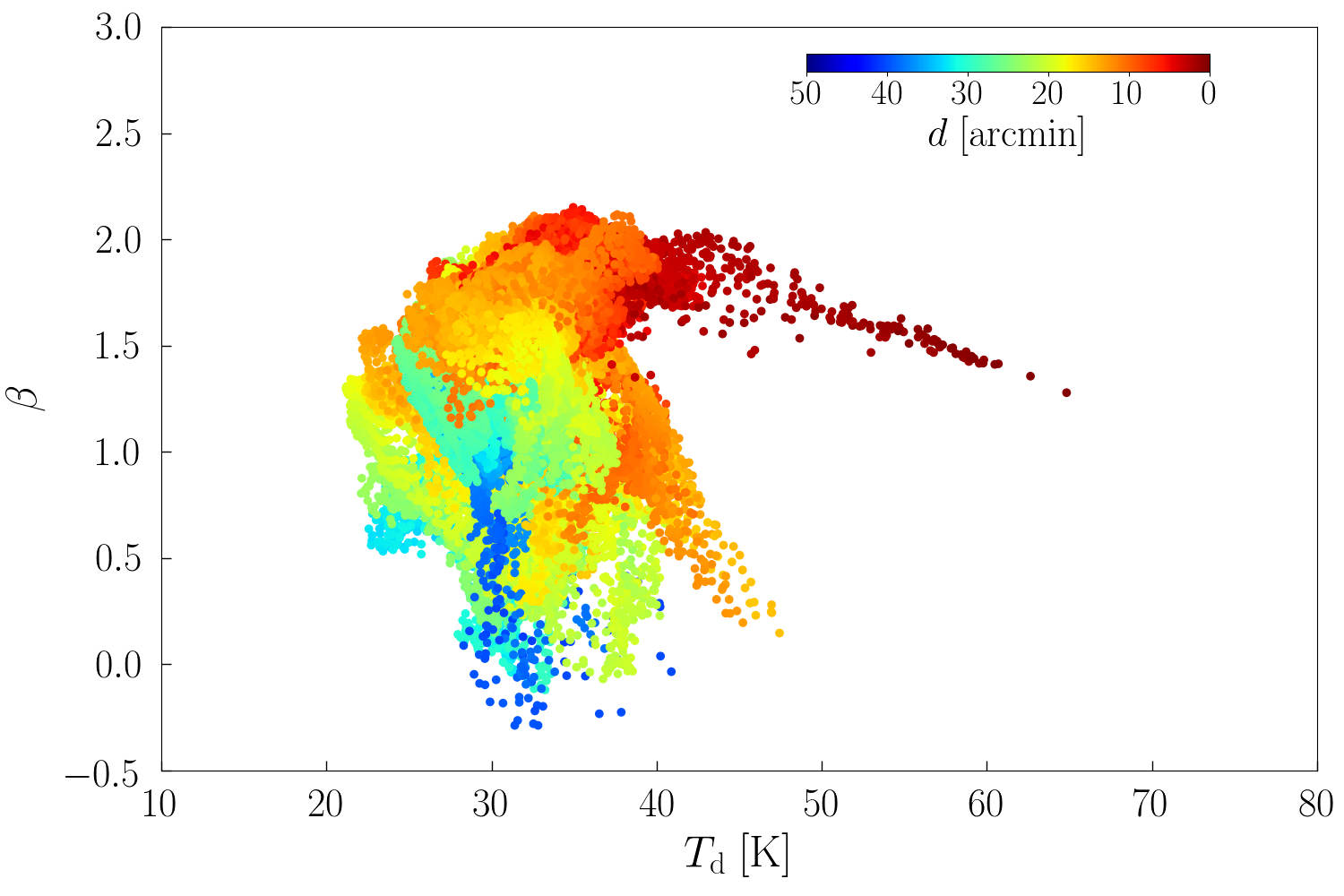}
\caption{Dust temperature $T_{\mathrm{d}}$ versus grain emissivity index $\beta$ for $\tau_{160}>5\times 10^{-4}$ ($\rho\simeq 0.00$). Colors code the distance from $\theta^1$ Ori C.}
\label{Fig.corr_Td_beta_free}
\end{minipage}
\hspace{0.015\textwidth}
\begin{minipage}[t]{0.49\textwidth}
\includegraphics[width=\textwidth, height=0.67\textwidth]{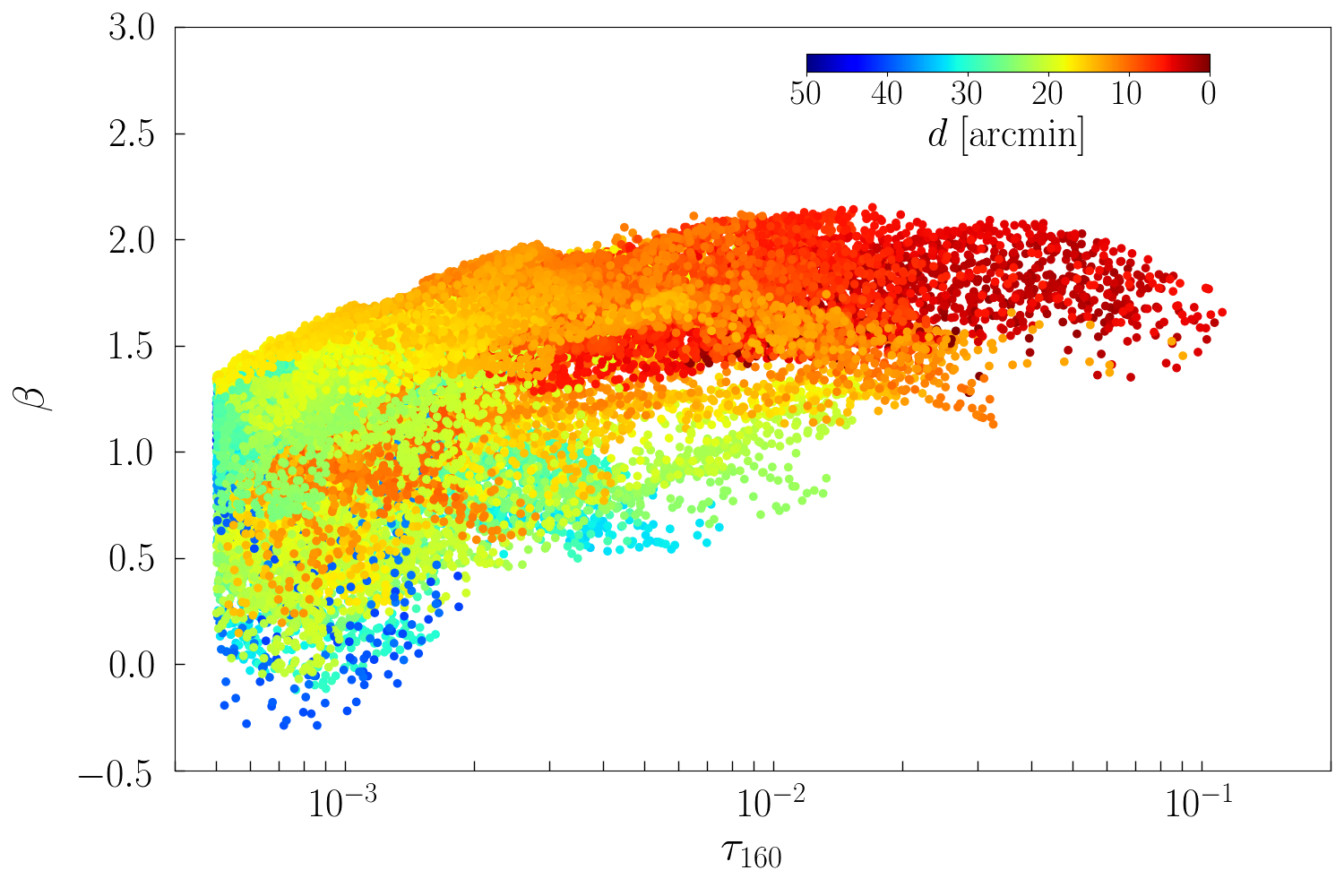}
\caption{Dust optical depth $\tau_{160}(>5\times 10^{-4})$ versus grain emissivity index $\beta$ ($\rho\simeq 0.32$). Colors code the distance from $\theta^1$ Ori C.}
\label{Fig.corr_tau_beta_free}
\end{minipage}
\end{figure*}

In Fig. \ref{Fig.corr_Td_d_free}, we clearly see that the dust temperature decreases with increasing distance from the presumably dominant heating source $\theta^1$ Ori C. Points corresponding to lower $\tau_{160}$ tend to be more distant and scatter around the mean dust temperature. 

From Fig. \ref{Fig.corr_Td_tau_free}, we conclude that also in a free-$\beta$ SED fit there is no correlation between the resulting dust temperature and optical depth. Independent of  $\tau_{160}$, most of the data points scatter around the mean dust temperature $T_{\mathrm{d}}\simeq 31.4\,\mathrm{K}$ with a standard deviation of $\sigma_{T_{\mathrm{d}}}\simeq 4.0\,\mathrm{K}$.

There is no correlation between $T_{\mathrm{d}}$ and $\beta$, as can be seen from Fig. \ref{Fig.corr_Td_beta_free}. The mean grain emissivity index is $\beta\simeq 1.33$, but with significant scatter around this value. 

The correlation between $\tau_{160}$ and $\beta$ has a significant slope (Fig. \ref{Fig.corr_tau_beta_free}), even though there is substantial scatter, which amounts to a correlation index of only $\rho\simeq 0.32$. Deeper structures tend to have higher $\beta$, but this is not reflected in dust temperature, as discussed above. This might be due to the fact that deeper structures also contain colder gas along with warmer (surface) gas, which leads to a relative suppression of the long-wavelength tail, resulting in a higher $\beta$.

\section{SED fits to single points}

\begin{figure*}[p]
\begin{minipage}{0.5\textwidth}
\includegraphics[width=\textwidth, height=0.67\textwidth]{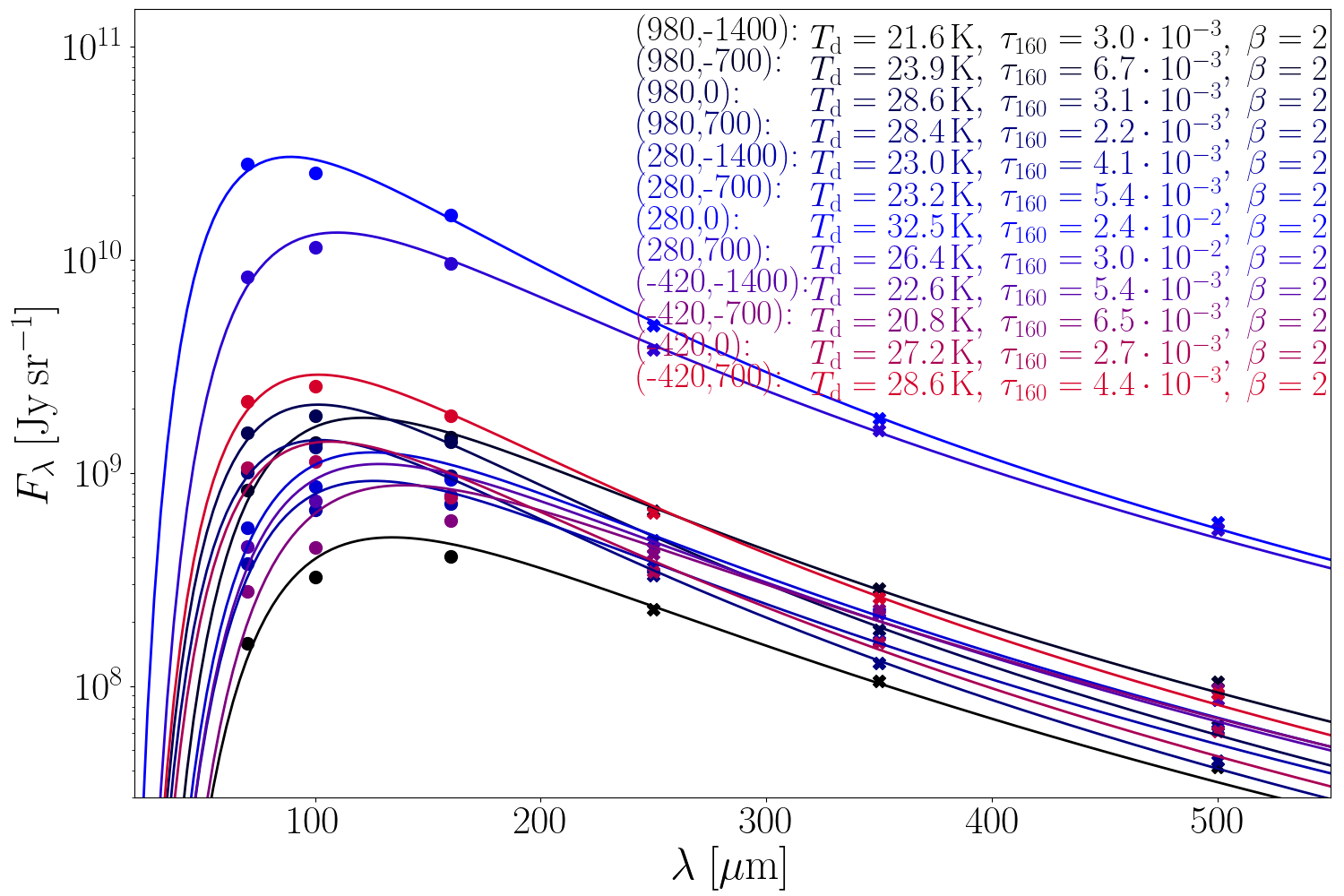}
\caption{SED fits to single pixels with $\beta=2$.}
\label{Fig.SED_points_2p0}
\end{minipage}
\begin{minipage}{0.5\textwidth}
\includegraphics[width=\textwidth, height=0.67\textwidth]{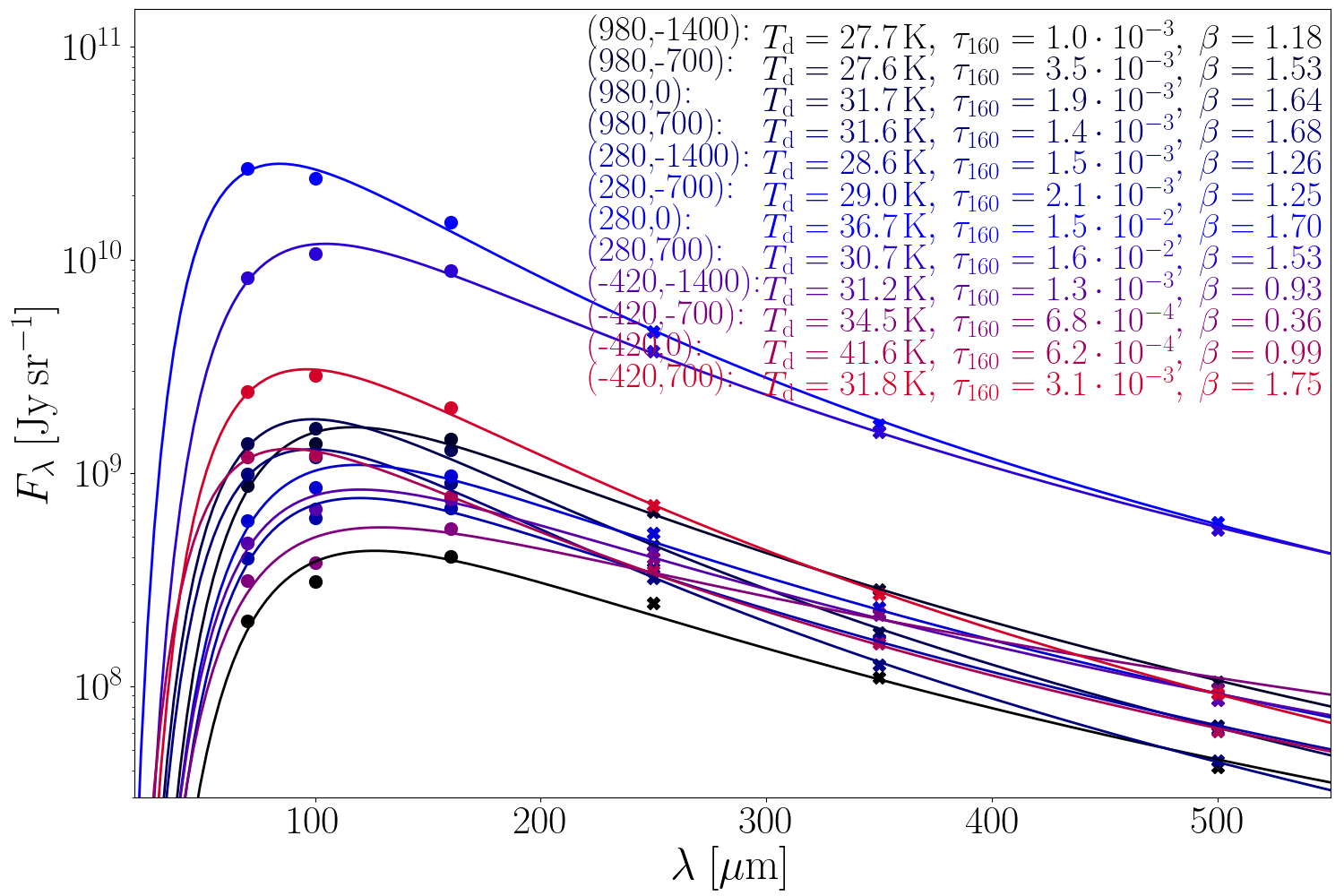}
\caption{SED fits to single pixels with free $\beta$.}
\label{Fig.SED_points_free}
\end{minipage}
\end{figure*}

Figure \ref{Fig.SED_points_2p0} shows SED fits with $\beta=2$ toward representative points. Fig. \ref{Fig.SED_points_free} shows SED fits where we let $\beta$ vary freely for the same selection of points.

\clearpage

\section{Spectral line fits}
\label{Sec.spectra}

\subsection{M42}
\label{Sec.sub-M42}

Figure \ref{Fig.spectra-high-CII} shows spectra toward six regions of high [CII] intensity in M42. Table \ref{Tab.CII-bright} gives the fit parameters of a multi-component Gaussian fit for each spectrum. Table \ref{Tab.13CII-bright-CII} gives the result of the [$^{13}$C\,{\sc ii}] analysis, where possible.
 
Spectrum 1 (Eastern Rim, north) exhibits two [C\,{\sc ii}] main components. Both can be observed in [${}^{13}$C\,{\sc ii}], as well. From the fit results, we obtain two rather different results: The $7.3\,\mathrm{km\,s^{-1}}$ yields a moderate excitation temperature, while the $10.5\,\mathrm{km\,s^{-1}}$ component seems to arise in very warm gas. The latter than traces only a very small column, while the former traces one third of the dust-traced column. Despite the high temperature, the second component displays a very narrow emission line, where the line broadening appears to be almost thermal. The origin of this component is unclear. In channel maps, it seems to be connected to the OMC1 core, that has a bright eastern extension at that velocity.

The bright [C\,{\sc ii}] emission in spectrum 2 (Eastern Rim, south) requires an incident radiation field of $G_0\simeq 10^4$ (from models). The excitation temperatures we derive from the [${}^{13}$C\,{\sc ii}] line (two components) are not extreme. Both components trace about 15\% of the dust column. While the [C\,{\sc ii}] components are relatively narrow, the high-velocity ${}^{12}$CO and ${}^{13}$CO components emit rather broad lines. We do not observe a significant broad [C\,{\sc ii}] component that might stem from the ionized gas, that is seen in H$\alpha$ toward this region. It is not certain how additional possible sources (e.g., T Ori) affect the excitation and dynamics in this region.

The bright [C\,{\sc ii}] emission of spectrum 3 (south of OMC1, north of OMC4) consists of two components. Despite the strong main [C\,{\sc ii}] line, the [${}^{13}$C\,{\sc ii}] line is very weak. This indicates rather low optical depth, but high temperature. Because of the relatively weak [${}^{13}$C\,{\sc ii}] line, it is unlikely that the main [C\,{\sc ii}] line is self- or foreground absorbed. Rather, we fit two emission components. We can only detect the brighter component in [${}^{13}$C\,{\sc ii}]. This [C\,{\sc ii}] component traces only a small fraction of the total dust column. The high temperature is reflected in the associated very bright CO emission (with a broad ${}^{13}$CO line), that traces a 4 times larger gas column than [C\,{\sc ii}] ($A_{\mathrm{V,los}}\simeq 40$). The other CO component, associated with the fainter [C\,{\sc ii}] main component, seems to arise in relatively more moderate conditions (narrow ${}^{13}$CO line) and seems to trace only a thin ($A_{\mathrm{V,los}}\simeq 3$) layer of gas. In addition to the strong main line, we observe a very broad component in the [C\,{\sc ii}] spectrum at negative velocities, that possibly stems from ionized gas expanding toward the observer.

The brighter [C\,{\sc ii}] components require very extreme conditions, according to the standard models of the PDR Toolbox \citep{Kaufman2006,PoundWolfire2008}. Also in the central regions of the Orion Nebula, the [C\,{\sc ii}] intensity is higher than expected from standard PDR models. Constant pressure models, as opposed to constant density models, might hold clues to the reason for this behavior.

In spectrum 4 (west of Orion S), we observe a single bright [C\,{\sc ii}] main component, well separated from two weak low-velocity components. Also the CO spectra reveal only one component. In the region of spectrum 4, we observe the lower branch of the [C\,{\sc ii}]-FIR correlation in the ISF (although this region lies adjacent to the main spine). Despite the face-on geometry in this area, the [C\,{\sc ii}] emission lies above the model predictions. We cannot exclude a tilted surface, however. If we assume optically thick [C\,{\sc ii}] emission, we compute an excitation temperature of $T_{\mathrm{ex}}=64\,\mathrm{K}$. The CO excitation temperature is only slightly lower and both emissions need rather high radiation field and density (from the models). The ${}^{13}$CO line is rather broad, which hints at elevated turbulence. The CO emission traces a large column of gas, $ A_{\mathrm{V,los}}\simeq 20$, which is 1/5 of the dust-traced column. The dust temperature is rather moderate ($T_{\mathrm{d}}=27\,\mathrm{K}$), the background consisting of a large column of cold dust. From the FIR intensity we estimate a radiation field of $G_0\simeq 4\times 10^3$, which is 2-3 times lower than required for the [C\,{\sc ii}] and CO intensity according to the models.

In spectra 5 and 6 (south of OMC4 in the ISF), we notice that the CO excitation temperature is further decreasing with distance from the Trapezium. The CO column density remains constant, CO tracing a column of $A_{\mathrm{V,los}}\simeq 10$. From comparison with the [C\,{\sc ii}] emission, we see that $G_0$ is dropping, but $n$ remains approximately constant, $n\simeq 1\text{-}5\times 10^4\,\mathrm{cm}^{-3}$.

\subsection{M43}
\label{Sec.sub-M43}

Figure \ref{Fig.spectra-M43-shell} shows six spectra toward the shell of M43. Table \ref{Tab.M43-shell} gives the fit parameters of a multi-component Gaussian fit for each spectrum. Table \ref{Tab.13CII-M43} gives the result of the [$^{13}$C\,{\sc ii}] analysis, where possible.

All spectra have too strong [C\,{\sc ii}] emission as to be consistent with face-on geometry. The column traced by [C\,{\sc ii}] corresponds to $A_{\mathrm{V,los}}\simeq 15$ in most cases. This clearly indicates an edge-on illuminated gas column, given the rather moderate radiation field in M43. Most spectra, both [C\,{\sc ii}] and CO, consist of at least two components. Spectrum 5 is the most enigmatic. The [C\,{\sc ii}] spectrum can be fitted only with four components plus one component that we attribute to [${}^{13}$C\,{\sc ii}] emission (it could be a red-shifted [$^{12}$C\,{\sc ii}] component, as well). However, the [${}^{13}$C\,{\sc ii}] component (if it is such and not an additional [$^{12}$C\,{\sc ii}] component) appears to be rather strong compared to the [$^{12}$C\,{\sc ii}] component, which yields a very high [C\,{\sc ii}] optical depth. The ${}^{12}$CO spectrum clearly comprises two components, but cannot be fitted satisfactorily with Gaussian profiles, while the ${}^{13}$CO spectrum can. We note that spectrum 5 lies in a region where the shell surrounding the M43 H\,{\sc ii} region seems to be interrupted, possibly allowing for ionized gas flowing out of the cavity. Outflows may lead to irregular line shapes in the surface layers of a molecular cloud. However, we do not observe a broad component in the [C\,{\sc ii}] spectrum, that might stem from the ionized gas itself. The fact that most spectra comprise more than one emission component may indicate that the gas layers along the line of sight experience shear due to stellar feedback. We judge that the spectra in M43 can be explained by emitting components only, rather than invoking additional absorbing layers.

\subsection{NGC 1977}
\label{Sec.sub-NGC1977}

Figure \ref{Fig.spectra-NGC1977-shell} shows six spectra toward the shell of NGC 1977. Table \ref{Tab.NGC1977-shell} gives the fit parameters of a multi-component Gaussian fit for each spectrum. Table \ref{Tab.13CII-NGC1977} gives the result of the [$^{13}$C\,{\sc ii}] analysis, where possible.

The CO component at $8\,\mathrm{km\,s^{-1}}$ in spectrum 2 corresponds to a weak [C\,{\sc ii}] component, that cannot be fitted properly because it is just above the noise level. Again, it is unlikely that ${}^{13}$CO is self-absorbed, hence we conclude that there are two emitting CO components within the line of sight.

The [C\,{\sc ii}] line in spectrum 3 is almost Gaussian, but slightly asymmetric. In a two-component fit the [C\,{\sc ii}] components line up with the CO components. This might indicate that the shell around NGC 1977 is indeed moving at differential velocities and that CO emission arises within the shell at low density. The cloud depth derived from the ${}^{13}$CO line is about $A_{\mathrm{V,los}}\simeq 0.1\text{-}0.4$.
 
The [C\,{\sc ii}] line in spectrum 4 is slightly asymmetric. The line widths in a two-component fit are close to the thermal line width, as in spectrum 3. We do not find a good fit with an absorbing component instead of a second emitting component. Since we expect the region to be in dynamic expansion, we may expect line asymmetries, and the Gaussian two-component fit may be misleading. 

Also the [C\,{\sc ii}] line in spectrum 5 has a relatively (close to thermal) line width, but an additional broad wing. Spectrum 5 is taken toward the [C\,{\sc ii}]-faint region, where the shell may be breaking open. The broad wing might be a signature of an ionized outflow or increased turbulence in the neutral gas due to the rupture (as in M43). The narrow line in spectrum 5 is blue-shifted with respect to the lines in other [C\,{\sc ii}] spectra toward NGC 1977, possibly reflecting an increased backward pressure in this region.

The [C\,{\sc ii}] line in spectrum 6 is slightly asymmetric, as well, and can be fitted with two Gaussian components. 

\subsection{Spectra behind the Bar}
\label{Sec.sub-behind-bar}

Fig. \ref{Fig.spectra-behind_bar} shows nine spectra toward the region southeast of the Orion Bar. Table \ref{Tab.behind-Bar} gives the fit parameters of a multi-component Gaussian fit for each spectrum. Table \ref{Tab.13CII-behind-Bar} gives the result of the [$^{13}$C\,{\sc ii}] analysis, where possible.

Spectra with bright [C\,{\sc ii}] emission, where we can detect the [$^{13}$C\,{\sc ii}] $F=2\text{-}1$ line, can be interpreted as exhibiting a significant absorbing component. However, invoking an absorbing component gives rise to rather extreme derived excitation condition, which we do not expect behind the Orion Bar and in the Eastern Rim. As the [$^{13}$C\,{\sc ii}] line is often only marginally detected, fit uncertainties are large. Deeper observations are needed to investigate this matter further (Kabanovic et al. in prep.). All spectra along the line cut behind the Orion Bar show multiple features, that are enigmatic in origin. A more detailed study, comparing with velocity-resolved data sets like CO rotational lines, optical and dust emission, would be needed to understand the structure of the background PDR and the Eastern Rim. We report here that the relation between [C\,{\sc ii}] and CO(2-1) emission is complex, as components in the respective spectra are not easy to match with each other.

\onecolumn

\begin{longtable}{llcccc}
\caption{Line fits to [C\,{\sc ii}]-bright spectra 1-6 in M42.}\\
\hline\hline
spectrum & line & comp. & $T_{\mathrm{P}}$ & $v_{\mathrm{P}}$ & $\Delta v_{\mathrm{FWHM}}$ \\
 &  & \# & [K] & [$\mathrm{km\,s^{-1}}$] & [$\mathrm{km\,s^{-1}}$] \\ \hline
 \endfirsthead
 \caption{Continued.}\\
 \hline\hline
spectrum & line & comp. & $T_{\mathrm{P}}$ & $v_{\mathrm{P}}$ & $\Delta v_{\mathrm{FWHM}}$ \\
 &  & \# & [K] & [$\mathrm{km\,s^{-1}}$] & [$\mathrm{km\,s^{-1}}$] \\ \hline
 \endhead
 \hline\\
\multicolumn{6}{c}{\parbox{\LTcapwidth}{{\bf Notes:} Line fit uncertainties are of the order $\sigma_{T_{\mathrm{P}}}\simeq 0.5\,\mathrm{K}$, $\sigma_{v_{\mathrm{P}}}\simeq 0.1\,\mathrm{km\,s^{-1}}$, and $\sigma_{v_{\mathrm{FWHM}}}\simeq 0.1\,\mathrm{km\,s^{-1}}$.}}\\
\endlastfoot 
1 &  [C\,{\sc ii}] & 1 & 35.1 & 7.4 & 2.3 \\  
 &   & 2 & 56.0 & 10.4 & 1.4 \\  
 &  \text{[$^{13}$C\,{\sc ii}]} & 1 & 1.3 & 7.1 & 5.2 \\
 &   & 2 & 0.7 & 10.6 & 3.0 \\
 &  $^{12}$CO & 1 & 12.8 & 7.3 & 2.5 \\ 
 &   & 2 & 47.1 & 11.0 & 3.0 \\ 
 &  $^{13}$CO & 1 & 4.2 & 6.7 & 2.3 \\
 &   & 2 & 28.9 & 11.2 & 1.8 \\ \hline 
2 &  [C\,{\sc ii}] & 1 & 4.3 & 2.1 & 1.6 \\
 &   & 2 & 50.7 & 7.9 & 1.5 \\  
 &   & 3 & 19.0 & 11.7 & 1.7 \\  
 &  \text{[$^{13}$C\,{\sc ii}]} & 2 & 1.4 & 7.7 & 2.3 \\
 &   & 3 & 0.6 & 10.6 & 2.0 \\
 &  $^{12}$CO & 1 & 23.4 & 8.8 & 3.9 \\ 
 &   & 2 & 28.2 & 12.5 & 3.6 \\ 
 &  $^{13}$CO & 1 & 4.5 & 7.5 & 1.4 \\
 &   & 2 & 3.7 & 11.9 & 3.7 \\ \hline
3 &  [C\,{\sc ii}] & 1 & 2.6 & -0.6 & 14.0 \\
 &   & 2 & 7.5 & 1.2 & 3.5 \\
 &   & 3 & 76.3 & 7.8 & 3.0 \\  
 &   & 4 & 32.9 & 10.6 & 2.5 \\  
 &  \text{[$^{13}$C\,{\sc ii}]} & 2 & 1.0 & 8.3 & 2.5 \\
 &  $^{12}$CO & 1 & 2.2 & -0.3 & 2.4 \\
 &   & 2 & 83.5 & 8.4 & 3.7 \\ 
 &   & 3 & 41.8 & 11.3 & 2.1 \\ 
 &  $^{13}$CO & 1 & 29.4 & 9.1 & 3.5 \\
 &   & 2 & 12.0 & 11.1 & 1.1 \\ \hline
4 &  [C\,{\sc ii}] & 1 & 4.5 & -1.0 & 4.4 \\
 &   & 2 & 2.2 & 3.2 & 2.4 \\
 &   & 3 & 25.7 & 9.1 & 3.3 \\  
 &  $^{12}$CO & 1 & 52.5 & 9.3 & 4.2 \\ 
 &  $^{13}$CO & 1 & 19.5 & 9.4 & 3.4 \\ \hline
5 &  [C\,{\sc ii}] & 1 & 10.7 & 7.3 & 2.2 \\ 
 &   & 2 & 3.4 & 8.9 & 0.8 \\ 
 &  $^{12}$CO & 1 & 36.7 & 7.8 & 3.3 \\ 
 &   & 2 & 9.5 & 10.6 & 2.0 \\
 &  $^{13}$CO & 1 & 15.0 & 8.0 & 2.9 \\ \hline 
6 &  [C\,{\sc ii}] & 1 & 1.2 & -2.9 & 4.6 \\ 
 &   & 2 & 6.3 & 8.0 & 1.8 \\ 
 &  $^{12}$CO & 1 & 27.4 & 8.2 & 3.8 \\ 
 &  $^{13}$CO & 1 & 15.8 & 8.0 & 2.8
\label{Tab.CII-bright}
\end{longtable}

\begin{table*}[!h]
\addtolength{\tabcolsep}{0pt}
\def\arraystretch{1.2}
\caption{Parameters and results of [${}^{13}$C\,{\sc ii}] analysis in M42.}
\begin{tabular}{l|ccccccccc}
\hline\hline
spectrum, & $T_{\mathrm{P}}([{}^{12}\mathrm{C\,\textsc{ii}}])$ & $T_{\mathrm{P}}([{}^{13}\mathrm{C\,\textsc{ii}}])$ & $\Delta v$ & $T_{\mathrm{d}}$ & $\tau_{160}$ & $\tau_{[\mathrm{C\,\textsc{ii}}]}$ & $T_{\mathrm{ex}}$ & $N_{\mathrm{C^+}}$ & $N_{\mathrm{C^+}}/\mathcal{A}_{\mathrm{C}}/N_{\mathrm{H}}$\\
comp. & [K] & [K] & [$\mathrm{km\,s^{-1}}$] & [K] & & & [K] & [$\mathrm{cm^{-2}}$] & \\ \hline
1, 1 & 35 & 1.3 & 2.3 & 37 & $1\times 10^{-2}$ & 3.9 & 81 & $3.0\times 10^{18}$ & 0.32 ($A_{\mathrm{V,los}}\simeq 10$) \\
1, 2 & 56 & 0.7 & 1.4 & 37 & $1\times 10^{-2}$ & 0.6 & 171 & $6.3\times 10^{17}$ & 0.07 ($A_{\mathrm{V,los}}\simeq 2$) \\
2, 2 & 51 & 1.4 & 1.5 & 34 & $1\times 10^{-2}$ & 2.8 & 100 & $1.7\times 10^{18}$ & 0.18 ($A_{\mathrm{V,los}}\simeq 5$) \\
2, 3 & 19 & 0.6 & 1.7 & 34 & $1\times 10^{-2}$ & 3.3 & 61 & $1.4\times 10^{18}$ & 0.15 ($A_{\mathrm{V,los}}\simeq 4$) \\
3 & 76 & 1.0 & 3.0 & 29 & $5\times 10^{-2}$ & 0.7 & 192 & $1.8\times 10^{18}$ & 0.04 ($A_{\mathrm{V,los}}\simeq 6$) \\ \hline
\end{tabular}
\tablefoot{Line fit uncertainties are of the order $\sigma_{T_{\mathrm{P}}}\simeq 0.5\,\mathrm{K}$. $T_{\mathrm{P}}([{}^{13}\mathrm{C\,\textsc{ii}}])$ is barely above the noise level and is uncertain by $\sigma_{T_{\mathrm{P}}}\simeq 0.3\,\mathrm{K}$.}
\label{Tab.13CII-bright-CII}
\end{table*}

\clearpage

\begin{longtable}{llcccc}
\caption{Line fits to spectra 1-6 in M43.}\\
\hline\hline
spectrum & line & comp. & $T_{\mathrm{P}}$ & $v_{\mathrm{P}}$ & $\Delta v_{\mathrm{FWHM}}$ \\
 &  & \# & [K] & [$\mathrm{km\,s^{-1}}$] & [$\mathrm{km\,s^{-1}}$] \\ \hline
 \endfirsthead
 \caption{Continued.}\\
 \hline\hline
spectrum & line & comp. & $T_{\mathrm{P}}$ & $v_{\mathrm{P}}$ & $\Delta v_{\mathrm{FWHM}}$ \\
 &  & \# & [K] & [$\mathrm{km\,s^{-1}}$] & [$\mathrm{km\,s^{-1}}$] \\ \hline
 \endhead
 \hline\\
\multicolumn{6}{c}{\parbox{\LTcapwidth}{{\bf Notes:} Line fit uncertainties are of the order $\sigma_{T_{\mathrm{P}}}\simeq 0.5\,\mathrm{K}$, $\sigma_{v_{\mathrm{P}}}\simeq 0.1\,\mathrm{km\,s^{-1}}$, and $\sigma_{v_{\mathrm{FWHM}}}\simeq 0.1\,\mathrm{km\,s^{-1}}$.}}\\
\endlastfoot 
1 &  [C\,{\sc ii}] & 1 & 73.9 & 8.7 & 3.9 \\   
 &   & 2 & 29.5 & 9.8 & 1.5 \\  
 &  \text{[$^{13}$C\,{\sc ii}]} & 1 & 2.7 & 8.7 & 2.0 \\
 &  $^{12}$CO & 1 & 19.9 & 8.7 & 2.6 \\ 
 &   & 2 & 9.4 & 11.2 & 2.8 \\ 
 &  $^{13}$CO & 1 & 3.1 & 9.2 & 3.4 \\
 &   & 2 & 0.4 & 11.1 & 2.1 \\ \hline
2 &  [C\,{\sc ii}] & 1 & 21.8 & 5.5 & 3.5 \\
 &   & 2 & 82.7 & 9.1 & 3.6 \\
 &  \text{[$^{13}$C\,{\sc ii}]} & 2 & 2.1 & 8.5 & 2.3 \\
 &  $^{12}$CO & 1 & 85.4 & 10.0 & 3.7 \\ 
 &  $^{13}$CO & 1 & 19.6 & 10.1 & 2.6 \\ \hline
3 &  [C\,{\sc ii}] & 1 & 17.5 & 5.6 & 3.6 \\
 &   & 2 & 67.3 & 10.0 & 3.7 \\ 
 &  \text{[$^{13}$C\,{\sc ii}]} & 2 & 1.0 & 9.5 & 3.4 \\
 &  $^{12}$CO & 1 & 3.1 & 6.8 & 1.9 \\
 &   & 2 & 59.4 & 10.8 & 2.5 \\ 
 &  $^{13}$CO & 1 & 0.8 & 7.8 & 1.7 \\
 &   & 2 & 15.2 & 10.6 & 1.6 \\ \hline
4 &  [C\,{\sc ii}] & 1 & 32.4 & 8.7 & 5.5 \\
 &   & 2 & 45.2 & 11.0 & 2.6 \\ 
 &  $^{12}$CO & 1 & 9.0 & 10.6 & 4.3 \\
 &   & 2 & 42.9 & 10.7 & 1.4 \\ 
 &  $^{13}$CO & 1 & 1.8 & 10.5 & 3.6 \\
 &   & 2 & 6.8 & 10.7 & 1.1 \\ \hline
5 &  [C\,{\sc ii}] & 1 & 20.9 & 6.9 & 4.1 \\ 
 &   & 2 & 22.9 & 10.3 & 3.0 \\  
 &  \text{[$^{13}$C\,{\sc ii}]} & 1 & 2.1 & 6.2 & 1.7 \\
 &  $^{12}$CO & 1 & 13.1 & 9.7 & 3.9 \\
 &   & 2 & 0.97 & 11.4 & 1.3 \\ 
 &  $^{13}$CO & 1 & 3.2 & 9.3 & 2.6 \\
 &   & 2 & 1.2 & 11.2 & 1.4 \\ \hline
6 &   [C\,{\sc ii}] & 1 & 3.7 & 5.4 & 1.6 \\ 
 &   & 2 & 72.1 & 9.7 & 3.5 \\  
 &   & 3 & 5.3 & 12.6 & 5.6 \\ 
 &  \text{[$^{13}$C\,{\sc ii}]} & 2 & 1.7 & 9.6 & 2.3 \\
 &  $^{12}$CO & 1 & 2.4 & 6.2 & 1.8 \\
 &   & 2 & 57.9 & 9.4 & 2.5 \\ 
 &   & 3 & 7.2 & 12.4 & 1.5 \\ 
 &  $^{13}$CO & 1 & 1.0 & 7.3 & 2.0 \\
 &   & 2 & 18.8 & 9.4 & 1.5 \\
 &   & 3 & 2.0 & 10.2 & 3.4
 \label{Tab.M43-shell}
 \end{longtable}
 
\begin{table*}[!h]
\addtolength{\tabcolsep}{0pt}
\def\arraystretch{1.2}
\caption{Parameters and results of [${}^{13}$C\,{\sc ii}] analysis in M43.}
\begin{tabular}{l|ccccccccc}
\hline\hline
spectrum & $T_{\mathrm{P}}([{}^{12}\mathrm{C\,\textsc{ii}}])$ & $T_{\mathrm{P}}([{}^{13}\mathrm{C\,\textsc{ii}}])$ & $\Delta v$ & $T_{\mathrm{d}}$ & $\tau_{160}$ & $\tau_{[\mathrm{C\,\textsc{ii}}]}$ & $T_{\mathrm{ex}}$ & $N_{\mathrm{C^+}}$ & $N_{\mathrm{C^+}}/\mathcal{A}_{\mathrm{C}}/N_{\mathrm{H}}$\\
& [K] & [K] & [$\mathrm{km\,s^{-1}}$] & [K] & & & [K] & [$\mathrm{cm^{-2}}$] & \\ \hline
1 & 74 & 2.7 & 3.9 & 37 & $1\times 10^{-2}$ & 3.8 & 124 & $7.7\times 10^{18}$ & 0.61 ($A_{\mathrm{V,los}}\simeq 25$) \\
2 & 83 & 2.1 & 3.6 & 36 & $2\times 10^{-2}$ & 2.5 & 139 & $5.2\times 10^{18}$ & 0.36 ($A_{\mathrm{V,los}}\simeq 15$) \\
3 & 67 & 1.0 & 3.7 & 38 & $9\times 10^{-3}$ & 1.1 & 149 & $2.6\times 10^{18}$ & 0.30 ($A_{\mathrm{V,los}}\simeq 10$) \\
5 & 21 & 2.1 & 1.7 & 37 & $5\times 10^{-3}$ & 11 & 64 & $5.0\times 10^{18}$ & 1.03 ($A_{\mathrm{V,los}}\simeq 15$) \\
6 & 72 & 1.7 & 3.5 & 34 & $1\times 10^{-2}$ & 2.2 & 127 & $4.2\times 10^{18}$ & 0.36 ($A_{\mathrm{V,los}}\simeq 15$) \\ \hline
\end{tabular}
\tablefoot{Line fit uncertainties are of the order $\sigma_{T_{\mathrm{P}}}\simeq 0.5\,\mathrm{K}$.}
\label{Tab.13CII-M43}
\end{table*}

\clearpage

\begin{longtable}{llcccc}
\caption{Line fits to spectra 1-6 in NGC 1977.}\\
\hline\hline
spectrum & line & comp. & $T_{\mathrm{P}}$ & $v_{\mathrm{P}}$ & $\Delta v_{\mathrm{FWHM}}$ \\
 &  & \# & [K] & [$\mathrm{km\,s^{-1}}$] & [$\mathrm{km\,s^{-1}}$] \\ \hline
 \endfirsthead
 \caption{Continued.}\\
 \hline\hline
spectrum & line & comp. & $T_{\mathrm{P}}$ & $v_{\mathrm{P}}$ & $\Delta v_{\mathrm{FWHM}}$ \\
 &  & \# & [K] & [$\mathrm{km\,s^{-1}}$] & [$\mathrm{km\,s^{-1}}$] \\ \hline
 \endhead
 \hline\\
\multicolumn{6}{c}{\parbox{\LTcapwidth}{{\bf Notes:} Line fit uncertainties are of the order $\sigma_{T_{\mathrm{P}}}\simeq 0.5\,\mathrm{K}$, $\sigma_{v_{\mathrm{P}}}\simeq 0.1\,\mathrm{km\,s^{-1}}$, and $\sigma_{v_{\mathrm{FWHM}}}\simeq 0.1\,\mathrm{km\,s^{-1}}$.}}\\
\endlastfoot 
1 &  [C\,{\sc ii}] & 1 & 141.2 & 11.7 & 2.9 \\
 &   & 2 & -101.1 & 12.0 & 2.2 \\  
 &  \text{[$^{13}$C\,{\sc ii}]} & 1 & 3.0 & 11.6 & 2.0 \\
 &  $^{12}$CO & 1 & 15.6 & 11.7 & 2.1 \\ 
 &   & 2 & 33.2 & 12.7 & 1.4 \\ 
 &  $^{13}$CO & 1 & 5.0 & 11.1 & 1.0 \\
 &   & 2 & 20.2 & 12.6 & 1.1 \\ \hline
2 &  [C\,{\sc ii}] & 1 & (1.0) & (8.0) & (2.0) \\
 &   & 2 & 32.8 & 11.4 & 2.6 \\
 &  $^{12}$CO & 1 & 0.3 & 8.0 & 1.5 \\ 
 &   & 2 & 1.0 & 11.0 & 1.1 \\ 
 &   & 3 & 1.1 & 11.5 & 3.0 \\ 
 &  $^{13}$CO & 1 & 0.09 & 8.7 & 2.6 \\ 
 &   & 2 & 0.38 & 11.1 & 1.2 \\
 &   & 3 & 0.22 & 12.7 & 1.2 \\ \hline
3 &   [C\,{\sc ii}] & 1 & 12.7 & 11.2 & 1.6 \\
 &   & 2 & 27.4 & 12.7 & 1.9 \\
 &  $^{12}$CO & 1 & 0.61 & 11.7 & 3.1 \\ 
 &   & 2 & 0.52 & 12.6 & 1.5 \\ 
 &  $^{13}$CO & 1 & 0.14 & 11.1 & 2.4 \\
 &   & 2 & 0.12 & 12.7 & 0.7 \\ \hline 
4 &  [C\,{\sc ii}] & 1 & 11.1 & 11.6 & 1.3 \\
 &   & 2 & 19.0 & 12.7 & 2.0 \\ \hline
5 &  [C\,{\sc ii}] & 1 & 2.1 & 11.6 & 5.1 \\
 &   & 2 & 3.8 & 13.7 & 1.5 \\ \hline
6 &  [C\,{\sc ii}] & 1 & 22.8 & 11.6 & 3.1 \\
 &   & 2 & 5.2 & 12.7 & 1.0
\label{Tab.NGC1977-shell}
\end{longtable}

\begin{table*}[!h]
\addtolength{\tabcolsep}{0pt}
\def\arraystretch{1.2}
\caption{Parameters and results of [${}^{13}$C\,{\sc ii}] analysis in NGC 1977.}
\begin{tabular}{l|ccccccccc}
\hline\hline
spectrum & $T_{\mathrm{P}}([{}^{12}\mathrm{C\,\textsc{ii}}])$ & $T_{\mathrm{P}}([{}^{13}\mathrm{C\,\textsc{ii}}])$ & $\Delta v$ & $T_{\mathrm{d}}$ & $\tau_{160}$ & $\tau_{[\mathrm{C\,\textsc{ii}}]}$ & $T_{\mathrm{ex}}$ & $N_{\mathrm{C^+}}$ & $N_{\mathrm{C^+}}/\mathcal{A}_{\mathrm{C}}/N_{\mathrm{H}}$\\
& [K] & [K] & [$\mathrm{km\,s^{-1}}$] & [K] & & & [K] & [$\mathrm{cm^{-2}}$] & \\ \hline
1 & 141 & 3.0 & 2.9 & 31 & $1\times 10^{-2}$ & 2.0 & 211 & $5.2\times 10^{18}$ & 0.42 ($A_{\mathrm{V,los}}\simeq 15$) \\ \hline 
\end{tabular}
\tablefoot{Line fit uncertainties are of the order $\sigma_{T_{\mathrm{P}}}\simeq 0.5\,\mathrm{K}$.}
\label{Tab.13CII-NGC1977}
\end{table*}

\clearpage

\begin{longtable}{llcccc}
\caption{Line fits to spectra 1-10 in the southeast of the Orion Bar.}\\
\hline\hline
spectrum & line & comp. & $T_{\mathrm{P}}$ & $v_{\mathrm{P}}$ & $\Delta v_{\mathrm{FWHM}}$ \\
 &  & \# & [K] & [$\mathrm{km\,s^{-1}}$] & [$\mathrm{km\,s^{-1}}$] \\ \hline
 \endfirsthead
 \caption{Continued.}\\
 \hline\hline
spectrum & line & comp. & $T_{\mathrm{P}}$ & $v_{\mathrm{P}}$ & $\Delta v_{\mathrm{FWHM}}$ \\
 &  & \# & [K] & [$\mathrm{km\,s^{-1}}$] & [$\mathrm{km\,s^{-1}}$] \\ \hline
 \endhead
 \hline\\
\multicolumn{6}{c}{\parbox{\LTcapwidth}{{\bf Notes:} Line fit uncertainties are of the order $\sigma_{T_{\mathrm{P}}}\simeq 0.5\,\mathrm{K}$, $\sigma_{v_{\mathrm{P}}}\simeq 0.1\,\mathrm{km\,s^{-1}}$, and $\sigma_{v_{\mathrm{FWHM}}}\simeq 0.1\,\mathrm{km\,s^{-1}}$.}}\\
\endlastfoot
1 &  [C\,{\sc ii}] & 1 & 35.3 & 9.1 & 5.1 \\
 &   & 2 & 117.8 & 10.9 & 3.0 \\
 &  \text{[$^{13}$C\,{\sc ii}]} & 2 & 3.4 & 11.0 & 2.5 \\
 &  $^{12}$CO & 1 & 115.6 & 9.9 & 4.0 \\
 &   & 2 & 35.3 & 11.3 & 1.9 \\ 
 &  $^{13}$CO & 1 & 7.6 & 9.9 & 4.2 \\
 &   & 2 & 35.3 & 10.1 & 2.1 \\ \hline
2 &  [C\,{\sc ii}] & 1 & 52.1 & 9.9 & 4.4 \\
 &   & 2 & 62.3 & 11.5 & 2.6 \\ 
 &  \text{[$^{13}$C\,{\sc ii}]} & 2? & 3.0 & 10.6 & 1.9 \\
 &  $^{12}$CO & 1 & 93.5 & 9.5 & 3.7 \\
 &   & 2 & 23.0 & 10.9 & 2.2 \\
 &  $^{13}$CO & 1 & 1.4 & 7.0 & 1.2 \\
 &   & 2 & 27.4 & 9.8 & 2.4 \\ \hline
3 &  [C\,{\sc ii}] & 1 & 54.0 & 10.6 & 4.0 \\
 &   & 2 & 30.4 & 12.0 & 2.0 \\
 &  $^{12}$CO & 1 & 55.3 & 9.0 & 2.1 \\
 &   & 2 & 49.0 & 10.9 & 2.3 \\
 &  $^{13}$CO & 1 & 8.0 & 8.9 & 1.2 \\ 
 &   & 2 & 10.1 & 10.6 & 2.0 \\ \hline 
4 &  [C\,{\sc ii}] & 1 & 3.1 & 2.2 & 15.9 \\
 &   & 2 & 61.7 & 11.1 & 3.1 \\
 &  $^{12}$CO & 1 & 50.6 & 10.0 & 3.6 \\
 &   & 2 & 12.0 & 11.6 & 1.8 \\
 &  $^{13}$CO & 1 & 6.4 & 10.2 & 3.2 \\ 
 &   & 2 & 4.2 & 11.3 & 1.5 \\ \hline 
5 &  [C\,{\sc ii}] & 1 & 2.3 & -0.9 & 16.2 \\
 &   & 2 & 56.4 & 10.9 & 2.9 \\
 &  $^{12}$CO & 1 & 27.7 & 8.9 & 4.3 \\
 &   & 2 & 24.5 & 11.5 & 2.5 \\
 &  $^{13}$CO & 1 & 3.2 & 9.4 & 4.3 \\ 
 &   & 2 & 2.2 & 11.7 & 2.0 \\ \hline 
6 &  [C\,{\sc ii}] & 1 & 1.8 & -0.6 & 19.6 \\
 &   & 2 & 28.7 & 10.3 & 3.7 \\
 &  $^{12}$CO & 1 & 7.1 & 7.2 & 2.3 \\
 &   & 2 & 25.8 & 10.6 & 4.5 \\
 &  $^{13}$CO & 1 & 0.6 & 6.9 & 2.1 \\ 
 &   & 2 & 2.3 & 10.4 & 3.7 \\ \hline 
7 &  [C\,{\sc ii}] & 1 & 4.4 & 7.5 & 3.2 \\ 
 &   & 2 & 19.7 & 10.5 & 2.7 \\ 
 &   & 3 & 2.3 & 14.3 & 1.1 \\ 
 &  $^{12}$CO & 1 & 7.0 & 6.9 & 2.1 \\
 &   & 2 & 28.6 & 10.4 & 2.6 \\
 &   & 3 & 24.2 & 14.4 & 4.3 \\
 &  $^{13}$CO & 1 & 0.8 & 7.0 & 1.9 \\ 
 &   & 2 & 2.0 & 10.3 & 2.9 \\ 
 &   & 3 & 2.7 & 14.8 & 2.9 \\ \hline  
8 &  [C\,{\sc ii}] & 1 & 14.1 & 8.3 & 4.8 \\ 
 &   & 2 & 13.1 & 10.7 & 1.8 \\ 
 &   & 3 & (1.0) & (13.0) & (4.0) \\ 
 &  $^{12}$CO & 1 & 7.3 & 7.7 & 3.0 \\
 &   & 2 & 45.8 & 10.8 & 1.9 \\
 &   & 3 & 21.4 & 13.1 & 3.5 \\
 &  $^{13}$CO & 1 & 0.8 & 7.8 & 2.9 \\ 
 &   & 2 & 7.7 & 10.9 & 1.5 \\ 
 &   & 3 & 2.5 & 13.5 & 2.6 \\ \hline 
9 &  [C\,{\sc ii}] & 1 & 9.5 & 5.9 & 7.8 \\
 &   & 2 & 26.7 & 6.1 & 1.8 \\ 
 &   & 3 & 49.6 & 8.0 & 1.7 \\ 
 &  \text{[$^{13}$C\,{\sc ii}]} & 2/3 & 2.3 & 6.8 & 4.1 \\
 &  $^{12}$CO & 1 & 30.6 & 7.2 & 1.6 \\
 &   & 2 & 4.8 & 10.6 & 4.8 \\
 &  $^{13}$CO & 1 & 6.9 & 7.3 & 1.1 \\ 
 &   & 2 & 0.5 & 9.7 & 6.0 \\ \hline 
10 &  [C\,{\sc ii}] & 1 & 20.2 & 7.0 & 4.5 \\
 &   & 2 & 26.4 & 8.0 & 1.8 \\
 &  \text{[$^{13}$C\,{\sc ii}]} & 2 & 3.0 & 7.9 & 1.6 \\
 &  $^{12}$CO & 1 & 66.0 & 7.3 & 1.8 \\ 
 &   & 2 & 4.2 & 11.1 & 3.0 \\
 &   & 1 & 23.1 & 7.4 & 1.3 \\ 
 &  $^{13}$CO & 2 & 0.6 & 10.9 & 2.6
\label{Tab.behind-Bar}
\end{longtable}

\begin{table*}[!h]
\addtolength{\tabcolsep}{0pt}
\def\arraystretch{1.2}
\caption{Parameters and results of [${}^{13}$C\,{\sc ii}] analysis of spectra to the southeast of the Orion Bar.}
\begin{tabular}{l|ccccccccc}
\hline\hline
spectrum & $T_{\mathrm{P}}([{}^{12}\mathrm{C\,\textsc{ii}}])$ & $T_{\mathrm{P}}([{}^{13}\mathrm{C\,\textsc{ii}}])$ & $\Delta v$ & $T_{\mathrm{d}}$ & $\tau_{160}$ & $\tau_{[\mathrm{C\,\textsc{ii}}]}$ & $T_{\mathrm{ex}}$ & $N_{\mathrm{C^+}}$ & $N_{\mathrm{C^+}}/\mathcal{A}_{\mathrm{C}}/N_{\mathrm{H}}$\\
& [K] & [K] & [$\mathrm{km\,s^{-1}}$] & [K] & & & [K] & [$\mathrm{cm^{-2}}$] & \\ \hline
1 & 118 & 3.4 & 2.5 & 38 & $3.0\times 10^{-2}$ & 2.9 & 175 & $5.4\times 10^{18}$ & 0.32 ($A_{\mathrm{V,los}}\simeq 15$) \\
2 & 62 & 1.9 & 2.9 & 37 & $1.8\times 10^{-2}$ & 3.1 & 113 & $4.3\times 10^{18}$ & 0.25 ($A_{\mathrm{V,los}}\simeq 15$) \\
9 & 50 & 1.8 & 1.7 & 33 & $0.9\times 10^{-2}$ & 3.8 & 95 & $2.6\times 10^{18}$ & 0.30 ($A_{\mathrm{V,los}}\simeq 10$) \\
10 & 26 & 1.1 & 1.8 & 30 & $1.0\times 10^{-2}$ & 4.4 & 67 & $2.2\times 10^{18}$ & 0.23 ($A_{\mathrm{V,los}}\simeq 10$) \\ \hline
\end{tabular}
\tablefoot{Line fit uncertainties are of the order $\sigma_{T_{\mathrm{P}}}\simeq 0.5\,\mathrm{K}$.}
\label{Tab.13CII-behind-Bar}
\end{table*}

\end{document}